\documentclass[12pt,twoside,english,a4paper]{article}
\usepackage{babel,amssymb,amsthm}
\usepackage{graphics}
\usepackage{amssymb,amsthm}
\usepackage{graphicx}
\usepackage{amsmath}
\usepackage{mathrsfs}
\usepackage{bm}
\usepackage[lmargin=1in,rmargin=1in,tmargin=1in]{geometry}
\usepackage{multicol}
\usepackage{bbold}
\usepackage{listings}
\usepackage{subfig}
\usepackage[show]{ed}

\usepackage{hyperref}
\hypersetup{colorlinks=true}
\hypersetup{citecolor=blue}

\newcommand{\diag}[1]{\mathbf{diag}\big(#1\big)}

\title{Pseudo-spectral collocation with Maxwell polynomials for kinetic equations with energy diffusion}
\author{Tonatiuh S\'anchez-Vizuet\footnote{Courant Institute of Mathematical Sciences, New York University. {\tt tonatiuh@cims.nyu.edu } }
   \;  and Antoine J. Cerfon\footnote{Courant Institute of Mathematical Sciences, New York University. {\tt cerfon@cims.nyu.edu } } }

\begin{document}
\maketitle

\begin{abstract}\noindent  We study the approximation and stability properties of a recently popularized discretization strategy for the speed variable in kinetic equations, based on pseudo-spectral collocation on a grid defined by the zeros of a non-standard family of orthogonal polynomials called Maxwell polynomials. Taking a one-dimensional equation describing energy diffusion due to Fokker-Planck collisions with a Maxwell-Boltzmann background distribution as the test bench for the performance of the scheme, we find that Maxwell based discretizations outperform other commonly used schemes in most situations, often by orders of magnitude. This provides a strong motivation for their use in high-dimensional gyrokinetic simulations. However, we also show that Maxwell based schemes are subject to a non-modal time stepping instability in their most straightforward implementation, so that special care must be given to the discrete representation of the linear operators in order to benefit from the advantages provided by Maxwell polynomials.

\end{abstract}

\noindent
{\bf Key words}: Pseudo-spectral methods, Fokker-Planck collisions, Kinetic calculations, Orthogonal polynomials.



\section{Introduction}
The high dimensionality of kinetic equations in plasma physics imposes strong constraints on continuum numerical solvers. For computationally tractable simulations, the number of grid points for each dimension is often limited to a very small number. In this context, high order discretization schemes are appealing, as they can lead to accurate representations even for small grid sizes. This aspect has long been recognized regarding the discretization of the spatial variables, for which Fourier based representations are commonly used. However, the plasma physics community has only started exploring the potential of high order schemes for the velocity variables recently. Specifically, Bratanov \textit{et al.} \cite{Bratanov} demonstrated that a truncated Hermite representation was superior to the more common finite difference schemes for the discretization of the parallel velocity. For the perpendicular velocity -- or the speed variable depending on the choice of the coordinate system for velocity space -- Landreman and Ernst \cite{ErLa:2013} rediscovered a family of non-classical polynomials often referred to as Maxwell polynomials, velocity polynomials, or speed polynomials \cite{CoSh:1963,CoPe:1965,ClSh:1993}, which are known to be particularly well suited for the accurate computation of the moments of distribution functions that are close to Maxwellian distributions \cite{Shizgal:1981}. They demonstrated that Maxwell polynomials are not only the method of choice for the accurate evaluation of integrals of the distribution function, but are also superior to the other discretization schemes for the calculation of derivatives of the distribution function, as long as the distribution function is well approximated by a function with a Maxwellian envelope. This is critical for simulations involving Fokker-Planck collisions, since the collision operator often has second order derivatives with respect to the speed variable \cite{Helander:2002}.

Landreman and Ernst limited their applications to time-independent drift kinetic equations, for neoclassical calculations of the bootstrap current and plasma flows in magnetic confinement devices \cite{ErLa:2013}. However, Maxwell polynomials were long known to also have favorable features for \textit{time-dependent} kinetic simulations with collisions \cite{Shizgal:1979,ShBl:1985}. In the context of kinetic simulations for plasma physics, this was confirmed by a numerical study of the performance of Maxwell polynomials for a simple one-dimensional partial differential equation describing energy diffusion in velocity space due to Fokker-Planck collisions with a Maxwellian distribution \cite{CeLaWi:2015}, and Maxwell polynomials defined on a truncated domain are used in a pseudospectral collocation scheme in the five-dimensional collisional gyrokinetic code CGYRO \cite{BeBrCa:2016}. 

In \cite{CeLaWi:2015}, the authors focused on a model one-dimensional diffusion equation because of its relevance to kinetic computations for plasma physics, and because they were able to construct a semi-analytic solution to the equation which can be evaluated at arbitrary points to arbitrary precision \cite{CeWi:2015}. This solution method thus served as benchmark for in depth accuracy studies, and highlighted unique features of these non classsical polynomials. However, the authors only analyzed the performance of Maxwell polynomials in a spectral scheme of the Galerkin type, which is traditionally challenging to implement in high dimensional kinetic simulations. Furthermore, the authors were able to integrate the equation exactly in time, and therefore put aside any issue associated with numerical time stepping. The purpose of the present work is to close this gap and characterize the strengths and limitations of Maxwell polynomials in a context that is directly relevant to kinetic simulations in plasma physics, namely a velocity discretization scheme based on a collocation framework, and advancing the solution in time with standard time stepping schemes.

The relative simplicity of the energy diffusion equation we focus on in this work allows us to investigate the key properties of discretization schemes based on Maxwell polynomials in more detail than is possible in multi-dimensional implementations of these schemes \cite{BeBrCa:2016,BeCa:2017}. In particular, we are able to consider grid sizes that are much larger than is currently feasible in five-dimensional simulations, to study various time-stepping schemes, and to analyze the spectral properties of the discretized operators. Our main findings are as follows: 1) the most straightforward pseudo-spectral collocation implementation of the energy diffusion operator with Maxwell polynomials can be subject to a non-modal instability which leads to bounded but significant numerical errors; 2) the schemes based on Maxwell polynomials we designed to avoid this non-modal instability outperform the more standard, non-Maxwell schemes in almost all cases, often by orders of magnitude; 3) the situation for which Maxwell based discretizations are not optimal corresponds to an initial condition with a singularity at the origin, for which we find a Chebyshev based scheme to be superior. 

The structure of the article is as follows. In Section \ref{sec:Preliminaries} we give a brief presentation of Maxwell polynomials, of the general concept of pseudo-spectral collocation and of other implementation-related considerations. Section \ref{sec:validation} introduces the model equation we will solve with our collocation schemes, and presents the results of the numerical experiments for the methods considered (Maxwell, Chebyshev, modified GS2 and a pseudo-spectral method based on the modified GS2 grid). In Section \ref{sec:stability} we discuss the stability properties of the time discretization of the discrete diffusion operator collocated on Maxwell nodes and the role played by the epsilon-pseudo spectra of the resulting matrix. We highlight a non-modal instability observed for the most direct discrete representation of the diffusion operator, and present a discrete operator which is not subject to the instability. We summarize our findings in Section \ref{sec:conclusions} and suggest directions for future work.

\section{Pseudo-spectral collocation schemes with Maxwell polynomials}\label{sec:Preliminaries}

\subsection{Maxwell Polynomials. \\}\label{sec:MaxwellReview}
We will use the term ``Maxwell polynomials" to denote the non-classical family of real-valued polynomials $p_n(x)$ over the positive real line defined to be orthogonal with respect to the inner product
\begin{equation}\label{eq:inner}
\langle p_n,p_m\rangle_w := \int_0^\infty p_n(x)\, p_m(x) w(x)\, dx, \qquad w(x): = x^2 e^{-x^2},
\end{equation}
which induces the natural norm
\[
\|f\|^2_w := \langle f,f\rangle_w.
\]
The weight $w$ for the inner product defining this family of polynomials was originally introduced as a natural choice for Gaussian quadrature rules tailored for the computation of moments of Maxwellian-like distribution functions \cite{CoSh:1963,CoPe:1965,Shizgal:1981}. Later work has demonstrated the desirable properties of these polynomials in much more general settings, involving differentiation and time evolution \cite{ErLa:2013,Shizgal:1979,ShBl:1985,CeLaWi:2015}. 

\begin{figure}[tb]
\begin{tabular}{ccc}
\includegraphics[width=.31\linewidth]{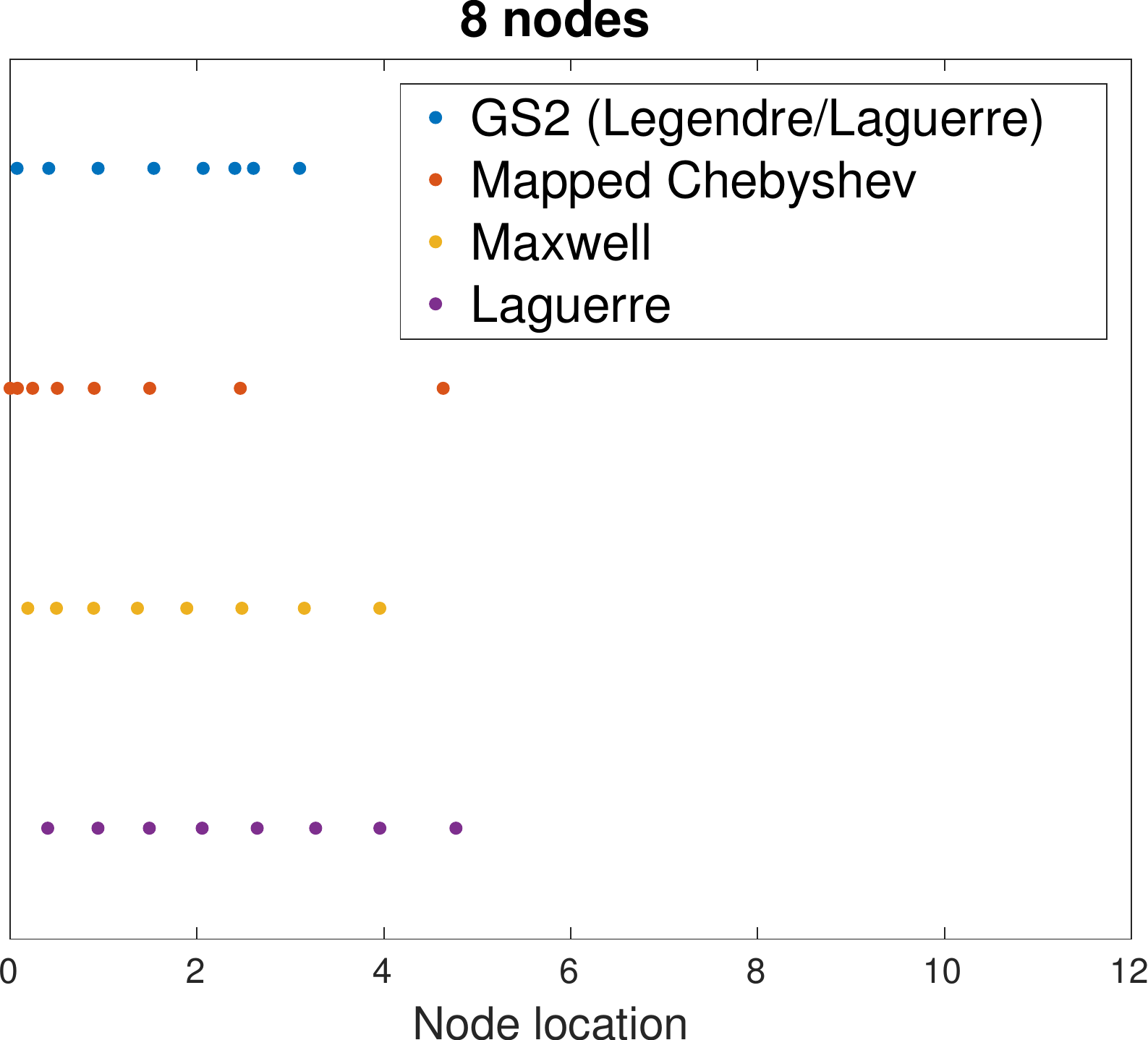} &
\includegraphics[width=.31\linewidth]{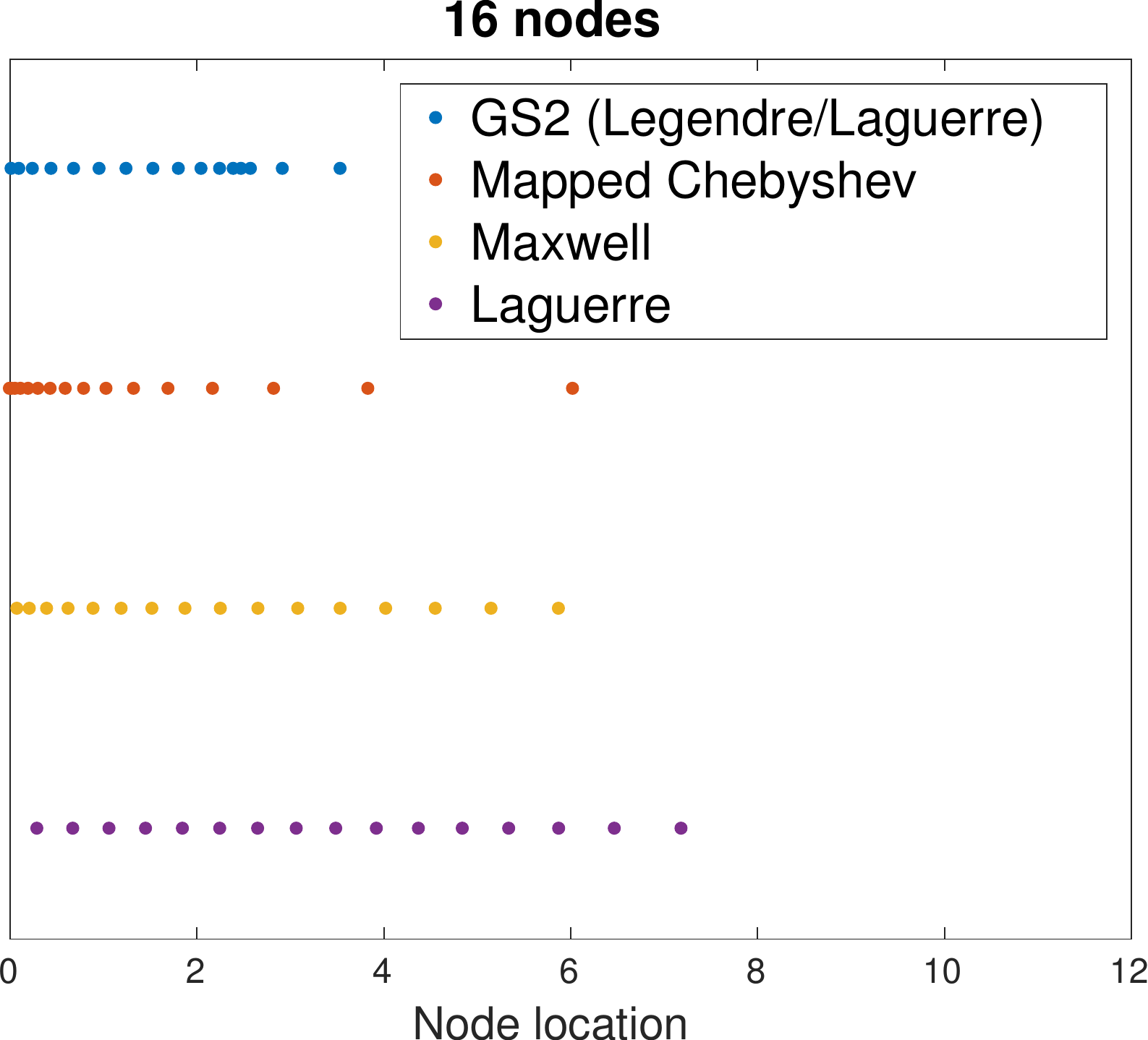}  &
\includegraphics[width=.31\linewidth]{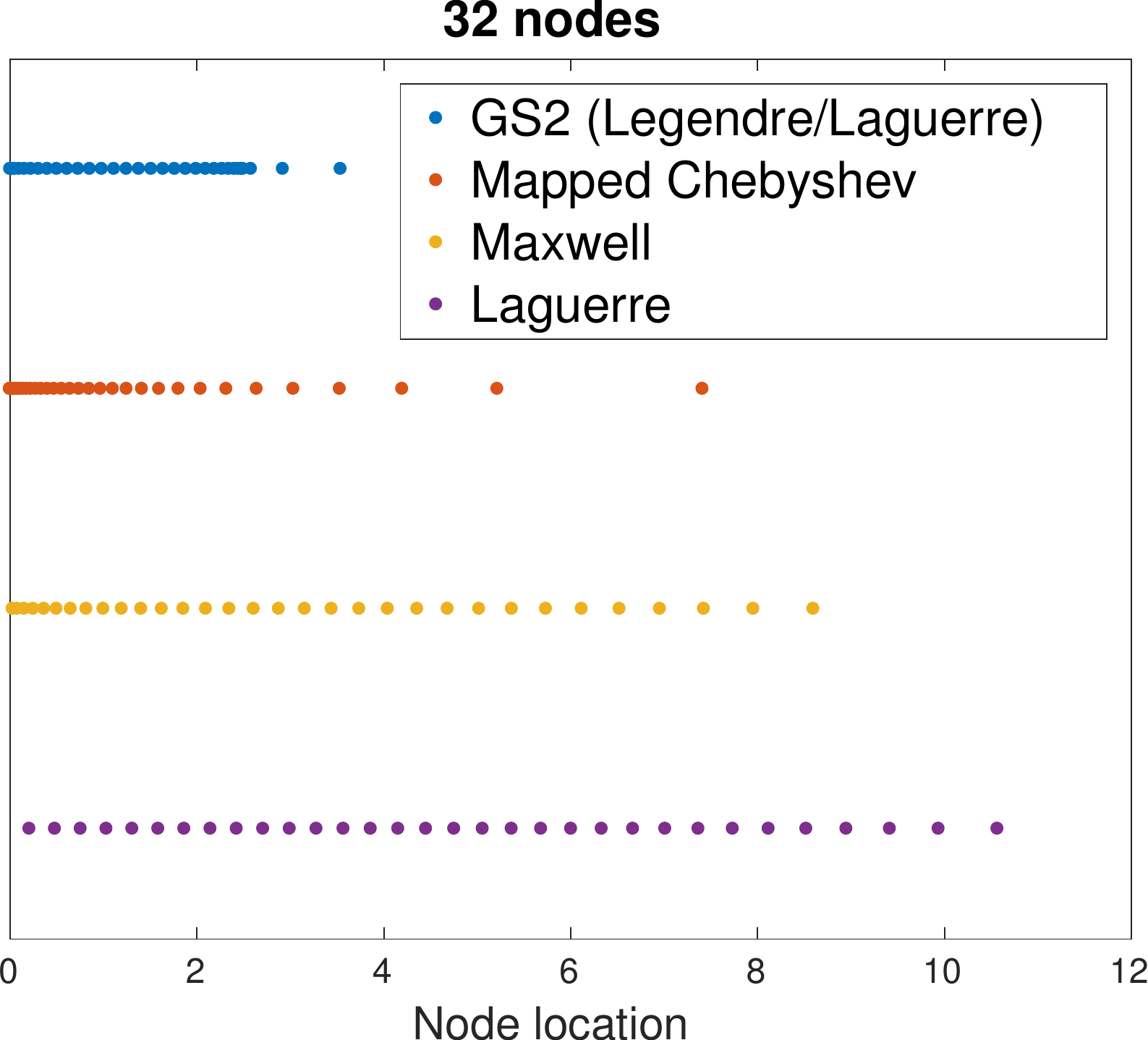}
\end{tabular}
\caption{{\scriptsize Computational grids based on the locations of the roots of different families of orthogonal polynomials: (top) Grid used in AstroGK and GS2, which combines the zeros of Legendre polynomials on the interval $[0,2.5]$ and the square roots of the zeros of Laguerre polynomials over $[2.5,\infty)$; (second from top) first-kind Chebyshev nodes over the interval $s\in[0,1]$ mapped to the half line with the mapping $x=-\log(1-s)$; (third from top) roots of the Maxwell polynomials defined by the weight $w(x)=x^2e^{-x^2}$; (bottom) square roots of the zeros of Laguerre polynomials.}}\label{fig:grids}
\end{figure}

Among the many features that make these polynomials desirable for pseudo-spectral discretization schemes for the speed variable, one may emphasize their ability to handle semi-infinite intervals, the convenient distribution of their roots that balances the clustering around zero, where increased resolution is often needed \cite{ErLa:2013}, with the sampling at increasingly larger distances from the origin, and their optimal location for the computation of integrals involving a Maxwell-Boltzmann distribution. Figure \ref{fig:grids} shows the grids obtained from the roots of Maxwell polynomials for 8, 16 and 32 nodes, and compares them with three grids which may also be considered for the discretization of the semi-infinite interval $[0,\infty)$: the grid used in the gyrokinetic code GS2 \cite{GS2:2009} and AstroGK \cite{NumataAstroGK:2010}, which combines the zeros of Legendre polynomials on the interval $[0,2.5]$ and the zeros of Laguerre polynomials mapped to the interval $[2.5,\infty)$ through the mapping $x_i \mapsto \sqrt{x_i +2.5^2}$ \cite{ErLa:2013,CeLaWi:2015}, a grid based on the zeros of the Chebyshev polynomials of the first kind, mapped from the interval $[0,1]$ to $[0,\infty)$ by the mapping $x=-\log(1-s)$, and a grid based on the square roots of the zeros of Laguerre polynomials. The grid sizes chosen for Figure \ref{fig:grids} correspond to standard grid sizes for the speed variable in kinetic simulations in plasma physics.   

As with any family of orthogonal polynomials, the entire monic sequence can be generated from the three-term recurrence relation 
\begin{subequations}\label{eq:recurrence}
\begin{align}
p_{-1}(x) =\,& 0, \\
p_0(x)    =\,& 1,    \\
p_{n+1}(x) =\,&(x-a_n)p_n(x) - b_np_{n-1}(x), \label{eq:recurrencec}
\end{align}
where the coefficients of the relation for $n=1,2,3\ldots$ are given by
\begin{equation}
\label{eq:recurrenced}
a_n = \frac{\langle x p_n,p_n\rangle_w}{\|p_n\|_w^2}, \qquad b_n = \frac{\|p_n\|_w^2}{\|p_{n-1}\|_w^2} \quad (b_0 := \|1\|_w^2).
\end{equation}
\end{subequations}
The direct implementation of this recursion, known as the Stieltjes method, is prone to underflow/overflow issues when the values of $\|p_n\|_w^2$ underflow/overflow the weights of the quadrature rule used to approximate the norms \cite{Gautschi:2004}. This difficulty can be mitigated by working with a normalized sequence of polynomials instead, as was done in \cite{CeLaWi:2015}, or by computing in extended precision arithmetic, as in \cite{BeBrCa:2016}. Other alternatives are algorithms like Lanczos and Stieltjes-Gautschi that avoid a direct implementation of the recurrence and address the problem of finding the coefficients indirectly by recasting it as a matrix factorization or an eigenvalue problem \cite{Gautschi:2004,Gautschi:1994}. We empirically observe that these difficulties occur for polynomials of degree 80 or higher. As we will show in this article, the particular appeal of Maxwell polynomials is that high accuracy is reached with polynomials with significantly lower degree ($n\leq 30$). That means that a simple implementation of \eqref{eq:recurrence} in double precision arithmetic is expected to be sufficient for a wide class of kinetic simulations. For simulations requiring higher resolution in the speed variable to resolve fine structures in velocity space \cite{WaSu:2006, TaEtAl:2009}, the method proposed in \cite{CeLaWi:2015} may be preferred.  
\subsection{Domain truncation.\\}
Given the rapid decay of the weight function in the inner product \eqref{eq:inner}, an alternative approach based on the truncation of the integration domain has been considered \cite{CeLaWi:2015,BeBrCa:2016}. The approximated polynomials are then defined to be orthogonal with respect to the inner product
\begin{equation}\label{eq:truncinner}
\langle p_{n,a},p_{m,a}\rangle_{w,\,a} := \int_0^a p_{n,a}(x)\, p_{m,a}(x) w(x)\, dx, \qquad w(x): = x^2 e^{-x^2},
\end{equation}
for an arbitrary value of $a$ that has been chosen to be as small as $a=\sqrt{8}$ in \cite{BeBrCa:2016} or $a=15$ in \cite{CeLaWi:2015}. The latter reference provides a very detailed comparison of the performance of the two approaches for the case of the diffusion equation \eqref{eq:model}, but focuses on a modal implementation which is often impractical for the nonlinear, high dimensional kinetic equations of plasma physics. In contrast, we will highlight in this work the differences between the two approaches for pseudo-spectral discretization schemes with a low number of grid points, which are currently more relevant to kinetic plasma simulations. 

Before we compare the performance of the two approaches in sections \ref{sec:ModelProblem} and \ref{sec:stability}, we can make the following general observations. First, the obvious appeal of domain truncation is that the inner products are defined on a fixed, finite interval. This simplifies the construction of the numerical quadratures, since the last quadrature interval can be treated in the same way as the other intervals, which is not the case for the ``true" polynomials defined on the interval $[0,\infty)$. 

\begin{figure}[tb]
\begin{tabular}{ccc}
\includegraphics[width=.31\linewidth]{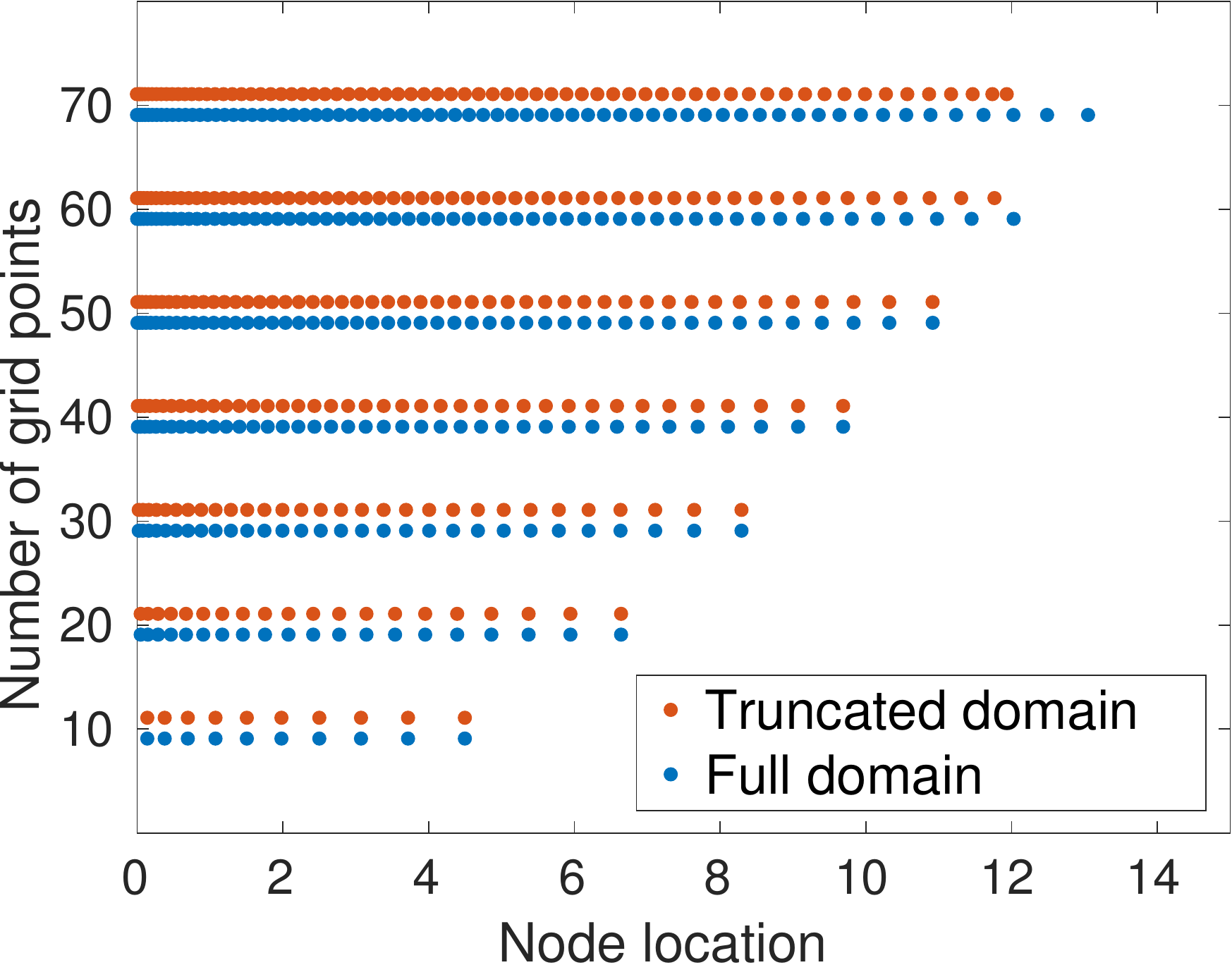} &
\includegraphics[width=.31\linewidth]{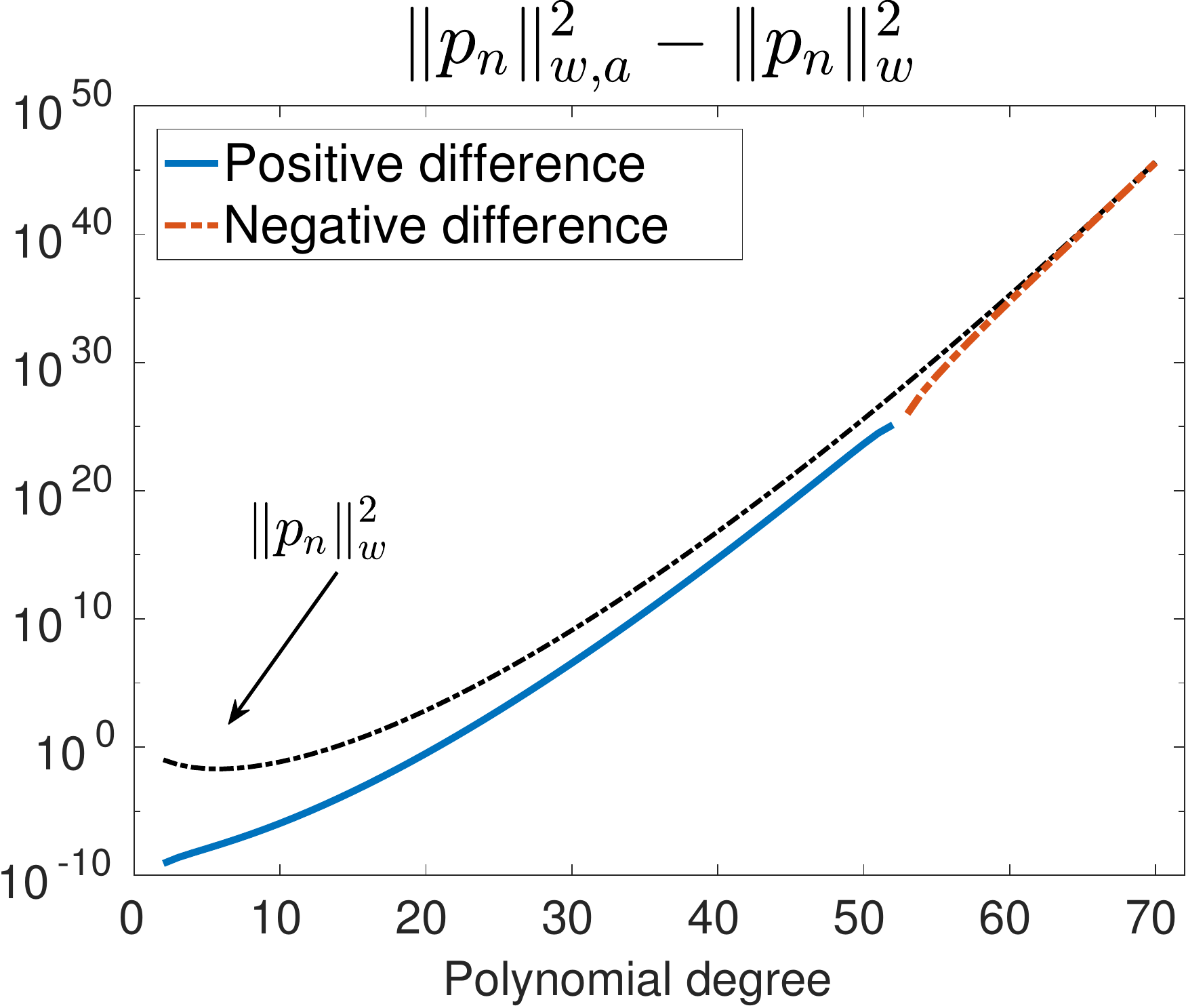} &
\includegraphics[width=.31\linewidth]{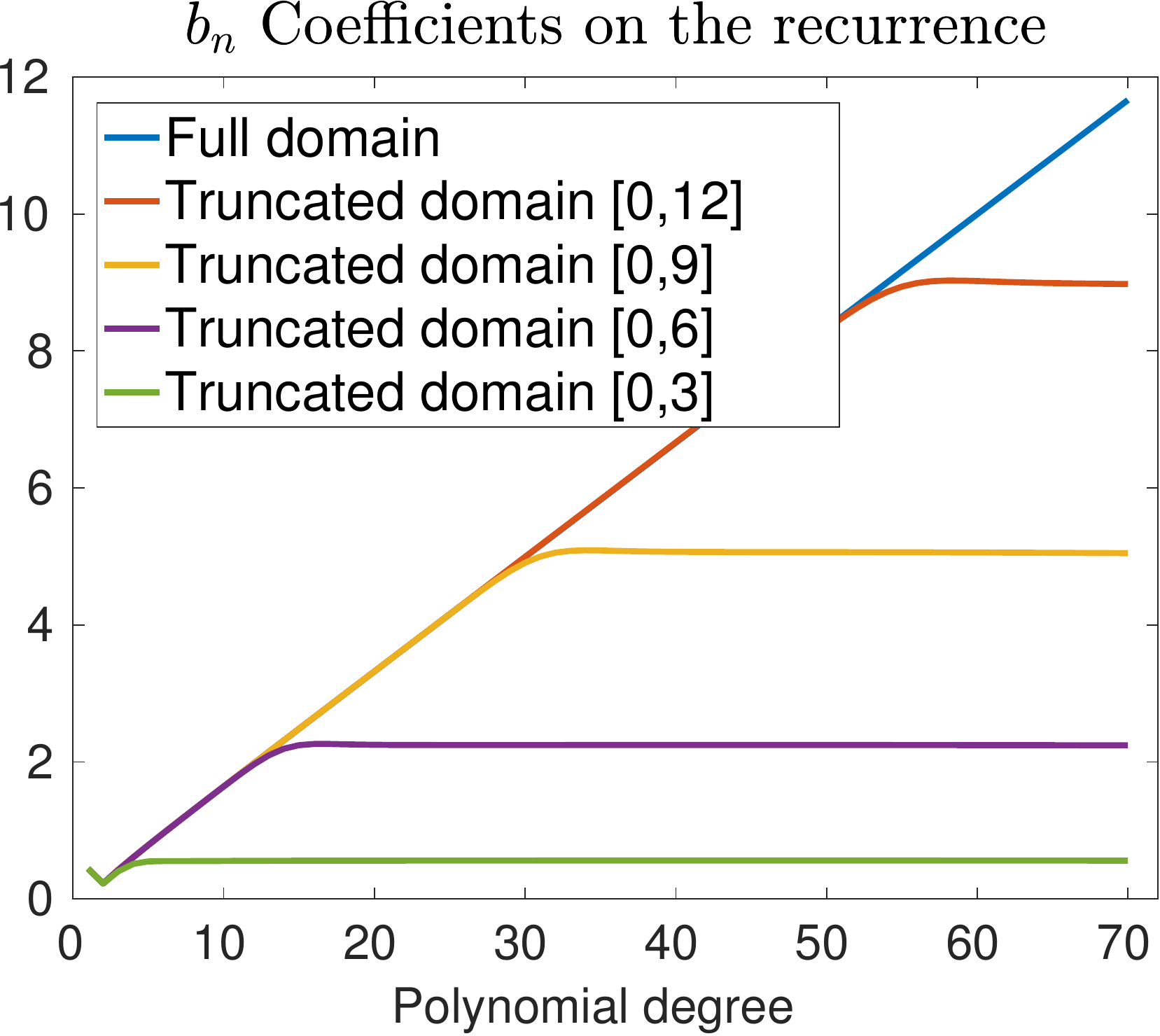}
\end{tabular}
\caption{{\scriptsize Left: Location of the grid points corresponding to the roots of the ``true" Maxwell polynomials, i.e. defined on the domain $[0,\infty)$ (blue) and those of the polynomials with the inner product truncated to $a=12$ (red). Center: Difference between the squared $L^2_{w,\,a}$ and $L^2_w$  norms of the monic polynomials for $a=12$ in the pre-asymptotic regime. The norm of the truncated polynomials is larger for degrees 52 and lower. The squared norm of the ``true" polynomials is shown in dashed line for comparison. Right: Growth of the $b_n$ coefficients of the recurrence relation for different values of the truncation point.}}\label{fig:FullVsTruncGrids}
\end{figure}

Second, since the truncated polynomials are restricted to $[0,a]$, the density of grid points on a fixed neighborhood of the origin increases more rapidly than that of the ``true" polynomials, whose largest root drifts away from the origin arbitrarily as the polynomial degree tends to infinity. As a consequence, domain truncation yields better results when resolving dynamics concentrated near the origin is critical \cite{CeLaWi:2015}. However, this advantage needs to be qualified in the context of kinetic simulations in plasma physics. Indeed, as we can see in Figure \ref{fig:FullVsTruncGrids}, which shows the grids for the two approaches, with the truncated domain chosen to be $[0,12]$, the locations of the nodes are virtually indistinguishable from one another until the polynomial degree becomes moderately large, $n\approx 60$. And we will show in the numerical experiments in this article that grids with such a large number of points may not be needed to achieve satisfactory performance.

In \cite{CeLaWi:2015}, it was shown that the polynomials on the truncated domain had the additional advantage that the coefficients $a_{n}$ and $b_{n}$ in the recurrence relation grow less rapidly than the coefficients for the true Maxwell polynomials. Specifically, the following asymptotic estimates were given:
\[
b_n=\frac{1}{6}(n+1) + \frac{17}{72}(n+1)^{-1}+\mathcal O(1/n^3) \quad \hbox{(Full)}, \qquad
b_{n,a} = c_a+\mathcal O(1/n^3) \quad \hbox{(Truncated)},
\]
where $c_a$ is a constant that depends only on the value of $a$ specifying the size of the truncated domain. Noting that $\|p_n\|_w^2 = \prod_{i=0}^{n}b_n$ for both the true and the truncated Maxwell polynomials, we have the following estimates for the growth of the $L^2$ norm of the monic polynomials:
\[
\|p_n\|_w^2 \sim (n/6)^n \quad \hbox{(Full)}, \qquad \|p_{n,a}\|_{w,\,a}^2 \sim c_a^n \quad \hbox{(Truncated)}.
\]
Observe the much faster growth of $\|p_n\|_w^2$ with $n$ than $\|p_{n,a}\|_{w,\,a}^2$. This rapid growth has important consequences in the ability to handle the computations in double precision arithmetic and is a compelling argument in favor of domain truncation if high polynomial degree is required. However, for small to moderately large polynomial degrees, the opposite holds: the norms of the full polynomials are in fact considerably smaller than the norms of the truncated polynomials. This is shown in Figure \ref{fig:FullVsTruncGrids}, which gives a comparison between the norms of the full and truncated polynomals, where the interval for truncation was chosen to be $[0,12]$. In that case, the norms of the full polynomial remain smaller until the degree has reached $n=52$.

The clear drawback of Maxwell polynomials defined on a truncated domain is that they are by definition unable to capture dynamical evolution beyond the end point $a$. That means that $a$ has to be chosen with a good a priori understanding of the dynamics of the equation under consideration. In the light of this, and the fact that the advantages of the truncated polynomials only become compelling for grid sizes well beyond the sizes that are currently realistic in kinetic solvers, we see Maxwell polynomials defined on the full open interval $[0,\infty)$ as more desirable for kinetic computations in the near future.
%
\subsection{Pseudo-spectral collocation.\\}\label{sec:Collocation}

\subsubsection{Quasi-polynomial interpolant}
Given a function $f$ and an appropriately chosen set of points $\{x_i\}_{i=1\ldots n+1}$, the idea behind pseudo-spectral methods is to approximate $f$ by a function $\pi$ that interpolates $f$ on the set of points, and use $\pi$ as a representation for $f$ over the entire domain of definition. Once the interpolant $\pi$ has been chosen, all the operations involving $f$ are performed on its interpolant and the result is used as an approximation to the exact computation. Depending on the choice of interpolation nodes and the smoothness of the function $f$, pseudo-spectral approximations tend to converge at faster rates than finite difference counterparts and provide a high order approximation over the entire interval on interest \cite{Fornberg:1996}.

The interpolant is often taken to be a polynomial of degree at most $n$ and the interpolation nodes are chosen to be the roots of another polynomial of degree $m>n$. However, for many plasma physics applications the function $f$ one solves for has a Maxwellian envelope: $f(x)\propto e^{-x^2}$ as $x\rightarrow\infty$. In this context, it is preferable to consider a slightly more general approach based on the quasi-polynomial interpolant \cite{Henrici:1982,WeRe:2000} given by
\begin{equation}\label{eq:interpolant}
\pi(x): = \sum_{j=1}^{n+1}\frac{\alpha(x)}{\alpha_j}L_j(x)f_j, \qquad \alpha_j:=\alpha(x_j)\qquad f_j:=f(x_j),
\end{equation}
where  the \textit{envelope} $\alpha$ is a smooth positive function, and $L_j$ is the usual Lagrange interpolating polynomial
\[
L_j(x):= \prod_{k\neq j}\frac{x-x_k}{x_j-x_k}, \qquad L_j(x_i) = \delta_{ji}.
\]
The use of an envelope different from the standard $\alpha(x)=1$ provides a way of including additional information for the global behavior of the function into the interpolant, if such information is available a priori. This results in a more efficient interpolant and greatly improves the approximation and convergence properties of the resulting scheme. If the solution $f$ of the equation is known to be well approximated by a function with a Maxwellian envelope, it is natural to take $\alpha(x)=e^{-x^2}$. This is precisely what we do to solve the model diffusion equation we present in Section \ref{sec:validation}.

As it has been pointed out \cite{BeTr:2004}, numerical interpolation can be stabilized and accelerated if the expression for $\pi$ is rearranged into what is known as the \textit{barycentric formula}
\begin{equation}\label{eq:barycentric}
\pi(x) = \frac{\alpha(x) \displaystyle\sum_{j=1}^{n+1}\frac{\lambda_j/\alpha_j }{x-x_j}f_j}{\displaystyle\sum_{j=1}^{n+1}\frac{\lambda_j}{x-x_j}}\,,\qquad \lambda_j :=\prod_{k\neq j}(x_j-x_k)^{-1} .
\end{equation}
The coefficients $\{\lambda_i\}_{i=1}^{n+1}$ are referred to as the \textit{barycentric weights} and depend only on the choice of interpolation nodes, hence can be precomputed once and reused as needed. We will show later that the process of calculating the barycentric weights $\lambda_i$ associated with the interpolation nodes for a Gaussian quadrature rule is closely related to that of finding the corresponding quadrature weights $w_i$. 

Once the barycentric interpolant has been determined, approximate differentiation of $f$ can be carried out by differentiating the interpolant, and evaluating its derivative at the interpolation nodes. This process can be written as a matrix-vector product
\[
\left(f^{(\ell)}(x_1),\ldots, f^{(\ell)}(x_{n+1})\right)^\top\approx \left(\pi^{(\ell)}(x_1),\ldots,\pi^{(\ell)}(x_{n+1})\right)^\top = \mathbf D^{(\ell)} \left(f_1,\ldots, f_{n+1}\right)^\top,
\]
where the matrix $\mathbf D^{(\ell)}$ is the $\ell$-th order \textit{spectral differentiation matrix}, whose $ij$-th entry consists of the $\ell$-th derivative of the $j$-th term in the sum \eqref{eq:interpolant} evaluated at the $i$-th interpolation node. Once the barycentric weights $\{\lambda_i\}_{i=1}^{n+1}$ have been determined, the entries of $\mathbf D^{(\ell)}$ can be computed from them through the following recursive procedure proposed by Welfert \cite{Welfert:1997}. The off-diagonal entries are given by
\begin{subequations}\label{eq:welfert}
\begin{alignat}{6}
\mathbf D^{(0)}_{ij} =\,& \mathbf \delta_{ij} ,& \\
\mathbf D^{(\ell)}_{ij} =\,& \frac{\ell}{x_i-x_j}\left(\frac{\lambda_j/\alpha_j}{\lambda_i/\alpha_i}\mathbf D_{ii}^{(\ell-1)}-\mathbf D_{ij}^{(\ell-1)}\right), & \quad i\neq j,
\end{alignat}
while the diagonal entries are generated by
\begin{alignat}{6}
\psi_{n,i}^{(\ell)}(x_i) = \,& \frac{\alpha^{(\ell)}(x_i)}{\alpha(x_i)} + \ell\!\!\!\sum_{m=1 (m\neq i)}^{n}{\!\!\! \frac{\psi_{m-1}^{(\ell-1)}(x_i)}{x_i-x_m} },& \\
\mathbf D^{(\ell)}_{ii}=\, &  \psi_{N,i}^{(\ell)}(x_i). &
\end{alignat}
\end{subequations}
This recursion is the procedure implemented by Weideman and Reddy in the package {\tt poldif.m} which we used to generate the differentiation matrices for our numerical experiments \cite{WeRe:2000}. It is interesting to note that the more straightforward recursion for the diagonal terms given by
\[
\mathbf D^{(\ell)}_{ii}= \frac{1}{\alpha_i}\left(\alpha^{(\ell)}(x_i)-\sum_{j=1}^{n+1}\widetilde{\mathbf D}^{(\ell)}_{ij}\alpha_j\right),
\]
leads to identical results for small values of $n$, but is strongly unstable as $n$ increases. 

%
\subsubsection{Computing the Gaussian nodes and weights \\}\label{sec:nodesnweights}
%
In order to be able to interpolate, differentiate and integrate using the pseudo-spectral methods described above, we must be able to determine the Gaussian nodes and weights associated with the Maxwell polynomials. As it was shown by Golub and Welsch \cite{GoWe:1969}, given the coefficients $(a_j,b_j)$ of the recurrence relation \eqref{eq:recurrencec}, it is possible to transform the problem of finding the Gaussian quadrature rule into an eigenvalue problem. If we denote the normalized polynomials by
\[
q_n(x) := \frac{p_n(x)}{\|p_n\|_w},
\]
then the system of equations described by \eqref{eq:recurrencec} can be recast in the matrix-vector form
\begin{equation}\label{eq:recurrmat}
x\mathbf q(x) = \mathbf J_{n} \mathbf q(x) + \sqrt{b_{n+1}}q_{n+1}(x)\mathbf e_{n},
\end{equation}
where
\begin{equation}\label{eq:jacmat}
\mathbf J_{n} = \left(\!\begin{array}{ccccc} a_0 & \sqrt{b_1}     &        &        & \\
				 \sqrt{b_1} & a_1     &  \sqrt{b_2}   &        & \\
				     &  \ddots & \ddots & \ddots & \\
				     &         & \sqrt{b_{n-1}} & a_{n-1} & \sqrt{b_{n}} \\
                                     &         &         & \sqrt{b_{n}} & a_n\end{array}\!\right), \; \mathbf q(x) = \left(\!\begin{array}{c}q_0(x)\\q_1(x) \\ \vdots \\ q_{n-1}(x)\\ q_{n}(x)\end{array}\!\right),\; \mathbf e_n=\left(\!\begin{array}{c}0\\0 \\ \vdots \\0\\ 1\end{array}\!\right). 				
\end{equation}
Since the Gaussian nodes of order $n$ are the values $x_i$ for which $p_{n+1}(x_i)=0$ we can substitute into \eqref{eq:recurrmat} to get
\begin{equation}\label{eq:gaussiannode}
x_i \mathbf q(x_i) = \mathbf J_{n}\mathbf q(x_i),
\end{equation}
Now, $\mathbf q(x_i)\neq\boldsymbol0$, since at least $q_0(x)$ is a non-zero constant, so we see that the Gaussian nodes are indeed eigenvalues of the so-called \textit{Jacobi matrix} $\mathbf J_n$. In fact, the solution of this eigenvalue problem yields the Gaussian and barycentric weights as well. Making use of the Christoffel-Darboux identity, it can be shown \cite[Sec. 2]{Wilf:1962} that the quadrature weights can be determined from the knowledge of the eigenvectors $\mathbf q(x_i)$ by the relation
\[
w_i = \left(\mathbf q(x_i)\cdot\mathbf q(x_i)\right)^{-1}.
\]
Moreover, if we denote by $\mathbf Q$ the matrix whose $i$-th column is $\mathbf q(x_i)$, then the quadrature weights are given by
\begin{equation}\label{eq:gaussianweight}
w_i = b_0(\mathbf e_1 ^\top \mathbf Q \cdot \mathbf e_i)^2.
\end{equation}
The Golub-Welsch algorithm exploits this facts in order to find both nodes and weights simultaneously through QR factorization of $\mathbf J_n$. Additionally, the Gaussian weights are connected to the barycentric weights through the relation
\begin{equation}\label{eq:weights}
\lambda_i =\frac{k_n}{\|p_n\|_w^2}p_n(x_i)w_i \quad (i=1,\ldots,n+1.),
\end{equation}
where $k_n$ is the leading coefficient of $p_n$ \cite[Eq. 2.10]{HuVaWa:2014}. Note that this formula requires the evaluation of $p_n(x)$ at the quadrature nodes and therefore requires $\mathcal O(n)$ operations for every $\lambda_i$, yielding an overall cost of $\mathcal O(n^2)$. Since the Golub-Welsch algorithm provides the means of finding the interpolation nodes and the quadrature weights simultaneously, equations \eqref{eq:gaussiannode}, \eqref{eq:gaussianweight}, and \eqref{eq:weights} along with \eqref{eq:welfert} readily yield all the ingredients required for a pseudo-spectral implementation.
\subsubsection{Remarks on computational complexity}
In closing, we mention that all the methods described in this section are applicable for general families of orthogonal polynomials but require $\mathcal O(n^2)$ floating point operations. This is not a problem if the target number of grid points is small, as is currently the case for plasma physics applications. Nevertheless, the availability of fast and stable algorithms requiring $\mathcal O(n)$ operations is desirable as the possibility of handling grids with an increasingly larger number of points becomes available. Regarding the computation of the Gaussian nodes and weights, several efficient algorithms are available for classical orthogonal polynomials \cite{Bogaert:2014, HaTo:2013, GlLiRo:2007} but there are very few methods available for non-standard nodes \cite{ToTrOl:2016}, and none specific to Maxwell nodes. As for the computation of the barycentric weights from their Gaussian counterparts, recent work by Wang, Huybrechs and Wandewalle \cite{HuVaWa:2014} provides simple explicit formulas for the barycentric weights for classical polynomials which, as opposed to a direct application of equation \eqref{eq:weights}, require only $\mathcal O(n)$ operations. The extension of similar results for Maxwell polynomials remains an open problem.

Turning to the application of this family of methods for the solution of kinetic equations, we observe that the matrices of discretized linear operators resulting from pseudo-spectral collocation are dense. Since numerical differentiation is performed as a matrix-vector product involving a dense matrix, it scales as $O(n^2)$. The solution of the linear system corresponding to the discretized kinetic equation has computational complexity scaling like $O(n^3)$. Finally, numerical integration is done by taking the dot product between the vector of function values at the grid points and the vector of Gaussian weights, which is an $O(n)$ operation. These apparently unfavorable scalings for the computational complexity are compensated in practice by the fact that high accuracy is achieved with few grid points, as we will show in this article. For small $n$, the complexity estimates are irrelevant, and dense linear algebra does not present a challenge. Moreover, even if the possibility of parallelization of parts of the computation exists, the computational overhead associated with parallelization is not justified for such small systems.

\section{Accuracy of Maxwell collocation schemes for energy diffusion equations}\label{sec:validation}

In this section, we present a detailed study of the accuracy and convergence properties of pseudo-spectral collocation schemes based on Maxwell polynomials for a one-dimensional energy diffusion equation relevant to kinetic computations in plasma physics. We first introduce this equation in Section \ref{sec:ModelProblem}, discuss its connection to the standard high-dimensional kinetic equations solved for plasma physics applications, and highlight the mathematical properties of this equation which are relevant to the numerical computation of its solution. In section \ref{sec:initial} we present the two initial conditions we will consider for our benchmarking studies, namely a singular initial condition we will refer to as Example 1 for the remainder of this article, and a smooth initial condition we will call Example 2. In section \ref{sec:comparison} we give a description of discretization schemes which are commonly used for the speed variable in plasma physics, and which we include in our benchmarking studies to show the performance of the Maxwell based collocation schemes \textit{relative} to these schemes. In Section \ref{sec:time-stepping} we briefly describe the implicit and explicit time-stepping schemes we implemented to solve the model diffusion equation, and we present the results of our accuracy and performance studies in Section \ref{sec:error_comp}.  

\subsection{Model Equation \\}\label{sec:ModelProblem}

\subsubsection{Energy diffusion due to Fokker-Planck collisions in one dimension}
In order to test the effectiveness of pseudo-spectral implementations based on Maxwell nodes for the solution of time-dependent problems in the context of kinetic calculations, we will consider the following equation, describing velocity space diffusion in one dimension \cite{CeLaWi:2015,CeWi:2015}:

\begin{equation}\label{eq:model}
\partial_t f = -Lf, \quad Lf:= -\frac{1}{x^2}\partial_x\left[\psi(x)w(x)\partial_x(e^{x^2}f)\right], \quad (x,t)\in\mathbb R_+^2,
\end{equation}
where $\partial_t$ and $\partial_x$ are the partial derivatives with respect to $t$ and $x$, respectively, and
\[
w(x):= x^2e^{-x^2}, \quad \psi(x):= \frac{1}{2x^3}\left(\text{erf}(x)-\frac{2}{\sqrt{\pi}}x e^{-x^2}\right)\,,\quad \text{erf}(x):= \frac{2}{\sqrt{\pi}}\int_0^x e^{-s^2}\,ds.
\]
Although Equation \eqref{eq:model} is one-dimensional, and only describes a diffusion process whose dynamics are far less rich than the dynamics described by the full multidimensional Fokker-Planck equation, this equation has many desirable properties for the benchmarking studies we present in this article.

Its first appeal is its relevance to kinetic equations with Fokker-Planck collisions. When pitch angle and speed are used as the velocity coordinates in kinetic solvers, the right-hand side of Equation \eqref{eq:model} corresponds to the only term which involves derivatives with respect to the speed variable in the linearized Fokker-Planck collision operator or in the model operators by Abel \textit{et al.}\cite{AbBaCoDoSc:2008} and by Sugama \textit{et al.} \cite{Sugama:2009}. In that sense, the right-hand side of Equation \eqref{eq:model} corresponds to the term that is the most numerically challenging as far as the speed variable is concerned, and determines accuracy with respect to this variable. 

The second desirable feature of Equation \eqref{eq:model} is that its solution has known physical properties which provide simple yet insightful tests for any numerical solver. Specifically, it is easy to prove analytically that Equation \eqref{eq:model} conserves mass, i.e. the value of $\int_{0}^{\infty} f x^2 dx$ is independent of time, and that according to Equation \eqref{eq:model}, $f$ relaxes to a Maxwellian of the form $f(x)=A e^{-x^2}$ as $t\rightarrow \infty$. 

The third point which makes Eq.\eqref{eq:model} particularly attractive is the fact that Wilkening and Cerfon were able to construct a semi-analytic formula for its solution \cite{CeWi:2015} subject to the appropriate boundary conditions, which are discussed in Section \ref{sec:boundary} below. This formula was obtained with a spectral transform method, and can be evaluated with arbitrary accuracy, for any initial condition. It thus serves as the reference solution against which the accuracy of all numerical schemes for Eq. \eqref{eq:model} can be tested.

Finally, since the equation is one-dimensional, it is inexpensive to solve numerically, which allows us to characterize in detail the overall performance of the pseudo-spectral schemes based on Maxwell polynomials over a wide range of grid sizes, for several choices of initial conditions and time stepping schemes, as we describe below.

\subsubsection{Boundary conditions}\label{sec:boundary}

Equation \eqref{eq:model} needs to be supplemented with boundary conditions and initial conditions on the distribution function $f$ in order to be solved. We focus on boundary conditions in this section, and initial conditions in Section \ref{sec:initial}.

In order to understand how to specify boundary conditions for Equation \eqref{eq:model}, it is most convenient to define the change of variable $u := e^{x^2} f$. Under this transformation, Equation \eqref{eq:model} takes the form
\begin{equation}\label{eq:operator}
u_{t}=-Bu\;\;,\;\; Bu := -\frac{1}{w}\partial_x\left(\psi\,w\,\partial_xu\right), \quad x\in[0,\infty).
\end{equation}
The linear operator $B$ is well defined over the Hilbert space 
\begin{equation}\label{eq:L2w}
L^2_w(\mathbb R_+) := \{ u : \|u\|_{w} < \infty \},
\end{equation}
where the norm $\|\cdot\|_w$ is the one induced by the inner product \eqref{eq:inner}. In this space, the operator is self-adjoint, has $\lambda=0$ as its only eigenvalue and a continuous spectrum consisting of the half line $(0,\infty)$. Furthermore, $B$ is singular at both endpoints, since $w$ vanishes at the origin and the domain extends to infinity to the right. The singularities at the endpoints can be shown to be of the limit-point type at infinity, and of limit-circle type at the origin \cite{CeWi:2015}. 

Limit-point type singularities do not require the prescription of a boundary condition in order to ensure uniqueness of the solution in the functional space where $B$ is well defined \cite{Stakgold:1998}. In this article, we focus on distribution functions with a Maxwellian envelope, i.e. of the form $f(x)=g(x)e^{-x^2/2}$. That means that the condition \eqref{eq:L2w} should be understood as a condition on $g$, the non-Maxwellian factor of the distribution function:
\[
\int_0^\infty u^2x^2e^{-x^2}dx = \int_0^\infty (e^{x^2}f)^2x^2e^{-x^2}dx = \int_0^\infty g^2x^2e^{-x^2}dx < \infty.
\]
This restriction allows for polynomial and even exponential growth of the non-Maxellian part of the distribution function, both of which are consistent with the global behavior $f\to0$ as $x\to\infty$. In the context of pseudo-spectral collocation, this implies that, as long as the interpolants used belong to the weighted Hilbert space defined in \eqref{eq:L2w}, no special care has to be taken at the final grid point. 

On the other hand, for singularities of limit-circle type, like the one presented by $B$ at the origin, boundary conditions are required. To guarantee uniqueness of the solution to $\eqref{eq:model}$, a condition must be prescribed at the origin. However, since the operator is not defined there, standard boundary conditions (Dirichlet, Neumann, Robin, etc.) cannot be imposed. In such situations, it has been proven \cite{HaKr:1992} that requiring boundedness of the solution at the origin is enough to ensure the well-posedness of the problem. Since our quasi-polynomial interpolant is bounded at the origin, the boundary condition is automatically satisfied in our pseudo-spectral schemes. Moreover, by choosing grid points associated with the interpolant that do not include the origin, we avoid difficulties associated with the singularity at the origin, and have a scheme which is remarkably simple to implement. This contrasts with a standard finite difference approach where further assumptions need to be made in order to deal simultaneously with the singularity at the origin and the need for a boundary condition.  
\subsection{Initial conditions \\}\label{sec:initial}

\begin{figure}[tb]
\begin{tabular}{cccc}
\includegraphics[width=.31\linewidth]{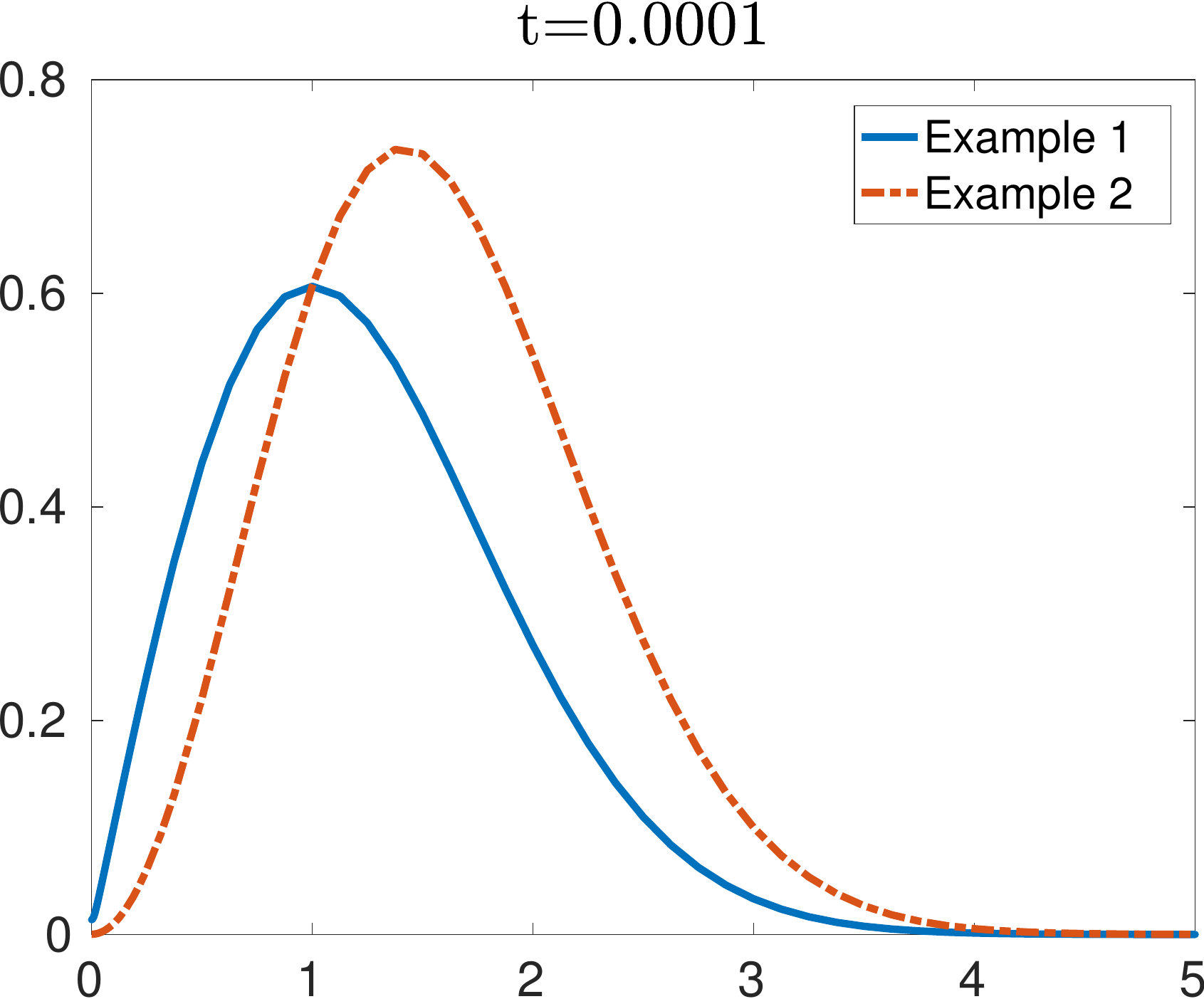} &
\includegraphics[width=.31\linewidth]{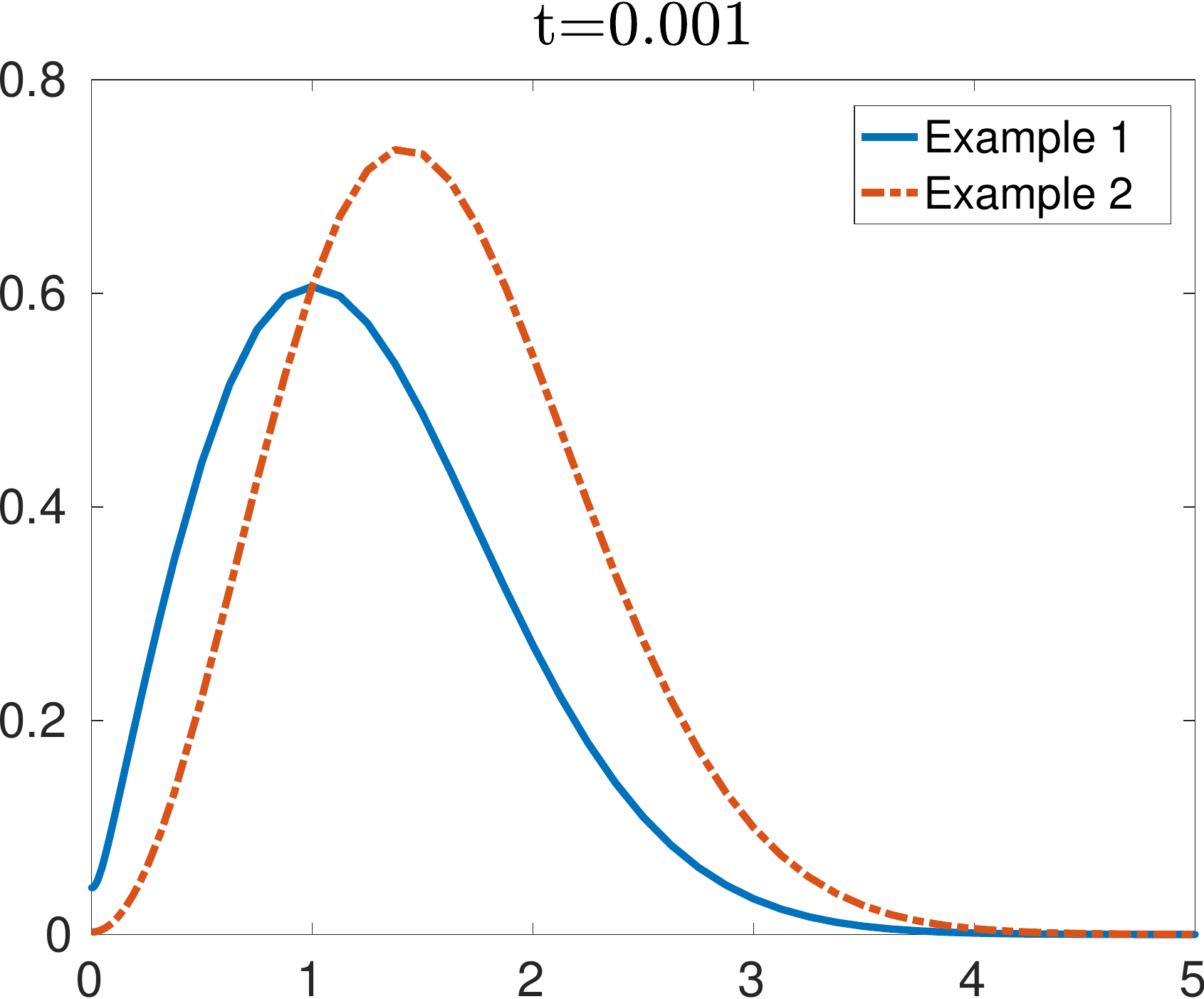}&
\includegraphics[width=.31\linewidth]{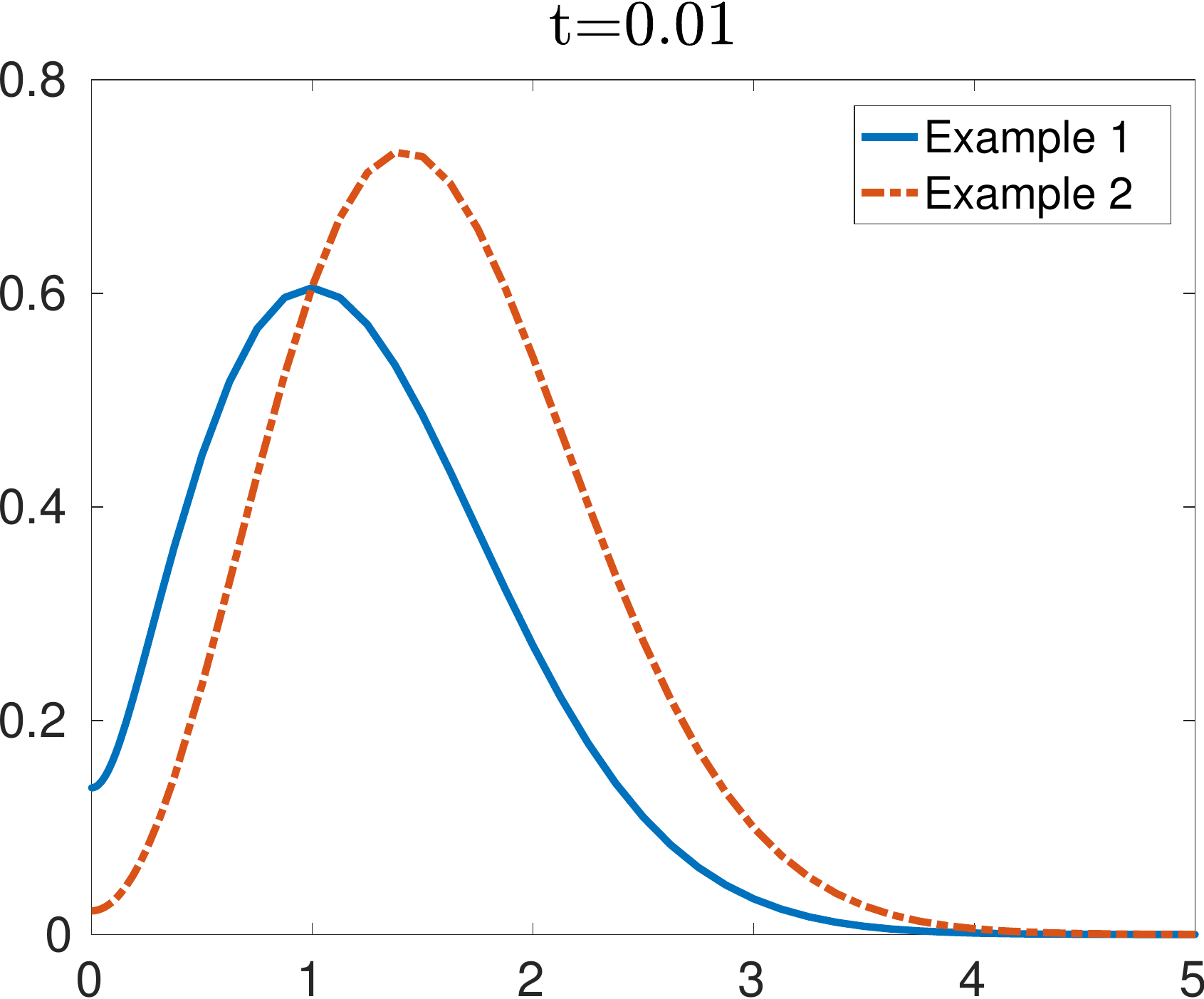} \\
\includegraphics[width=.31\linewidth]{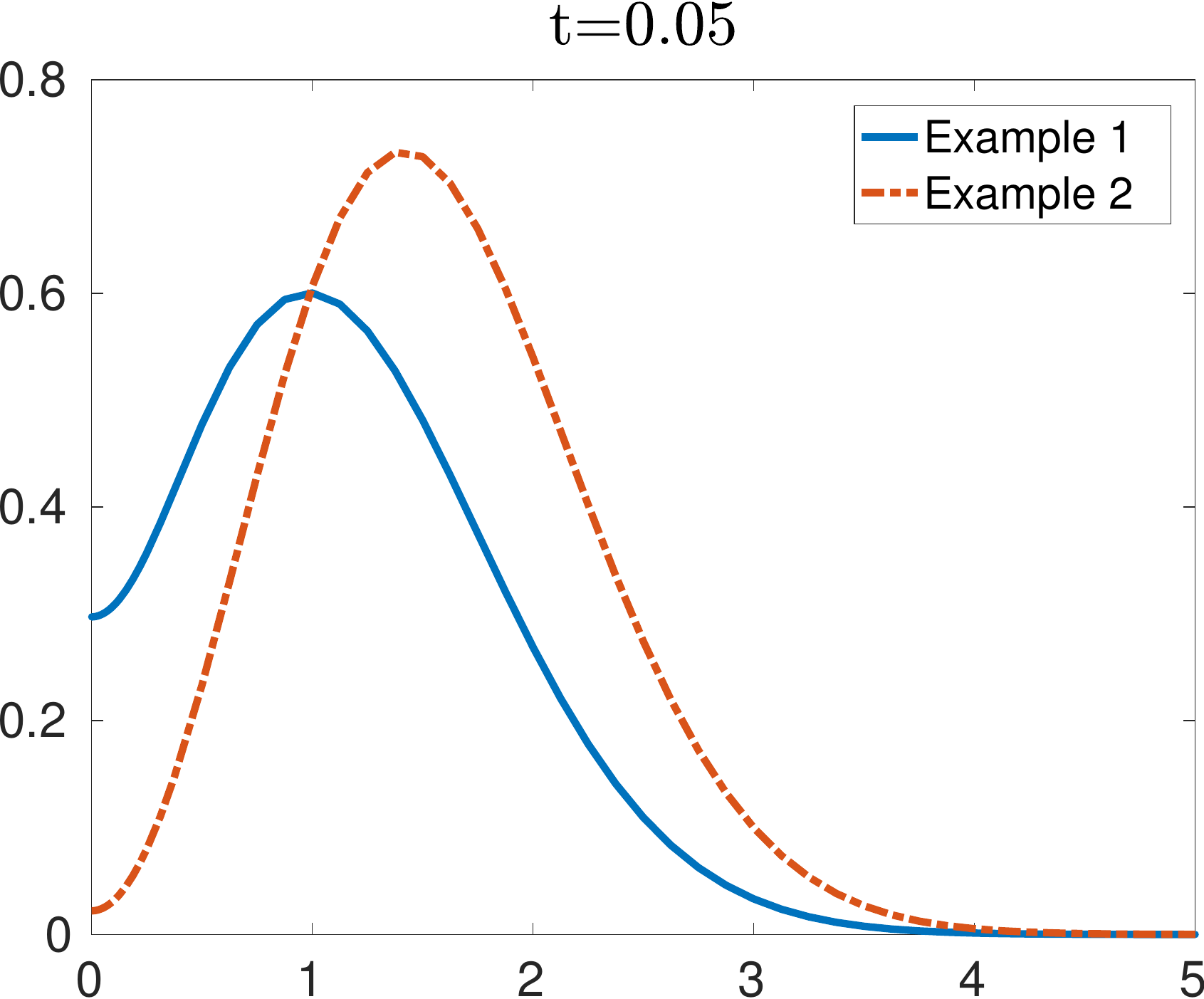} &
\includegraphics[width=.31\linewidth]{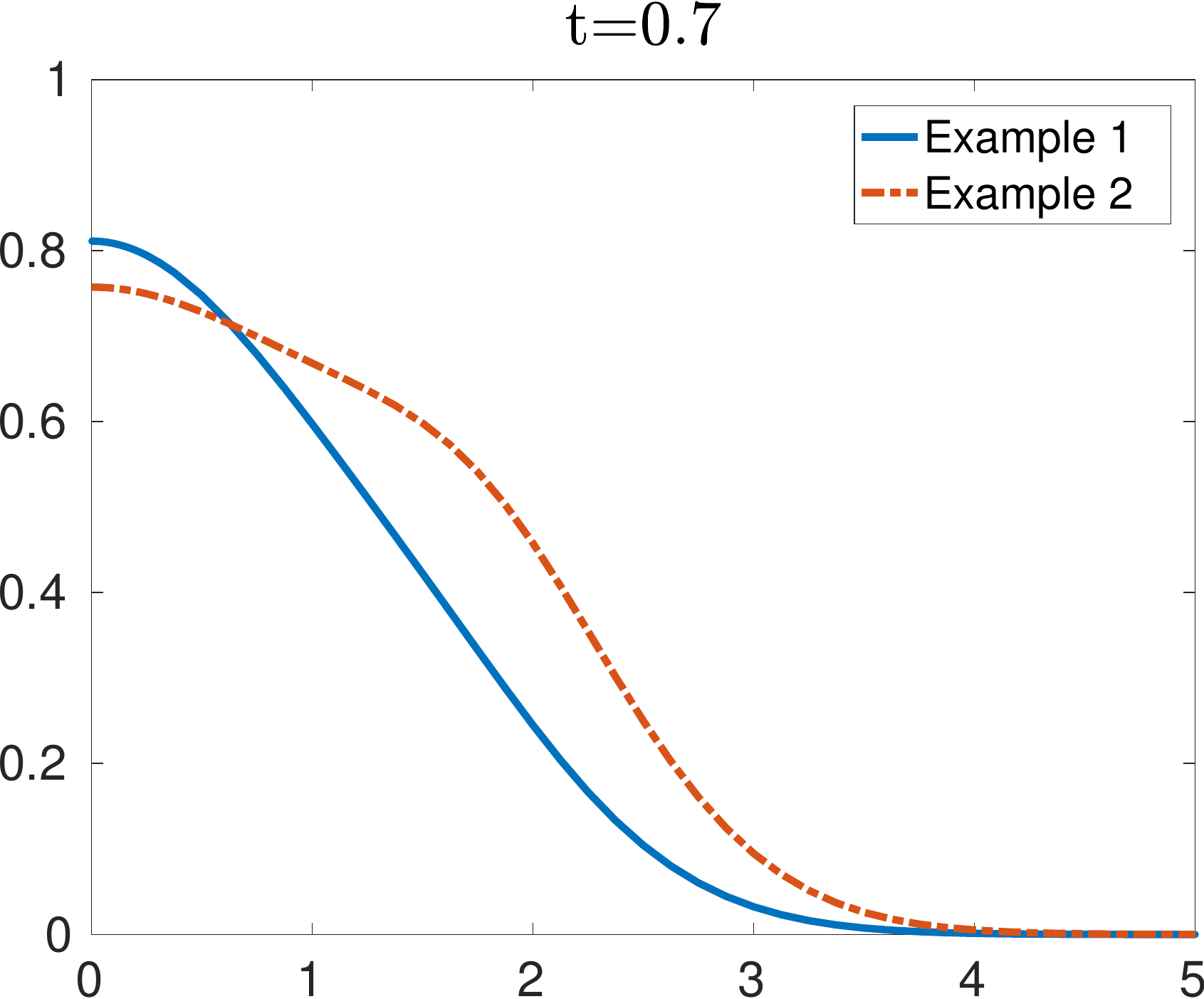} &
\includegraphics[width=.31\linewidth]{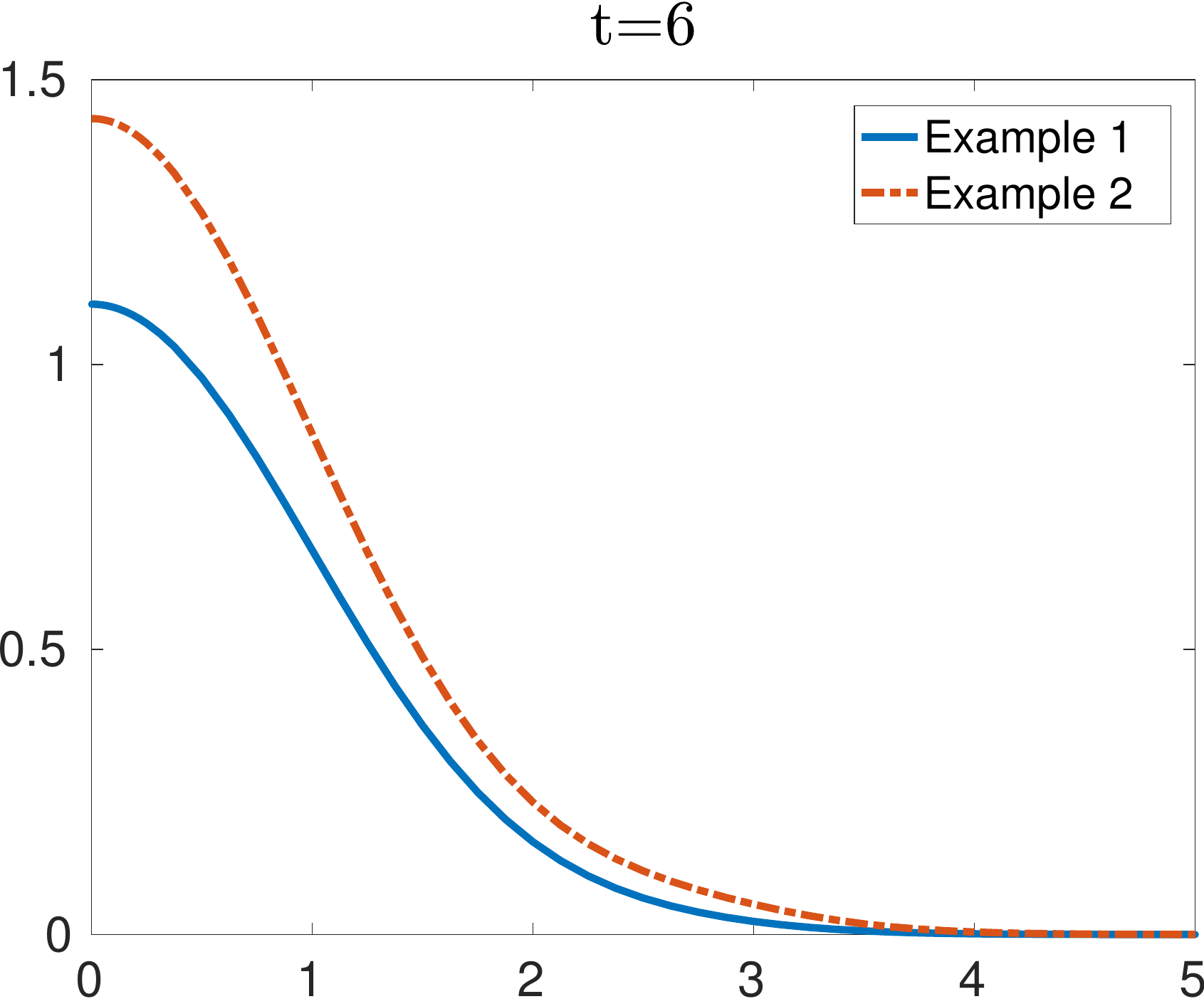} 
\end{tabular}
\caption{{\scriptsize Dynamics of the exact solutions of equation \eqref{eq:model} for the two benchmark initial conditions \eqref{eq:examples}. The reference solutions were obtained using the spectral transform for times $t= 0.0001,0.001,0.01,0.5,0.7,6$. For the sake of clarity, the graph depicts the rescaled solutions $e^{\frac{1}{2}x^2}f_i(x,t), \quad i=1,2.$}}\label{fig:exactsolution}
\end{figure}

For our tests of the performance of the pseudo-spectral discretizations based on full Maxwell polynomials and on Maxwell polynomials defined by truncating the semi-infinite interval to $[0,12]$, we consider two different initial conditions for equation \eqref{eq:model},
\begin{subequations}\label{eq:examples}
\begin{alignat}{6}
\hbox{\textbf{Example 1:}} \qquad& f_1(x,0) =\, & xe^{-x^2}, \\
\hbox{\textbf{Example 2:}}\qquad&  f_2(x,0) =\, & x^2e^{-x^2}. 
\end{alignat}
\end{subequations}
Example 1 represents a singular initial state, since the action of the operator on $xe^{-x^2}$ yields a $1/x$ behavior at the origin. We chose this singular state intentionally, because it leads to specific challenges for discretization schemes based on Maxwell polynomials, as we will show. Situations with singularities of this type may occur in plasma physics. For example, it does occur when one calculates the resistivity of a plasma \cite{ErLa:2013} by solving the time-dependent kinetic equation and letting the distribution function relax to a steady-state. In contrast, Example 2 corresponds to a smooth initial condition. Examples 1 and 2 were also the initial conditions studied in \cite{CeLaWi:2015}, and their evolution was evaluated using the semi-analytic spectral transform method to be used as a reference solution. For both examples, the initial distribution function undergoes a rapid transient period before relaxing into a steady state Maxwellian distribution, as can be seen in Figure \ref{fig:exactsolution}.

\subsection{Comparison with other discretization schemes}\label{sec:comparison}

In this study, we do not only compare the accuracy of the discretization scheme based on Maxwell polynomials with the semi-analytic spectral transform solution, but also with other standard schemes that are used in existing solvers or have been considered in the literature. The comparison of the performance of the schemes with respect to one another can guide the choice of discretization for the speed variable in new versions of existing codes or in new solvers.

The zeros of first or second kind Chebyshev polynomials have been the preferred choices of collocation nodes for general purpose pseudo-spectral codes. This is not surprising, given their outstanding interpolation properties together with the availability of fast and stable algorithms for the generation of the collocation nodes and barycentric weights, and for sampling of the interpolant \cite{DrHaTr:2014}. It is therefore natural to include in our study a comparison with a Chebyshev-based grid. There are several ways of mapping the Chebyshev nodes from the interval $[0, 1]$ into a semi-infinite interval. We opt here for the mapping $x = -\log(1-s)$ which was also considered by Landreman and Ernst \cite{ErLa:2013}, and use first kind Chebyshev nodes to avoid the singularity at the origin.

A second natural grid to include in our comparisons is the scheme implemented in the gyrokinetic code GS2 which relies on a conservative finite difference approach on staggered grids, very much in the style of cell-centered finite volume methods. The grid in GS2 combines the zeros of Legendre polynomials mapped to the interval $[0, 2.5]$ with a fixed number of Laguerre points mapped by the transformation $x=\sqrt{s + 2.5^2}$  into the interval $[2.5, \infty)$. If the number of total grid points exceeds $n=12$, GS2 uses two grid points on the unbounded part of the interval; otherwise it only uses one grid point \cite{GS2:2009}. Given that the mapped Laguerre nodes on the right endpoint are few and spaced out, we enhanced the differentiation by adding one additional point and by considering third order stencils on the Laguerre nodes.

Finally, we also consider a scheme based on the GS2 grid, but relying on a pseudo-spectral method instead of the low order finite-difference stencils used in GS2. We expect the global differential operators to provide improved accuracy. Nevertheless, the fact that the GS2 grid is composite leads to block diagonal differentiation matrices that act on each of the grid components independently. Given the low number of Laguerre-based nodes in the GS2 grid and the block-diagonal nature of the pseudo-spectral differentiation matrices, the pseudo-spectral version of the GS2 discretization can have at most a second order accurate differentiation operator in that part of the grid. This implies that the enhanced GS2 implementation, with a third order accurate differentiation operator on the Laguerre grid, may outperform its pseudo-spectral counterpart on occassion.

\subsection{Time stepping}\label{sec:time-stepping}

If we define the numerical discretizations $\mathbf{f}(t)$ and $\mathbf{L}$ of the solution $f$ and the operator $L$ in Eq. \eqref{eq:model}, then this equation takes the simple form
\begin{displaymath}
\frac{d\mathbf{f}}{dt}=-\mathbf{L}\mathbf{f}(t)
\end{displaymath}
which can be readily integrated in time to yield the general solution
\begin{equation}
\mathbf{f}(t)=\mathbf{f}(0)e^{-\mathbf{L} t}.
\label{eq:exact_time}
\end{equation}
In other words, the solution to our model problem can be advanced in time \textit{exactly}. The advantage in doing so in benchmarking studies is that one avoids numerical errors associated with numerical time stepping, and can focus on the error due to the space discretization scheme. This is the reason why this method was chosen in \cite{CeLaWi:2015}. However, time integration as given in Eq. \eqref{eq:exact_time} is not very relevant to the usually nonlinear, multidimensional kinetic simulations in plasma physics, where such a simple solution is not available. Since one of the central goals of this article is to provide observations and guidelines that are directly applicable to kinetic solvers, we choose instead to solve Eq. \eqref{eq:exact_time} with time-stepping schemes which are used in current solvers. Specifically, we consider two second order time stepping algorithms: the implicit two-step backward differentiation algorithm known as BDF2, and the explicit second order predictor-corrector algorithm. 

The time step size used was
\begin{equation}\label{eq:time-step}
\Delta t = 5\times 10^{-5} \quad \hbox{(Implicit)}, \qquad \qquad
\Delta t = \min\{5\times 10^{-5}, 0.8/|\lambda_{max}|\} \quad \hbox{(Explicit)},
\end{equation}
where $\lambda_{max}$ is the eigenvalue of the discretization of $L$ with the largest magnitude. This choice was made in order to enforce stability conditions by forcing the eigenvalues of the discretized operator into the stability region of the time-stepping method. It sometimes resulted in an extremely small time step, leading to time consuming simulations. In such cases, we accelerated the computation with a Runge-Kutta-Chebyshev procedure \cite{AlAmGr:1996,VeHuSo:1990}. Note that we have found that $|\lambda_{max}|$ is slightly larger for Maxwell polynomials defined on the truncated interval than it is for Maxwell polynomials defined on the full interval. The ability to use slightly larger time steps with the latter can be seen as another small advantage in their favor. 

It is important to stress that the performance of our numerical solver depends sensitively on the way the discretization of the operator $L$ is constructed, as we will show in detail in Section \ref{sec:stability}. For all the figures discussed in the present section, we used the discrete operator $\mathbf{L}_{b}$, defined in Eq. \eqref{eq:Lb}, as the discrete representation of $L$.

\subsection{Error comparison results}\label{sec:error_comp}
For our comparisons between the different discretization schemes, we consider the error measured with the following two norms  
\begin{align}
\hbox{$L^\infty([0,12])$ Error}  \qquad& \max_{x\in\mathcal I} |f_h(x,t)-f_{ST}(x,t)|, \\
\hbox{$L^1_\rho([0,12])$ Error}   \qquad&  \int_0^{12}|f_ h(x,t)-f_{ST}(x,t)|x^2\, dx.
\end{align}
where the set $\mathcal I$ consists of 120 equispaced points in the interval $[0,12]$ at which the semi-analytic spectral transform solution $f_{ST}$ was computed, and $f_h$ denotes the approximate solution. We measured the errors for six different times: $t= 0.0001, 0.001, 0.01, 0.05, 0.7, 6$. The $L^{\infty}$ norm is a natural norm to consider from a numerical point of view, as it measures the error in the computed solution $f$ directly. Furthermore, it serves as a diagnosis of the accuracy of differentiation for all the discretization schemes considered here. The $L^{1}_\rho$ norm is more physically motivated, as it measures the error in the zeroth order moment of the distribution function, i.e. the physical density. Since it involves integration, the error in the $L^{1}$ norm tests the performance of the schemes for both differentiation and integration. Errors in the $L^{1}$ norm are consistently smaller than errors in the $L^{\infty}$ norm because integration is a smoothing operation.

\begin{figure}[tb]
\begin{tabular}{ccc}
\includegraphics[width=.31\linewidth]{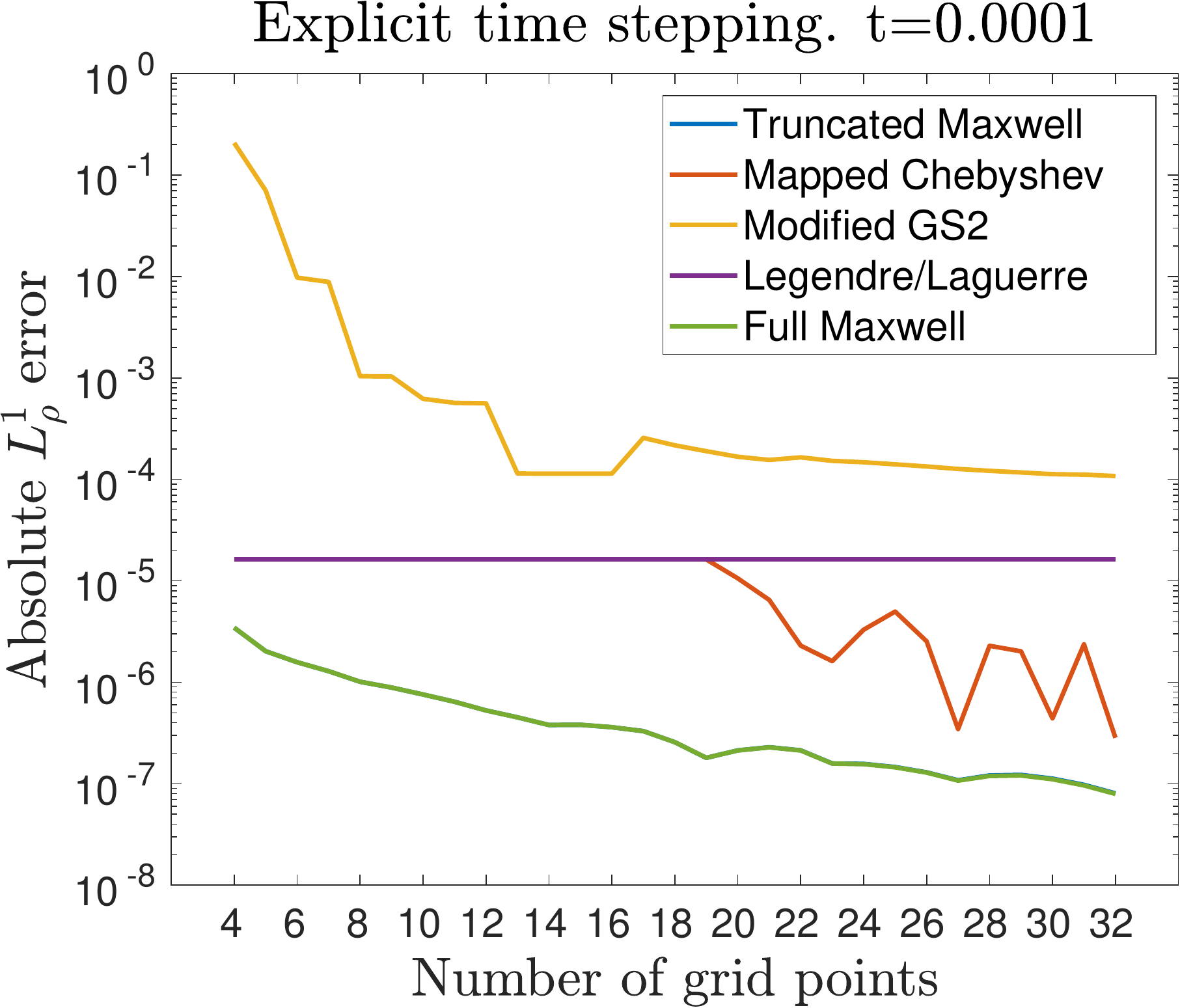} &
\includegraphics[width=.31\linewidth]{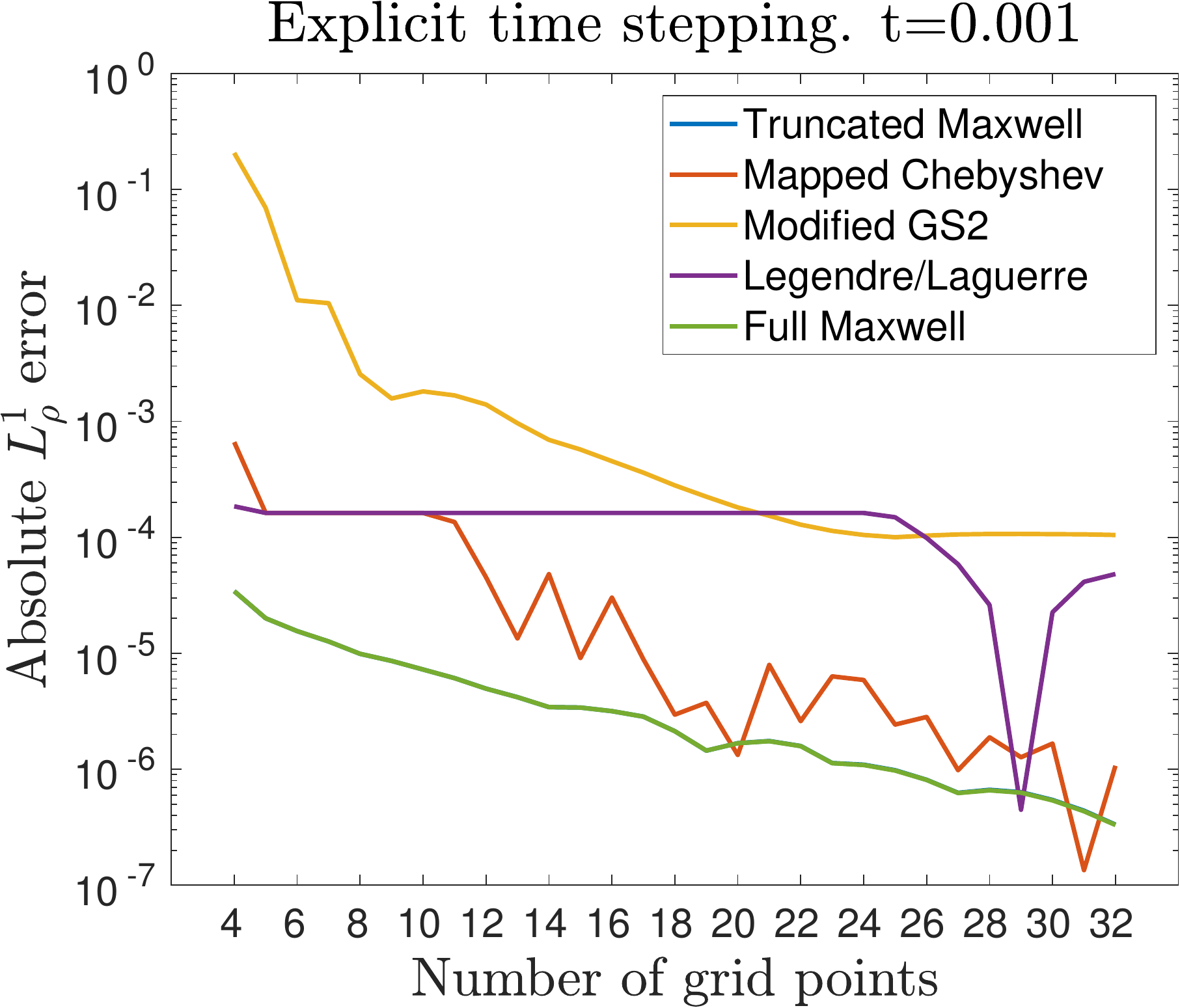} &
\includegraphics[width=.31\linewidth]{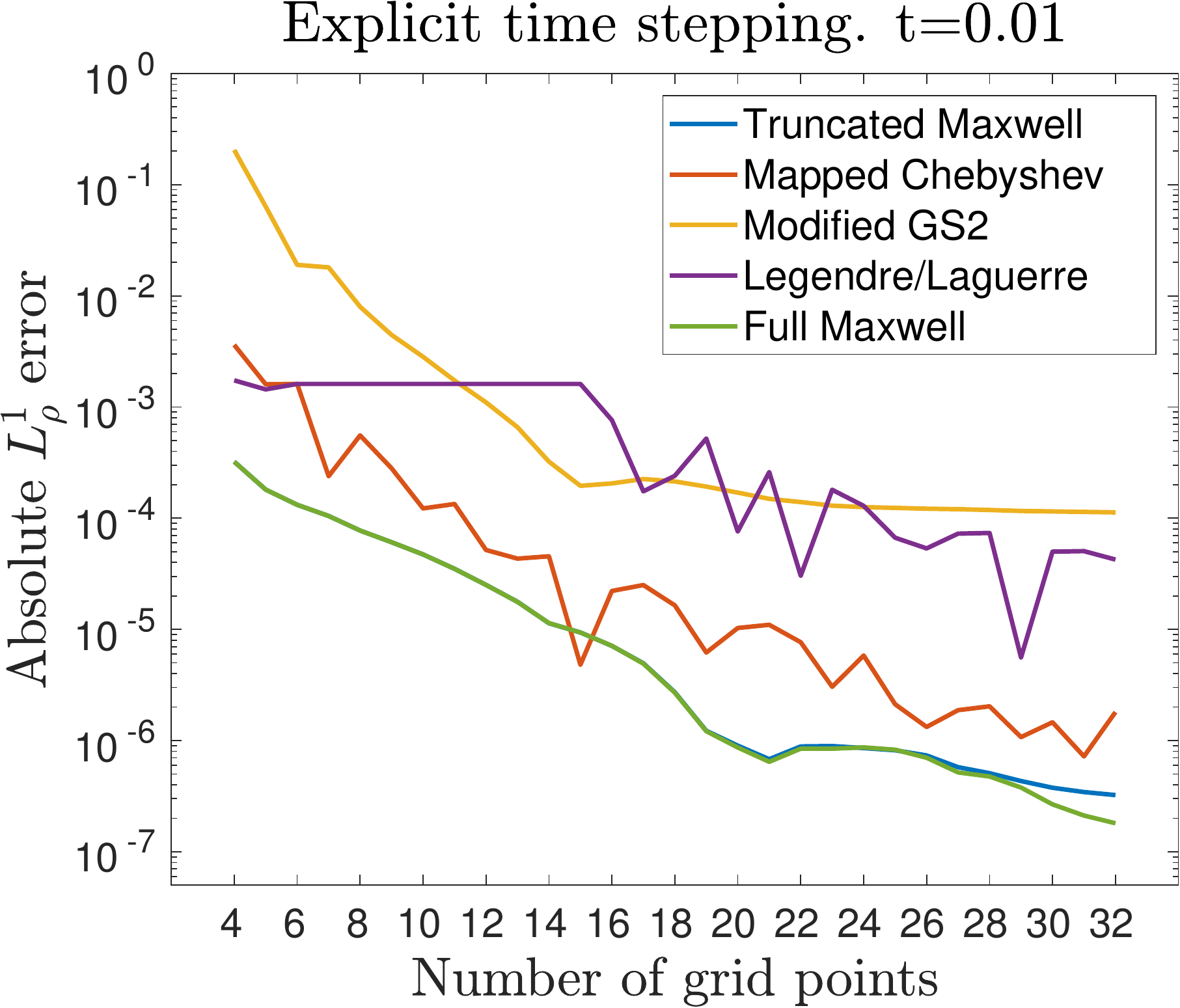} \\
\includegraphics[width=.31\linewidth]{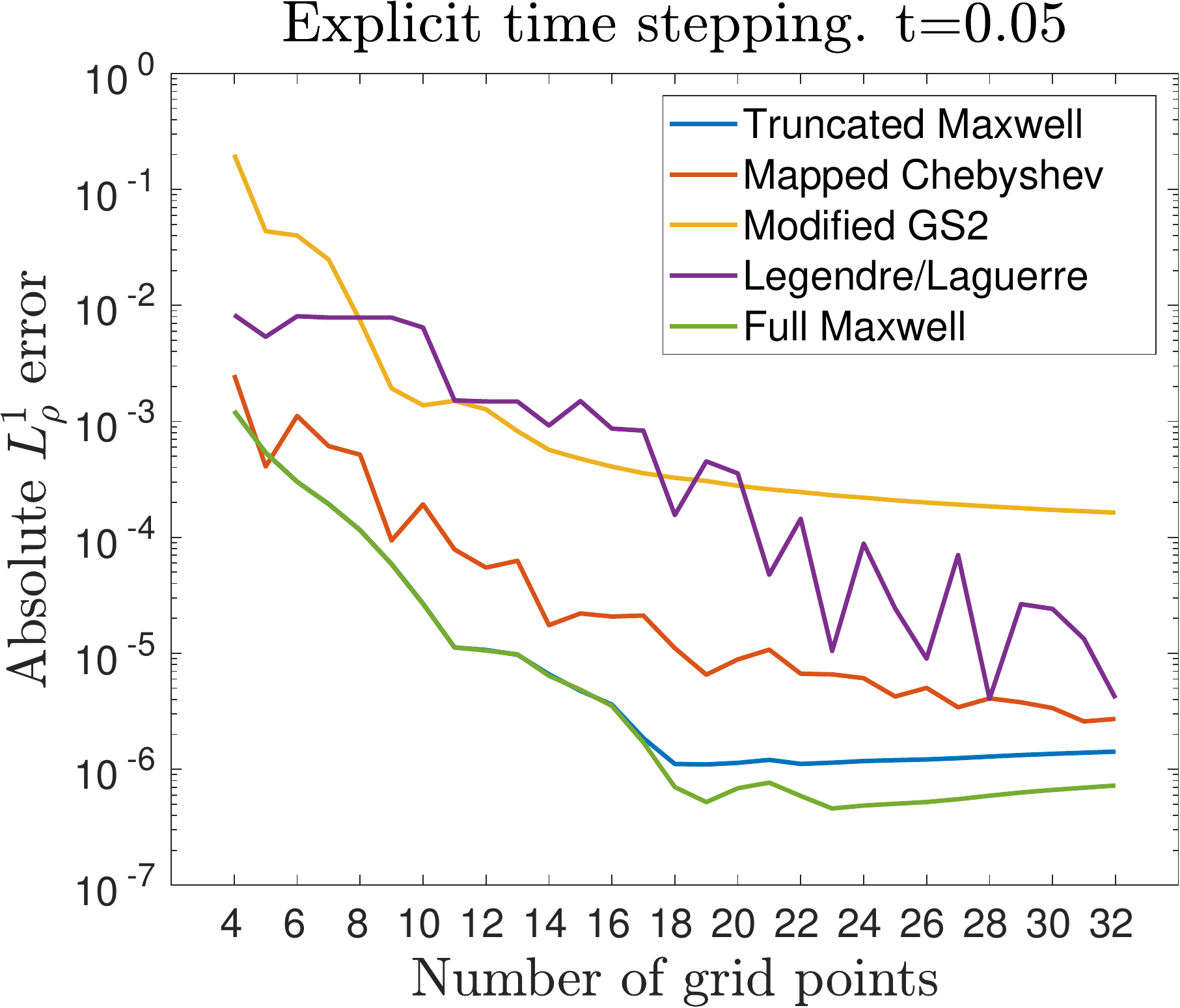} &
\includegraphics[width=.31\linewidth]{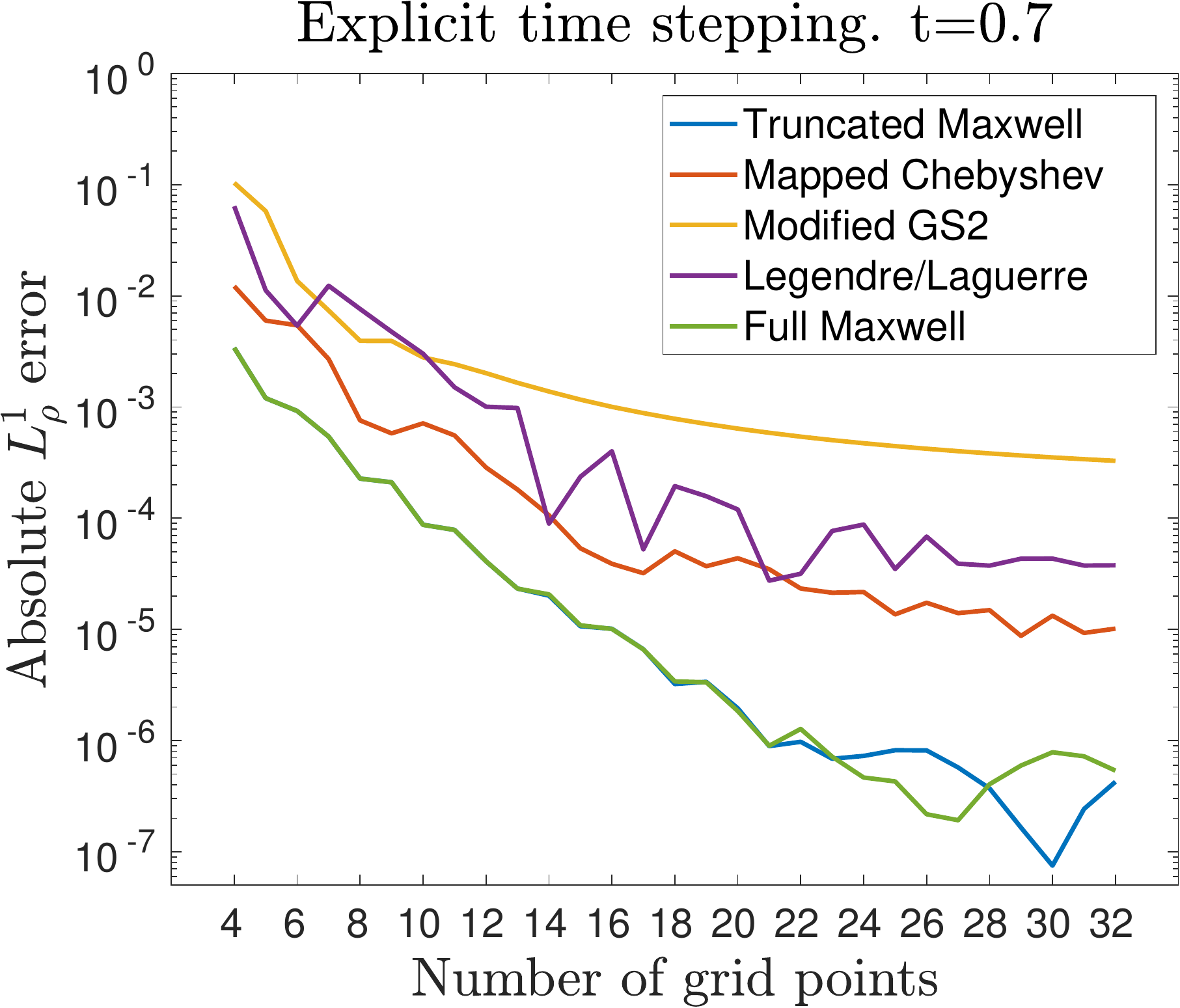} &
\includegraphics[width=.31\linewidth]{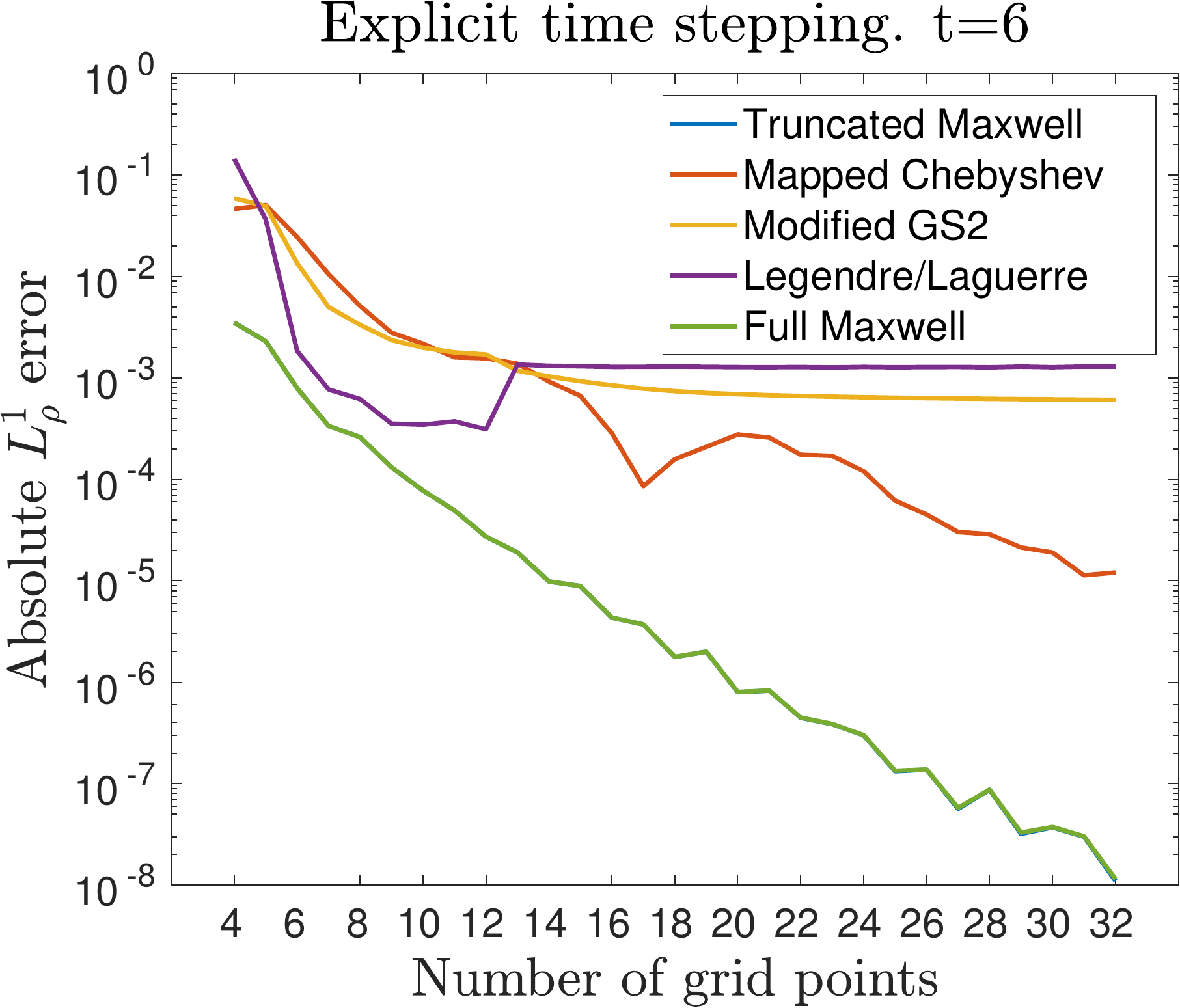} 
\end{tabular}
\caption{{\scriptsize Convergence plots in the $L^1_\rho$ norm for Example 1 with explicit time stepping. The behavior of the full and truncated versions of Maxwell nodes is virtually indistinguishable.}}\label{fig:E1ExplicitL1error}
\end{figure}

\begin{figure}[tb]
\begin{tabular}{ccc}
\includegraphics[width=.31\linewidth]{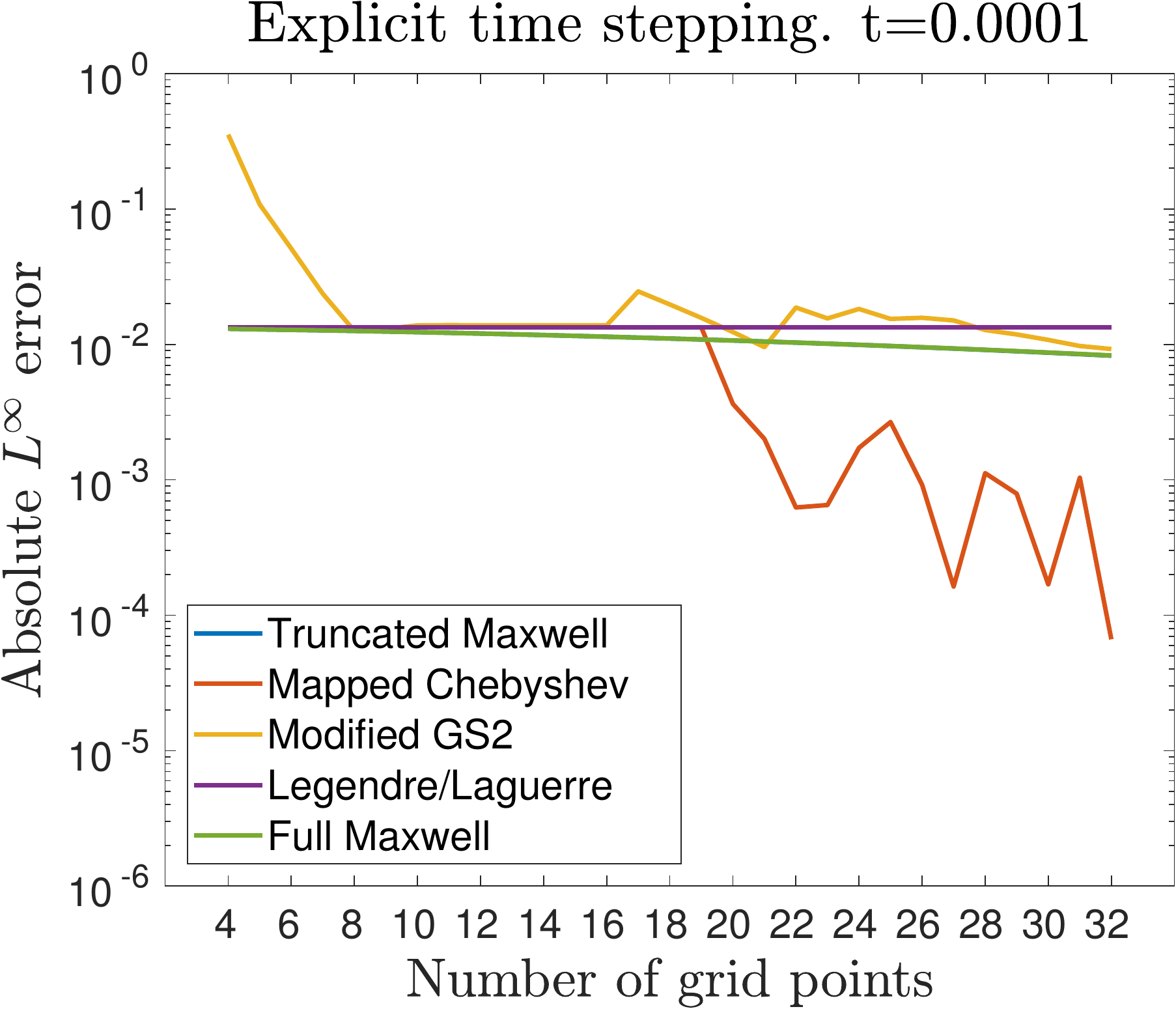} &
\includegraphics[width=.31\linewidth]{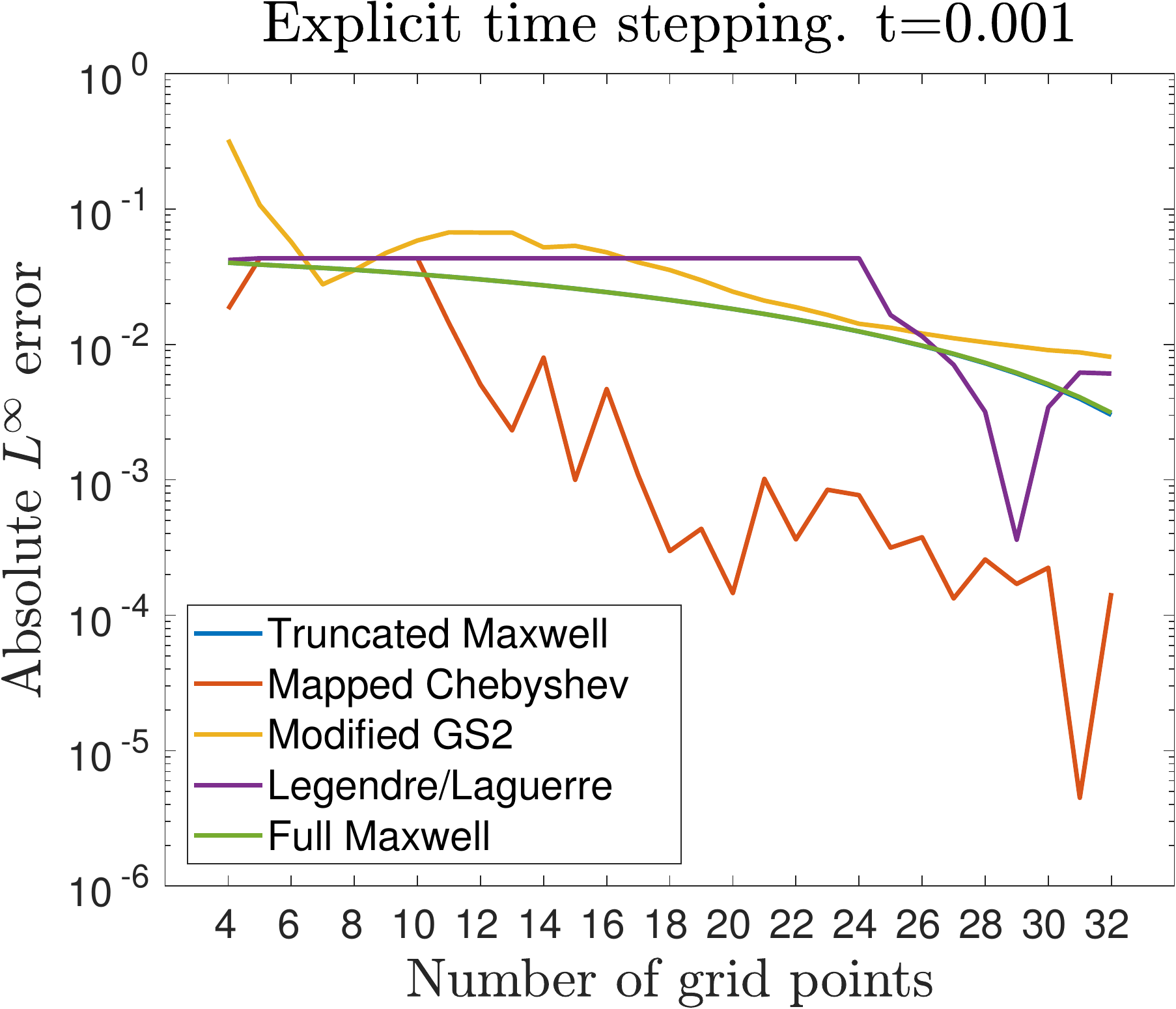} &
\includegraphics[width=.31\linewidth]{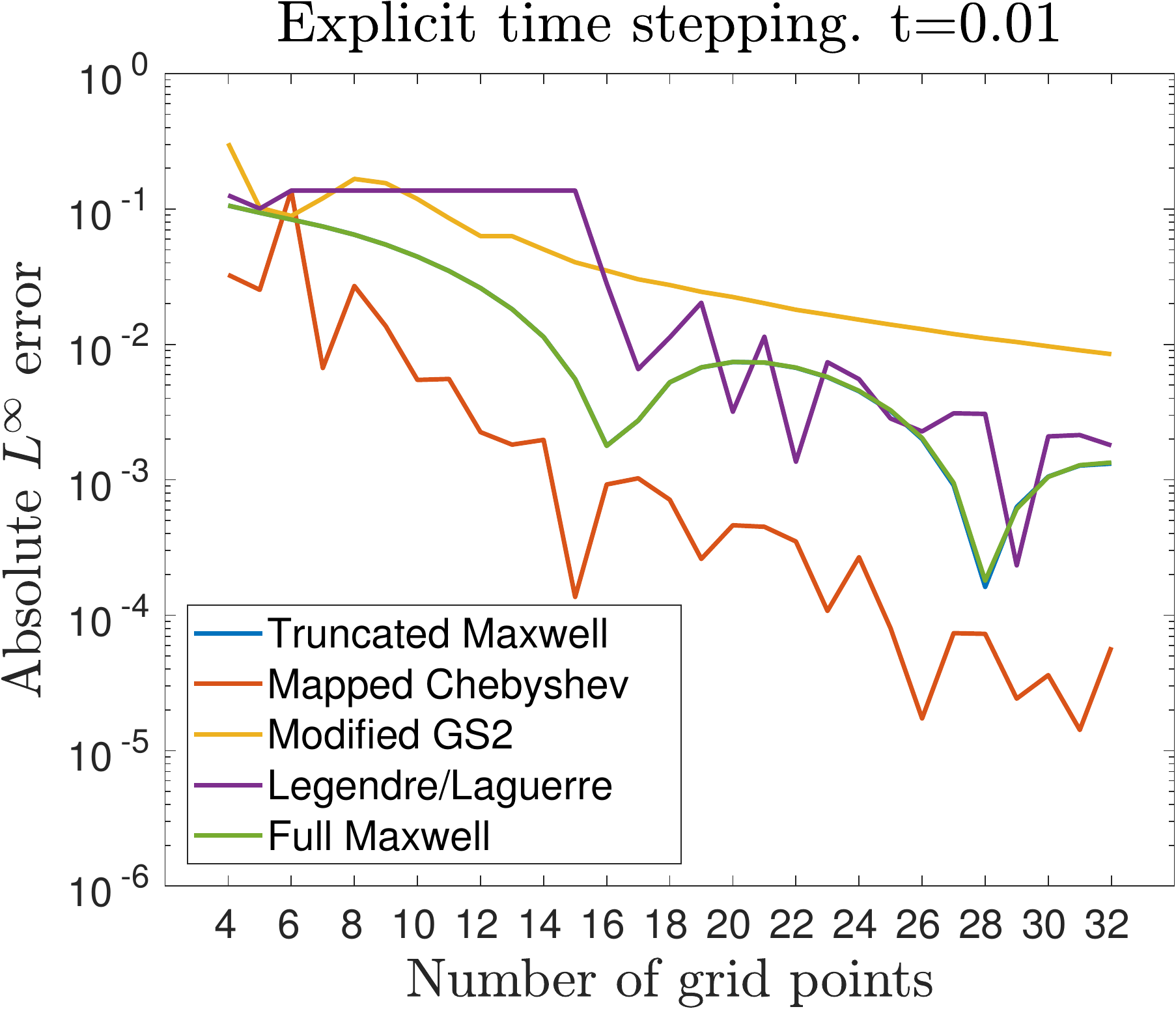} \\
\includegraphics[width=.31\linewidth]{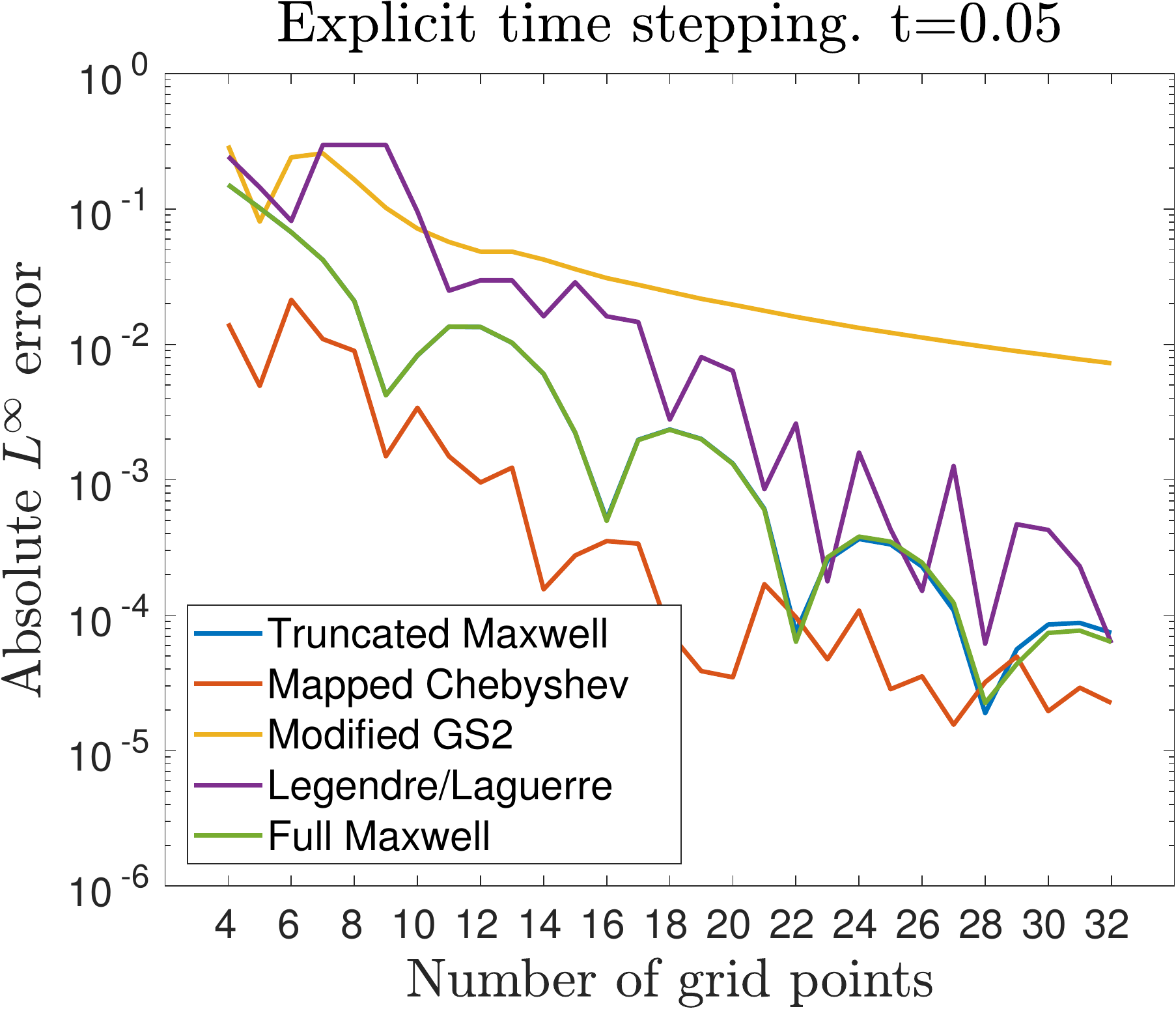} &
\includegraphics[width=.31\linewidth]{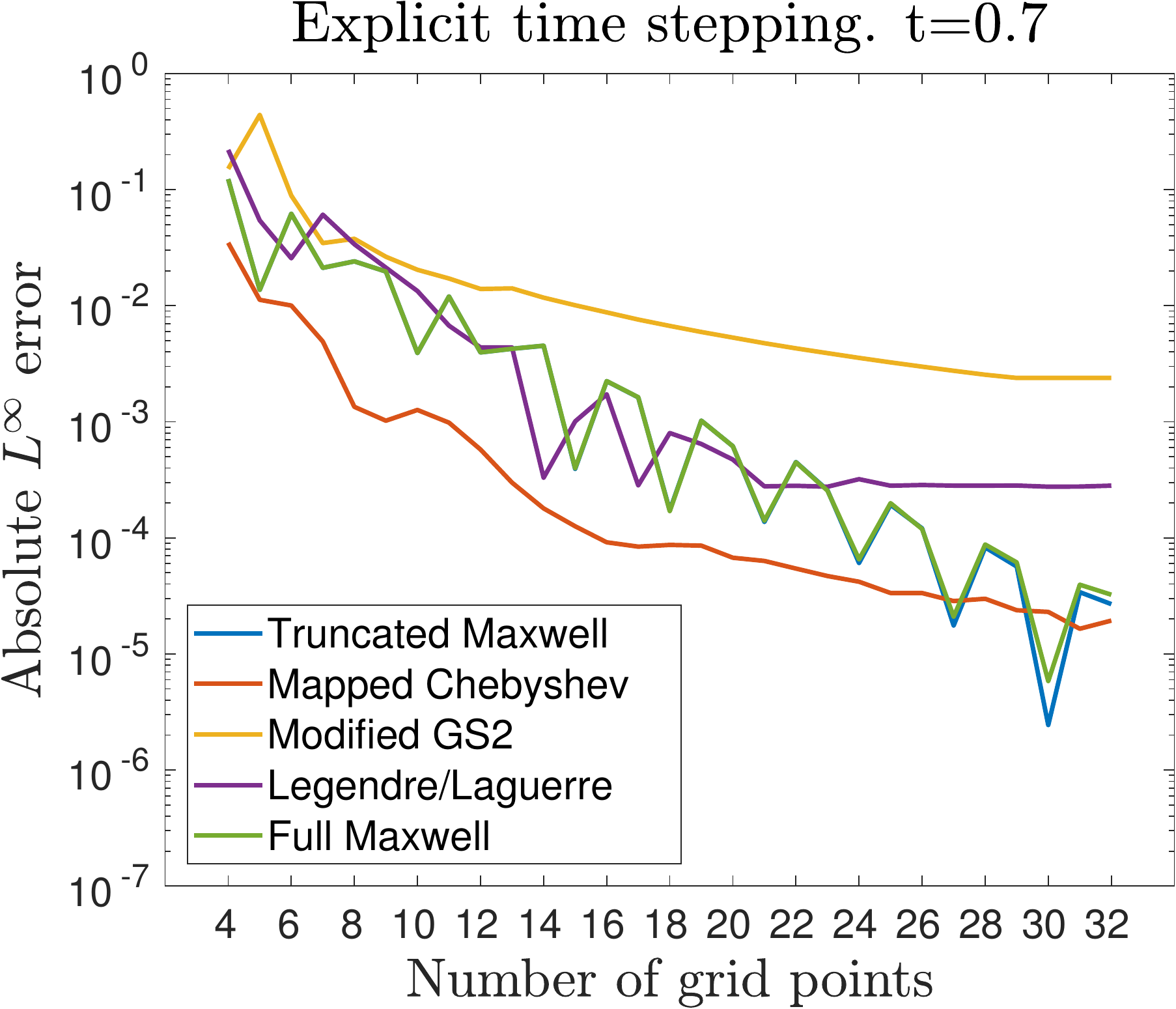} &
\includegraphics[width=.31\linewidth]{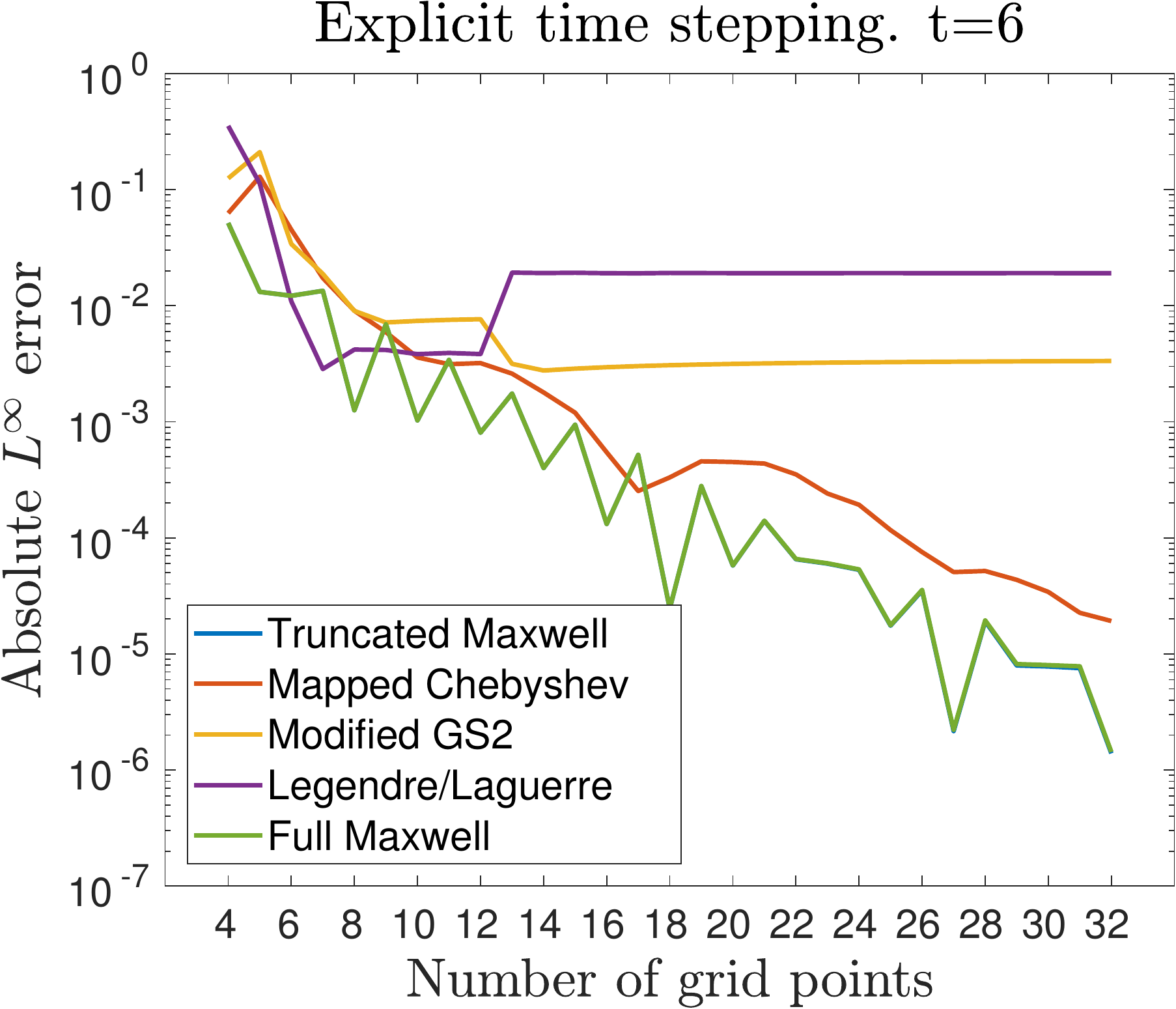} 
\end{tabular}
\caption{{\scriptsize Convergence plots in the $L^\infty$ norm for Example 1 with explicit time stepping. The behavior of the full and truncated versions of Maxwell nodes is virtually indistinguishable.}}\label{fig:E1ExplicitMAXerror}
\end{figure}

\begin{figure}[tb]
\begin{tabular}{ccc}
\includegraphics[width=.31\linewidth]{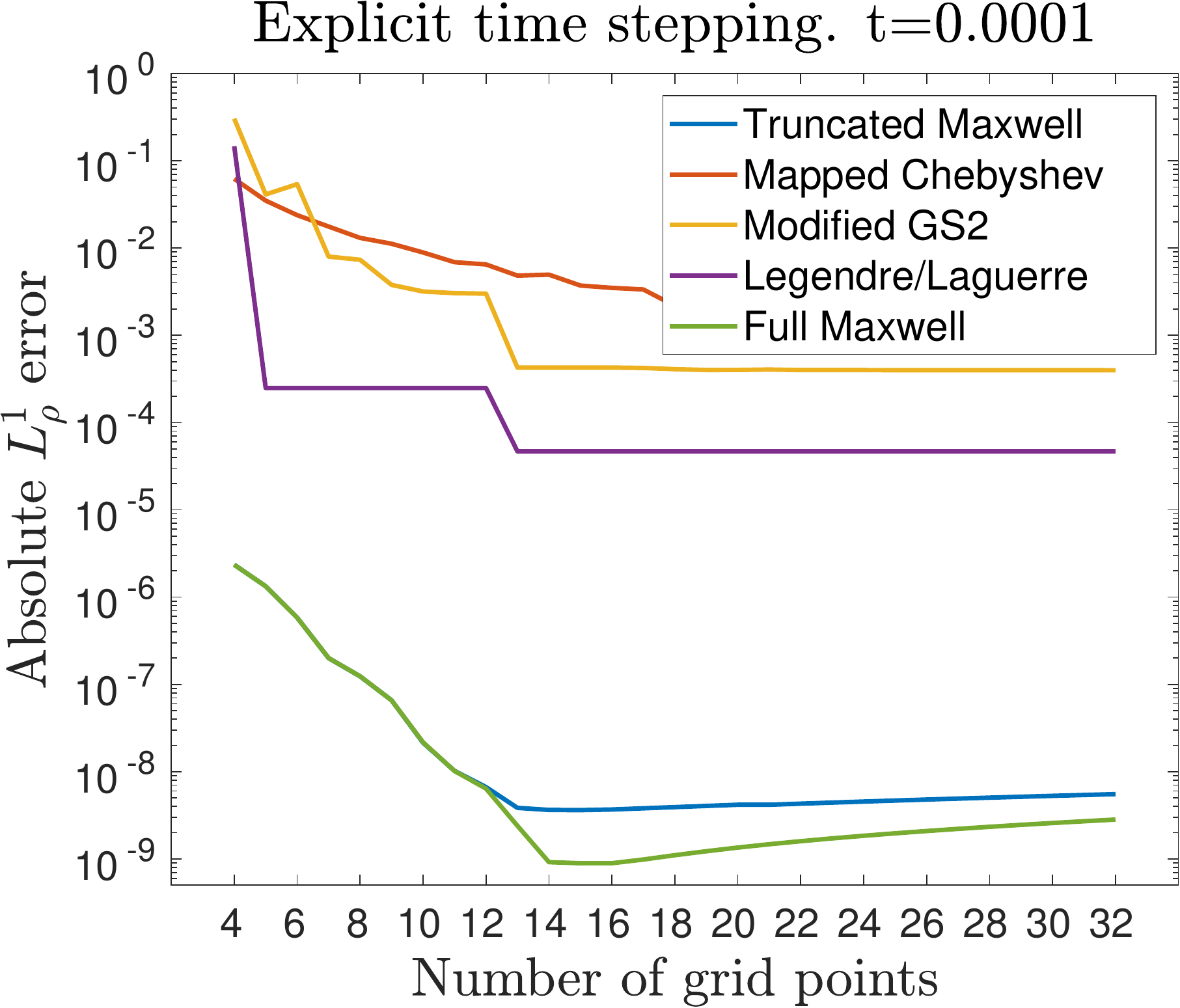} &
\includegraphics[width=.31\linewidth]{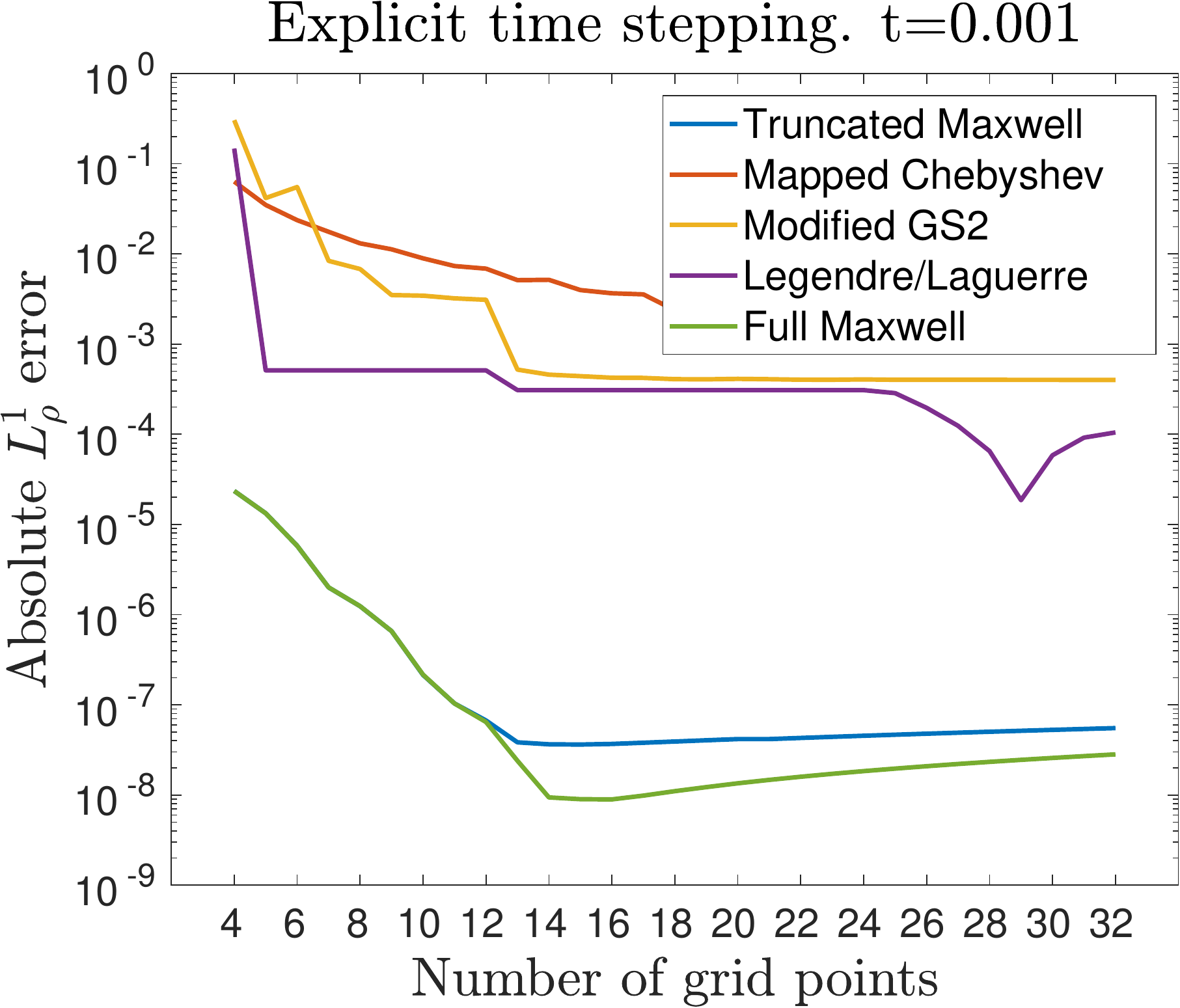} &
\includegraphics[width=.31\linewidth]{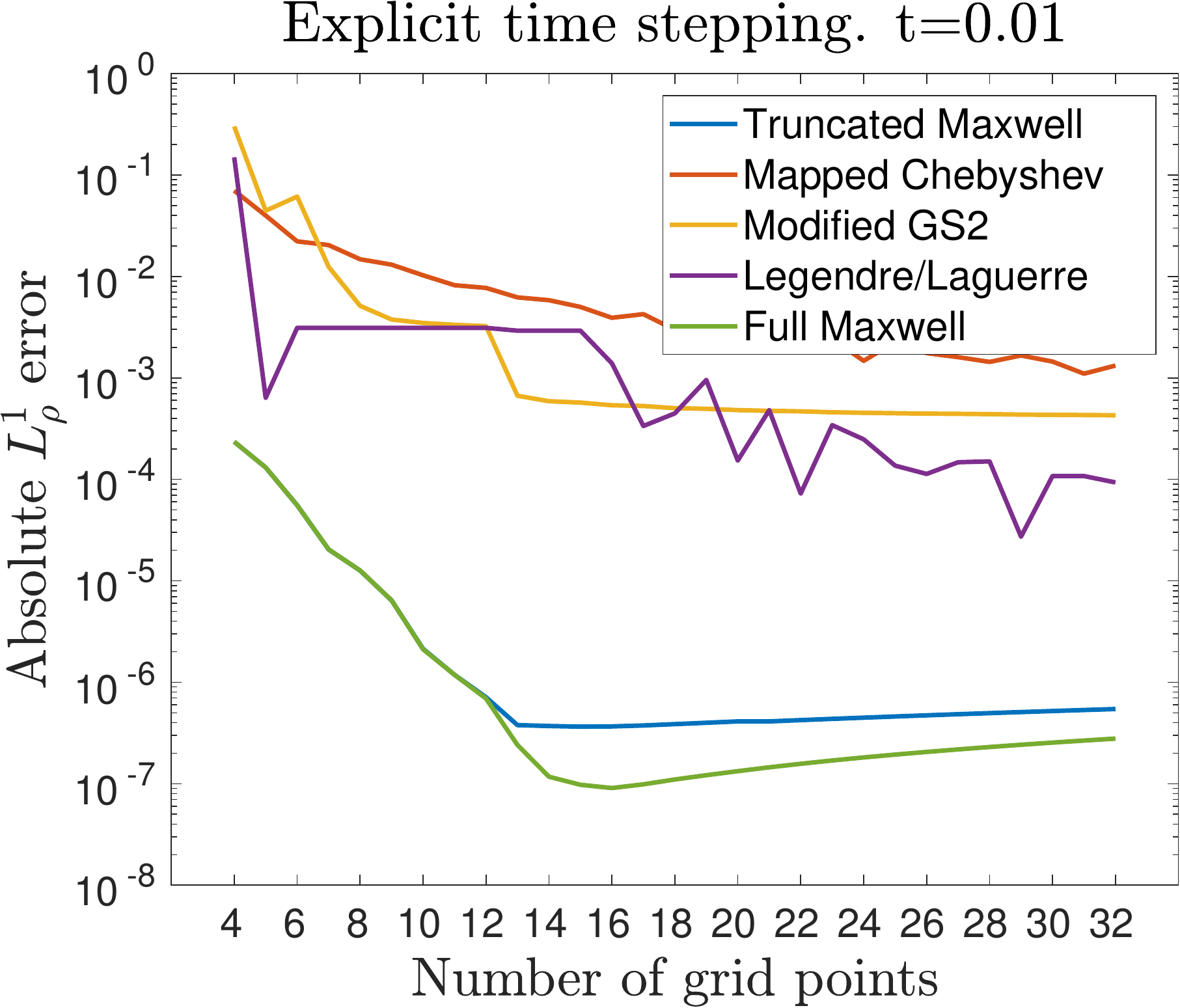} \\
\includegraphics[width=.31\linewidth]{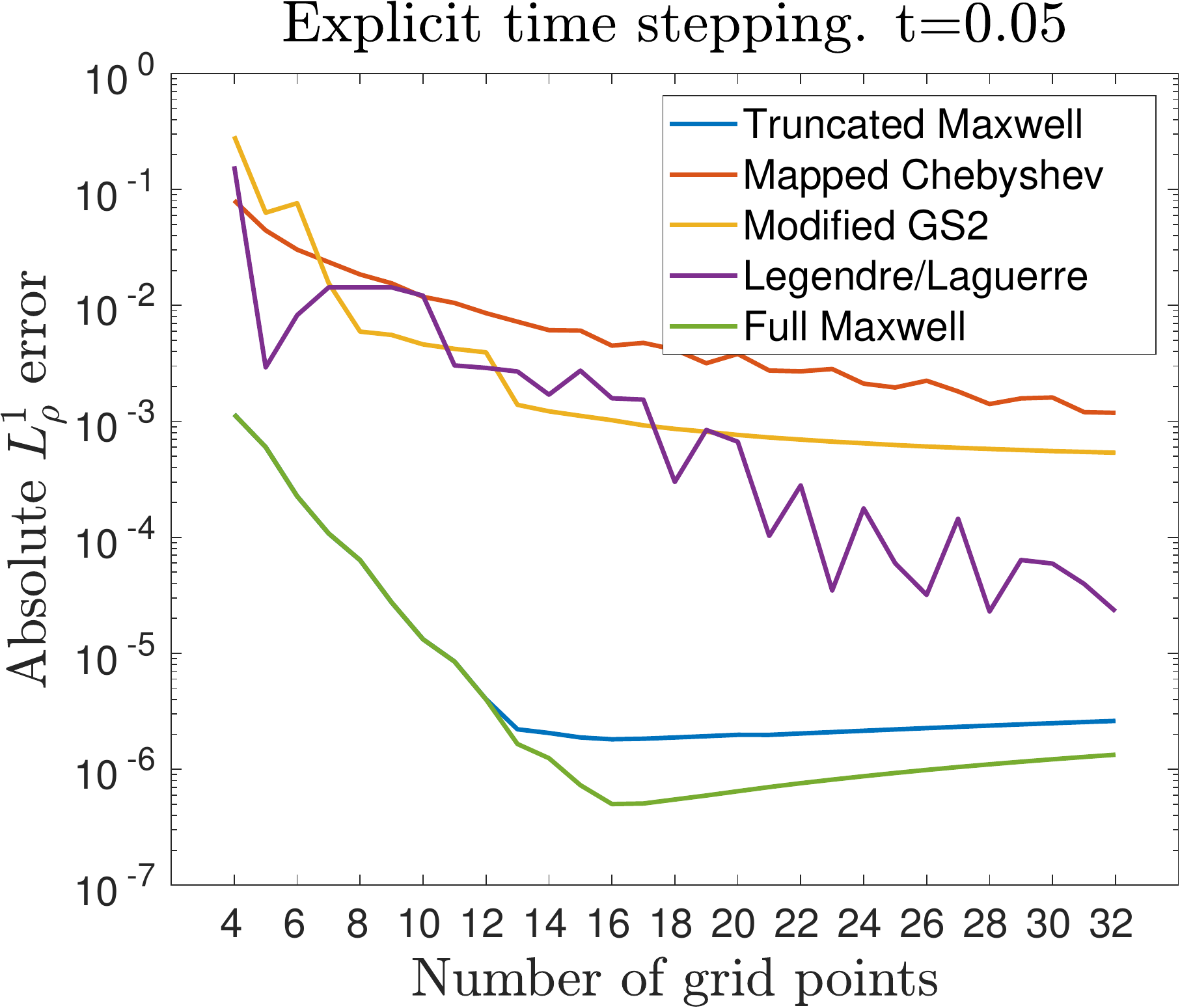} &
\includegraphics[width=.31\linewidth]{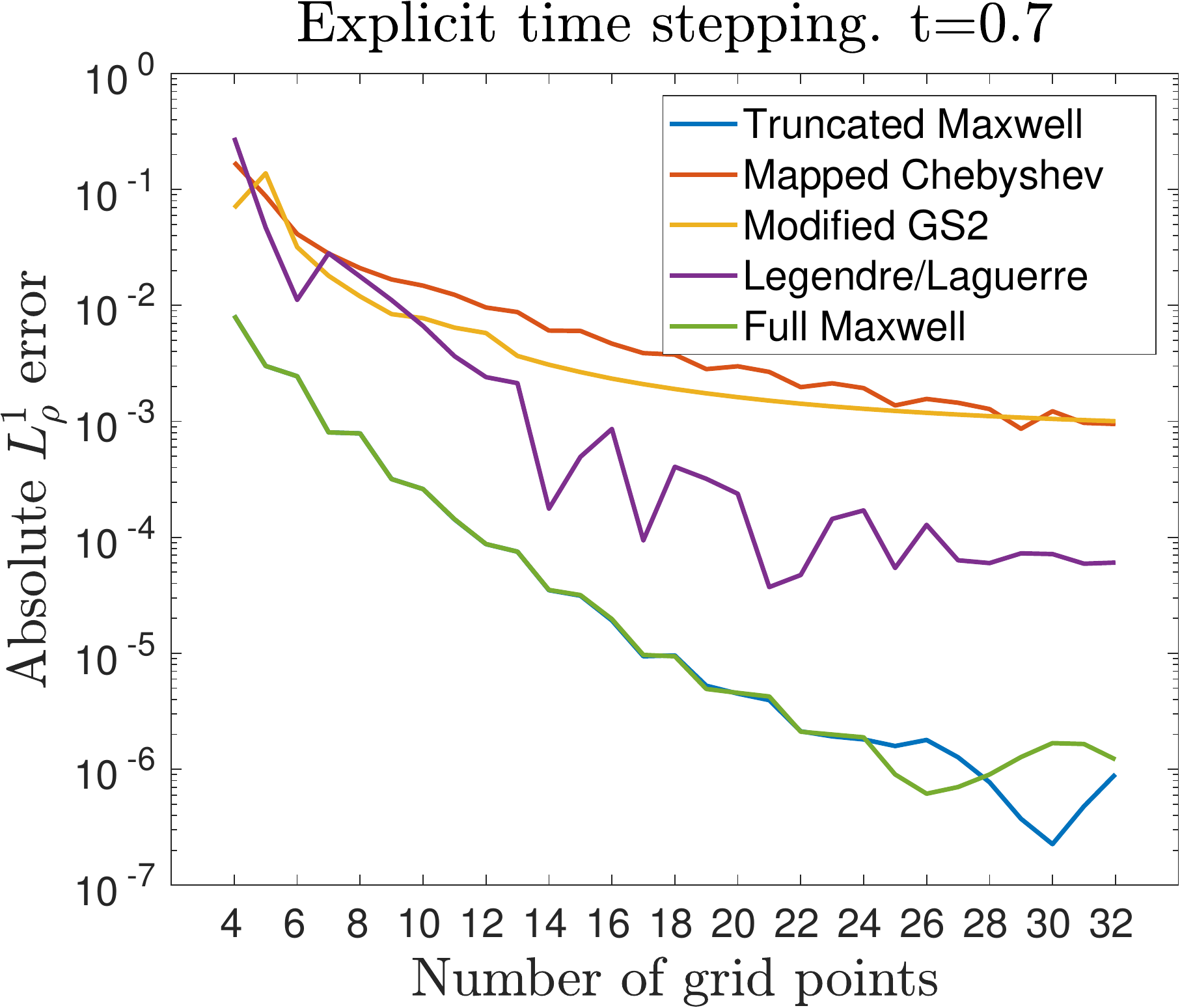} &
\includegraphics[width=.31\linewidth]{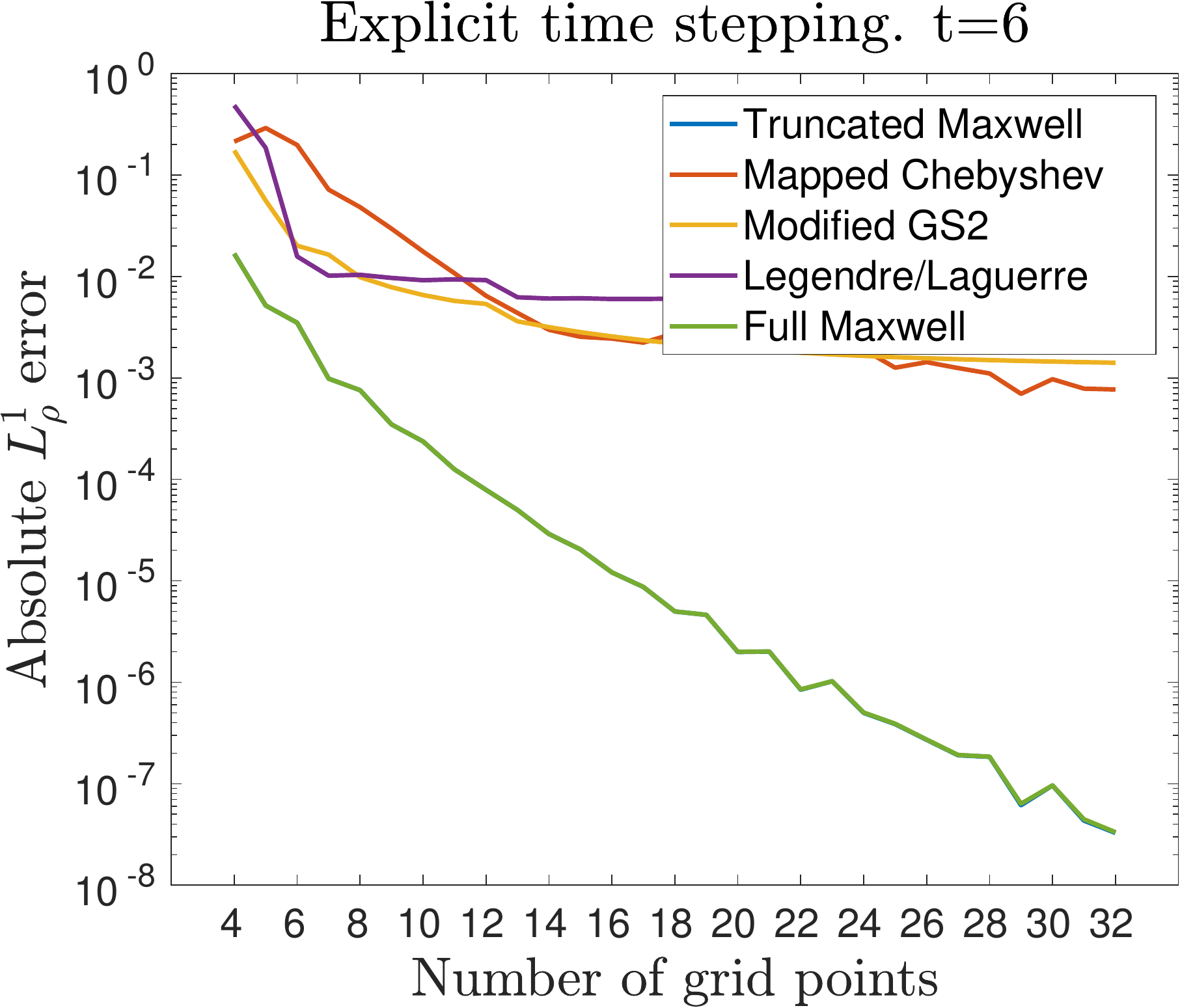} 
\end{tabular}
\caption{{\scriptsize Convergence plots in the $L^1_\rho$ norm for Example 2 with explicit time stepping. The behavior of the full and truncated versions of Maxwell nodes is virtually indistinguishable.}}\label{fig:E2ExplicitL1error}
\end{figure}

\begin{figure}[tb]
\begin{tabular}{ccc}
\includegraphics[width=.31\linewidth]{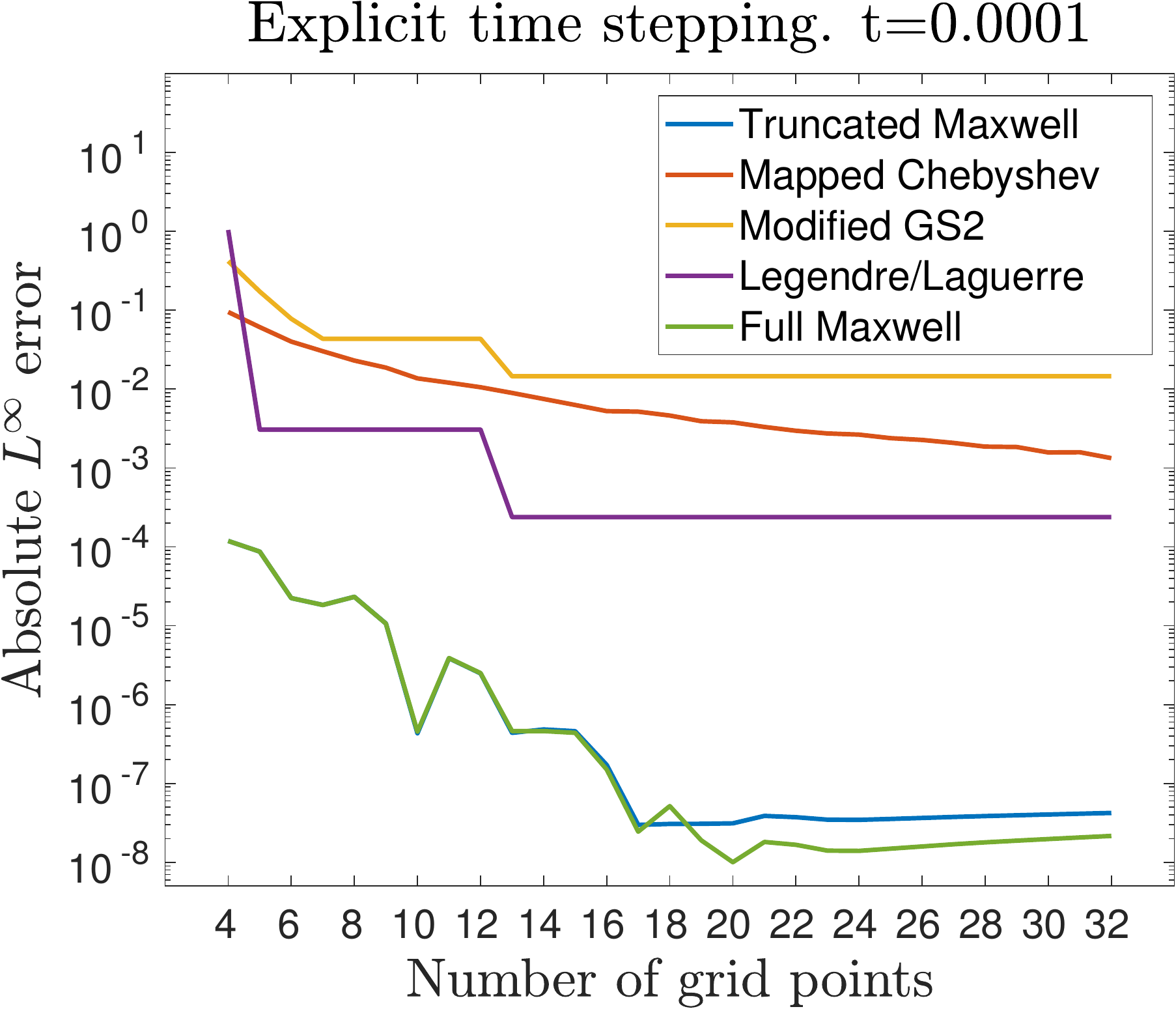} &
\includegraphics[width=.31\linewidth]{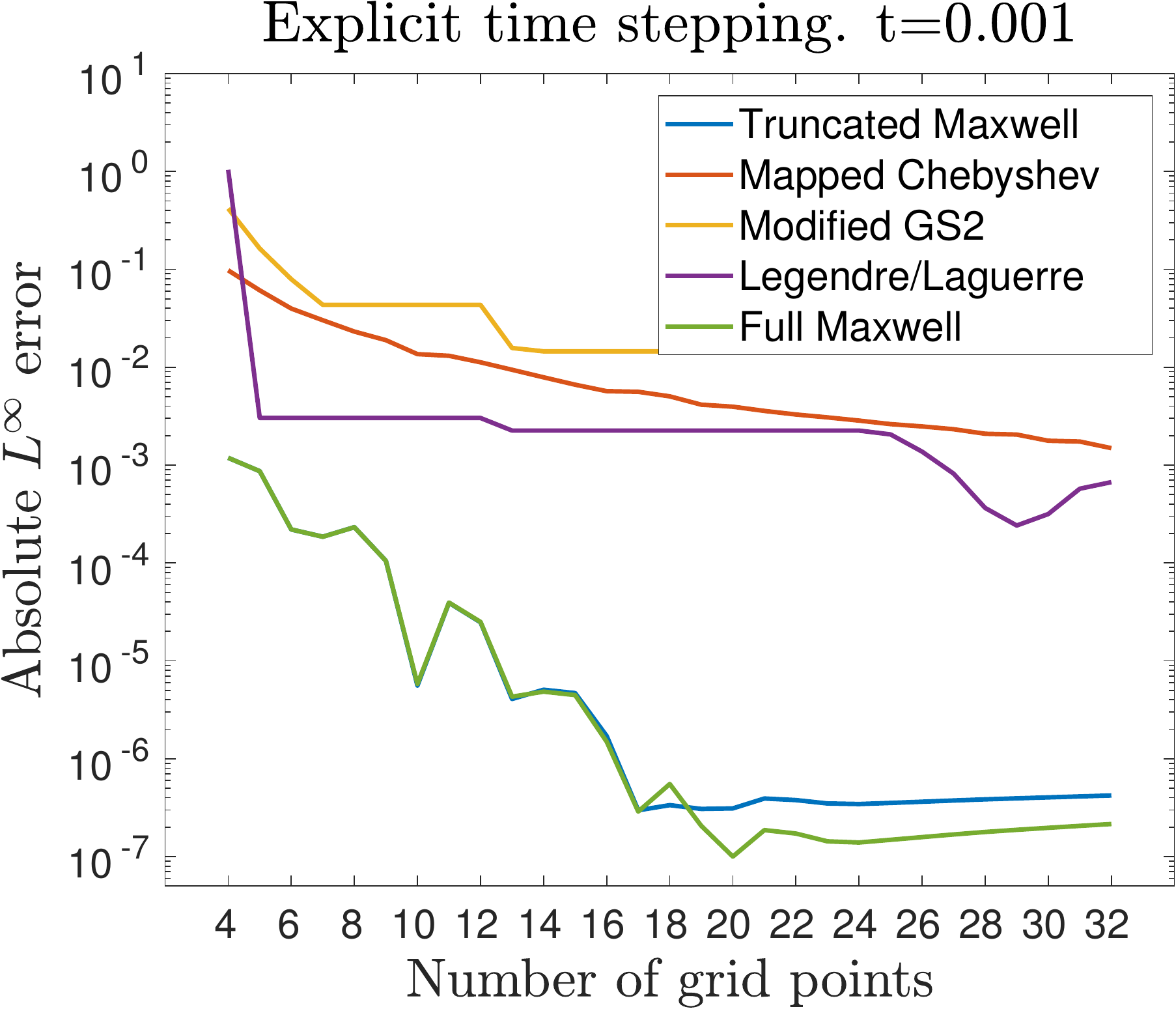} &
\includegraphics[width=.31\linewidth]{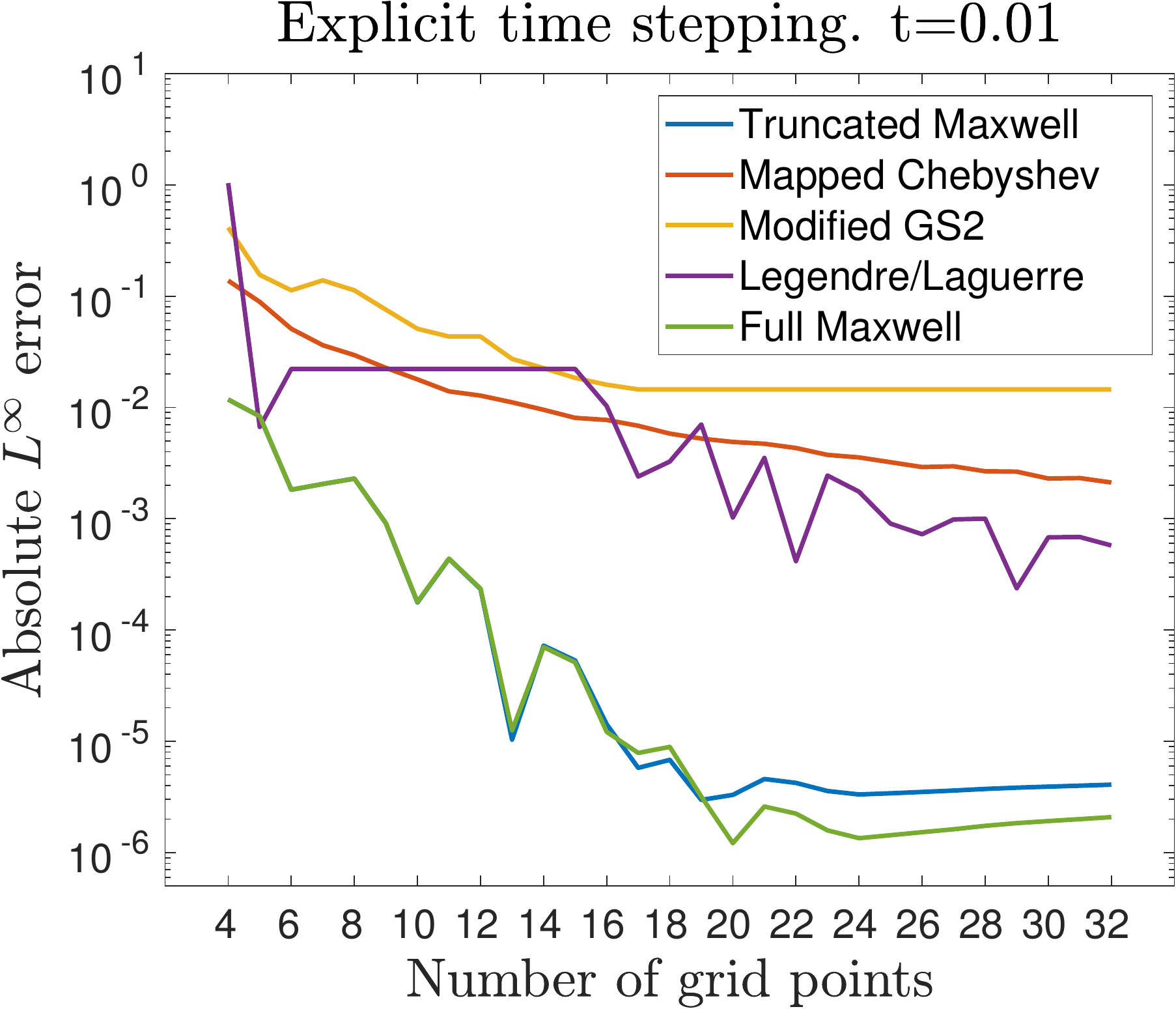} \\
\includegraphics[width=.31\linewidth]{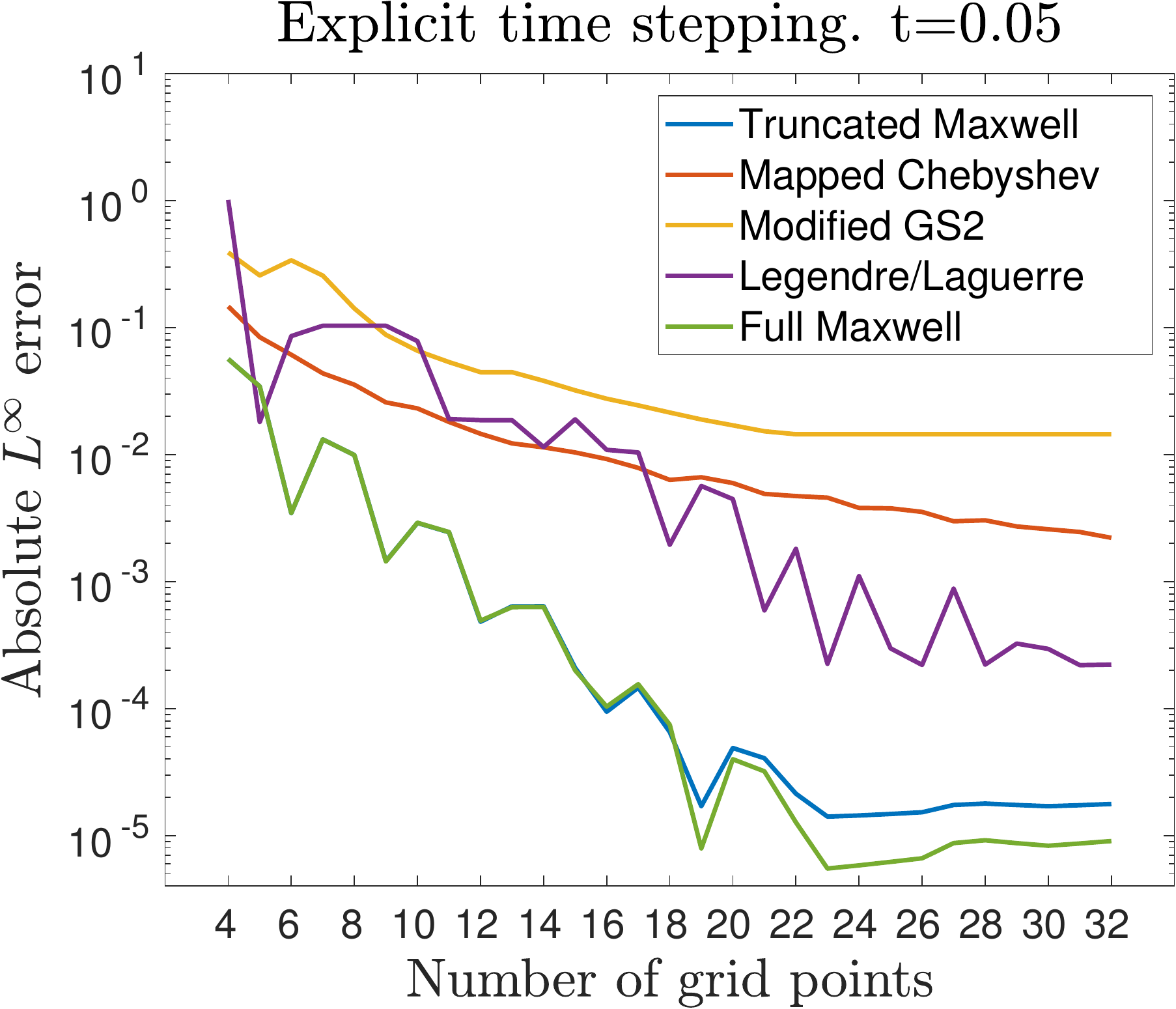} &
\includegraphics[width=.31\linewidth]{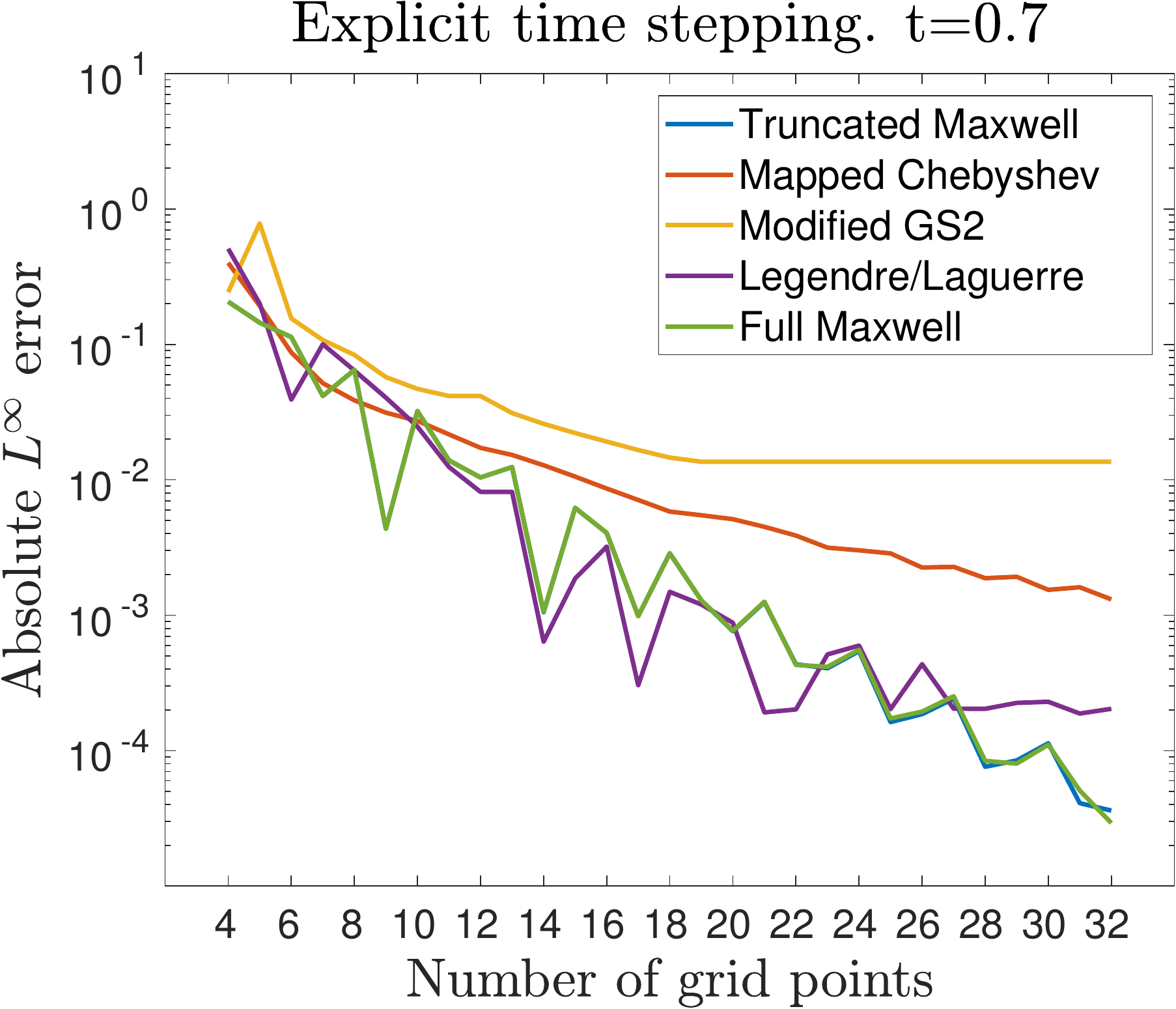} &
\includegraphics[width=.31\linewidth]{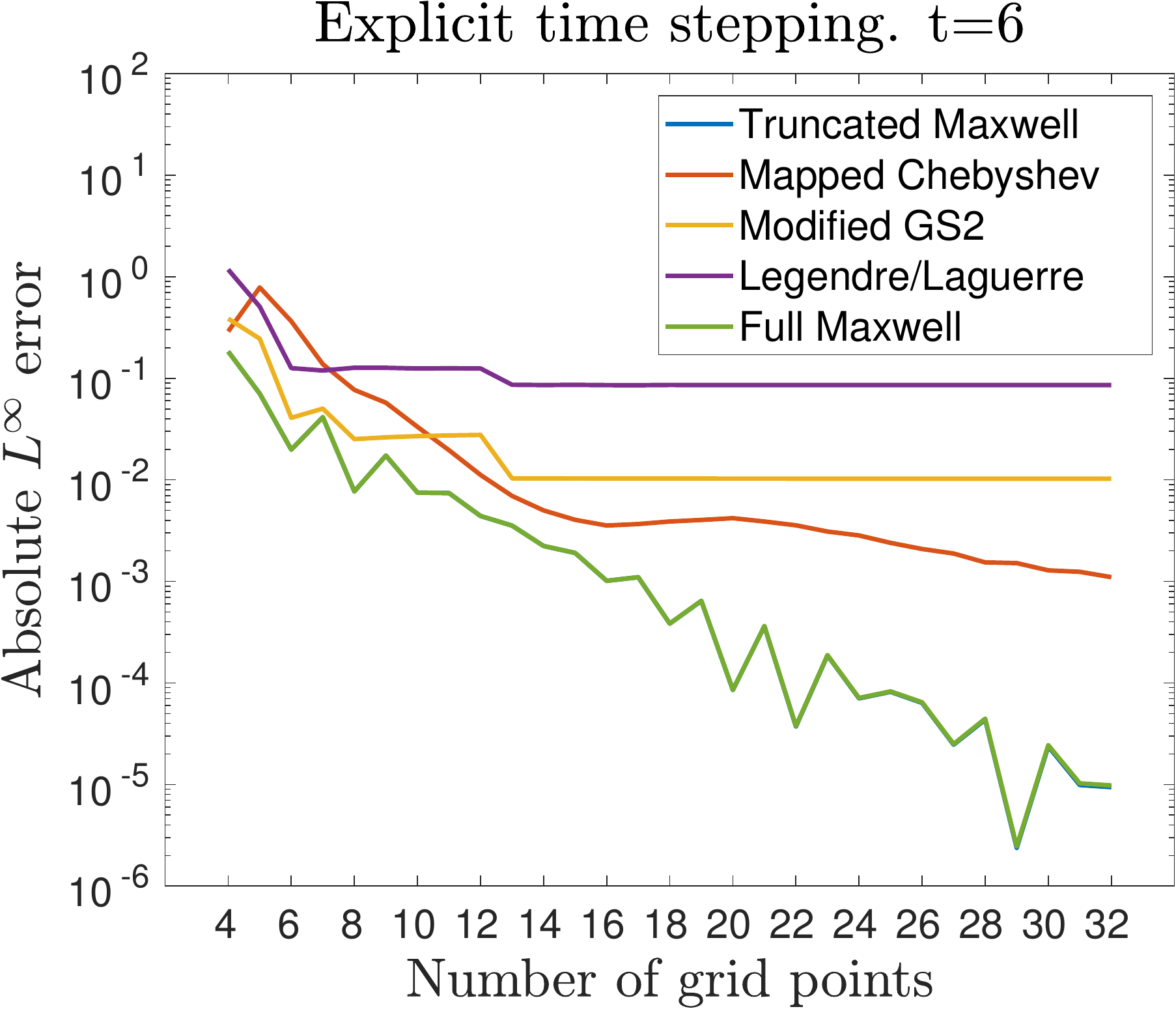} 
\end{tabular}
\caption{{\scriptsize Convergence plots in the $L^\infty$ norm for Example 2 with explicit time stepping. The behavior of the full and truncated versions of Maxwell nodes is virtually indistinguishable.}}\label{fig:E2ExplicitMAXerror}
\end{figure}

\begin{figure}[tb]
\begin{tabular}{ccc}
\includegraphics[width=.31\linewidth]{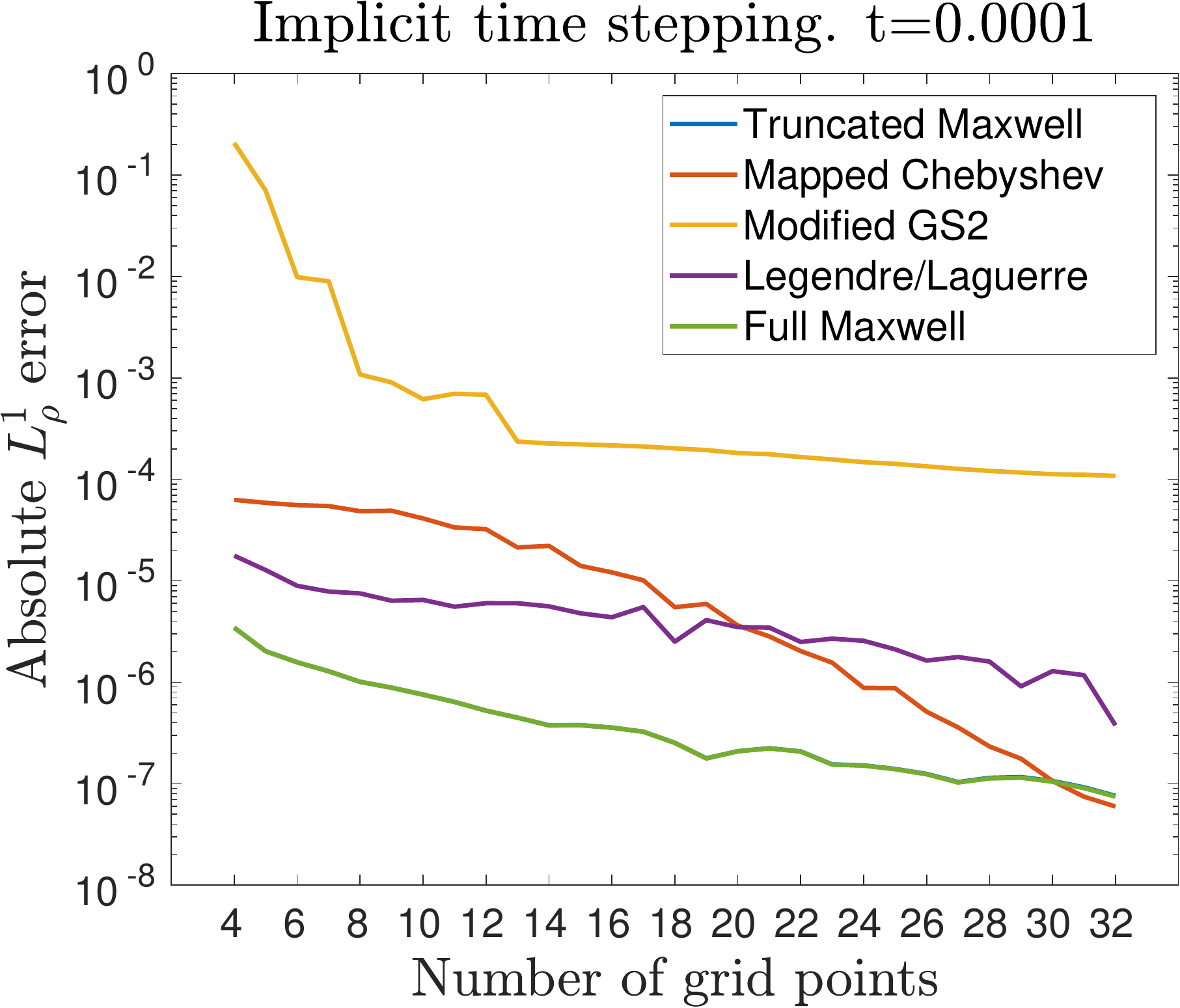} &
\includegraphics[width=.31\linewidth]{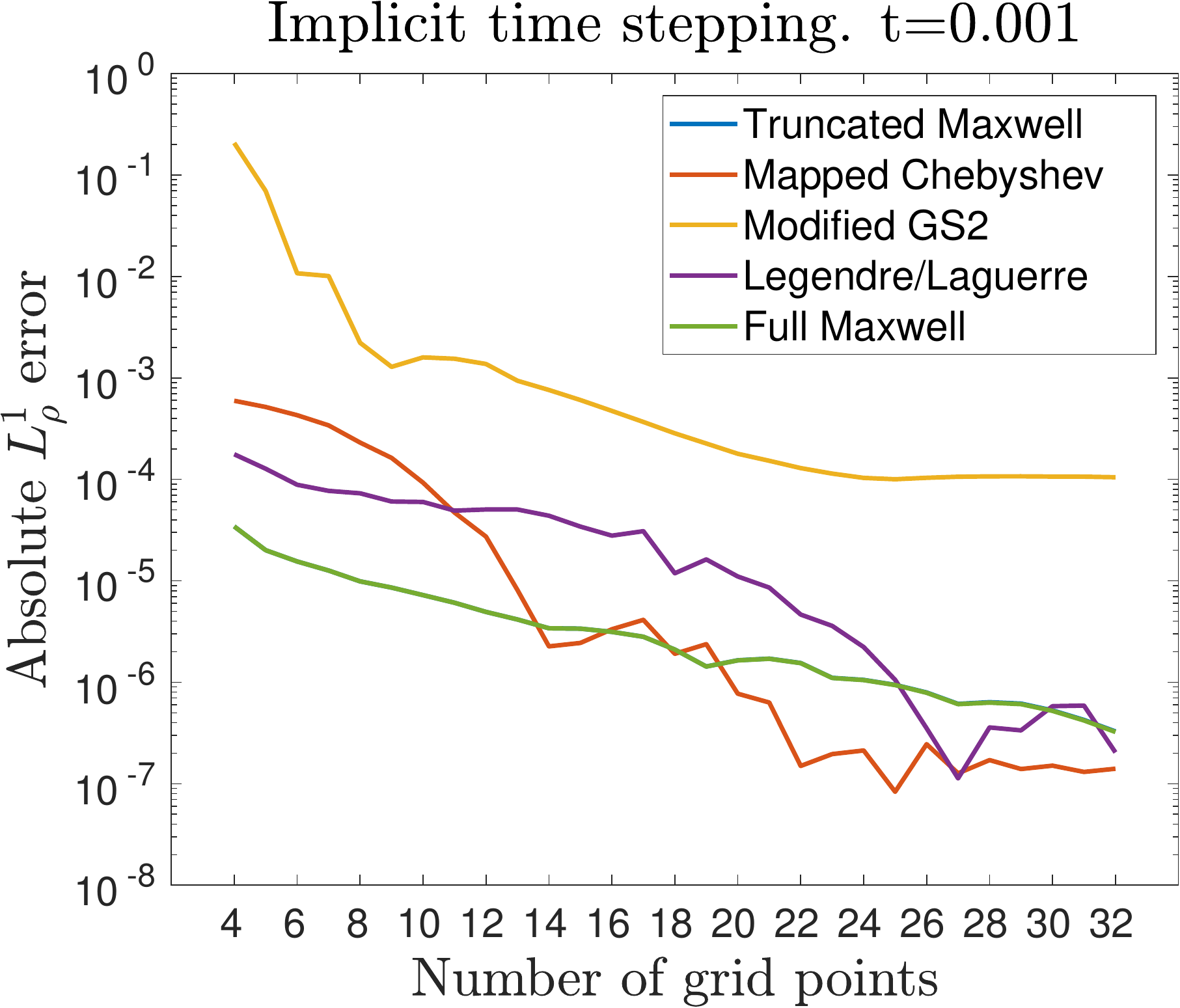} &
\includegraphics[width=.31\linewidth]{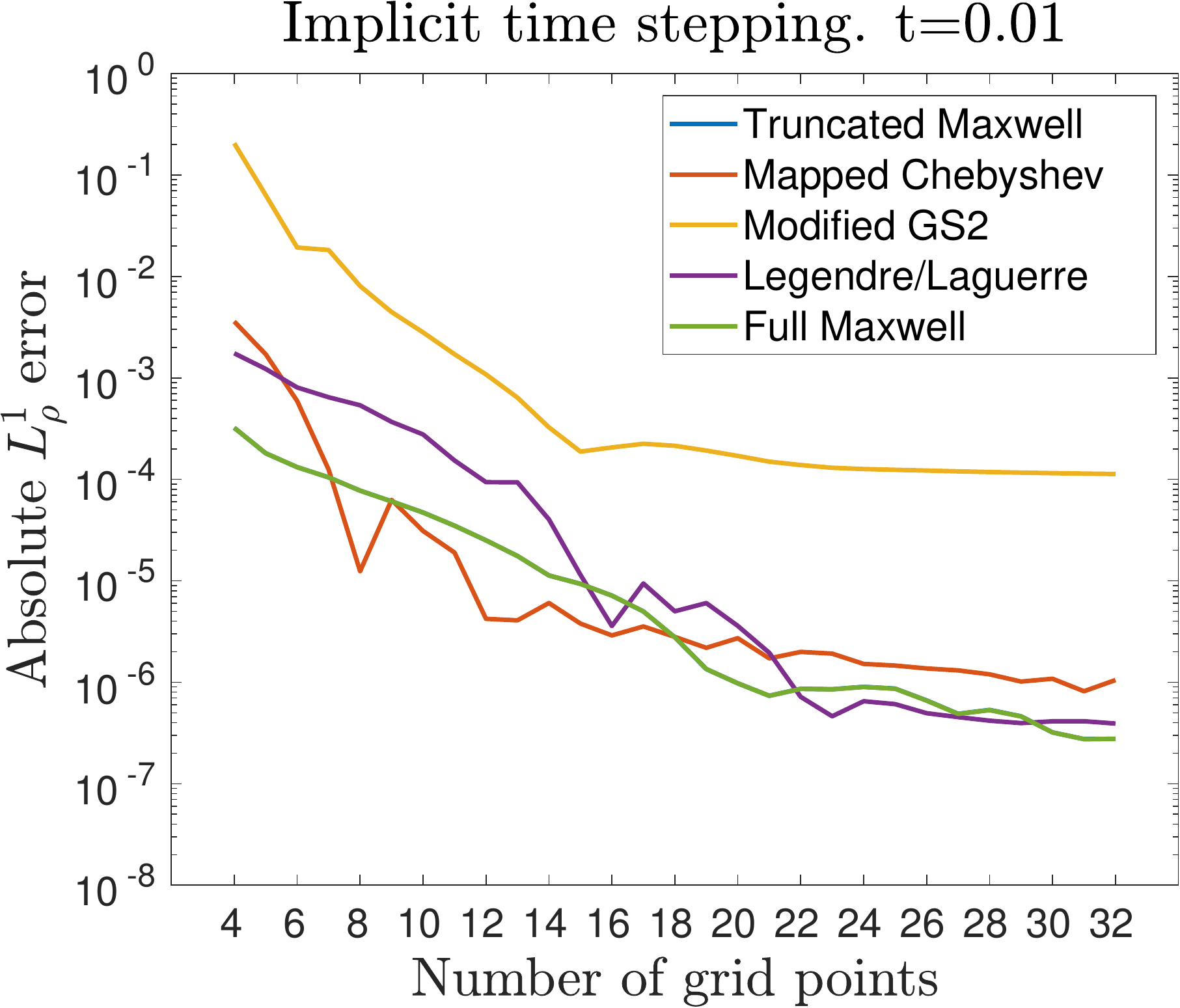} \\
\includegraphics[width=.31\linewidth]{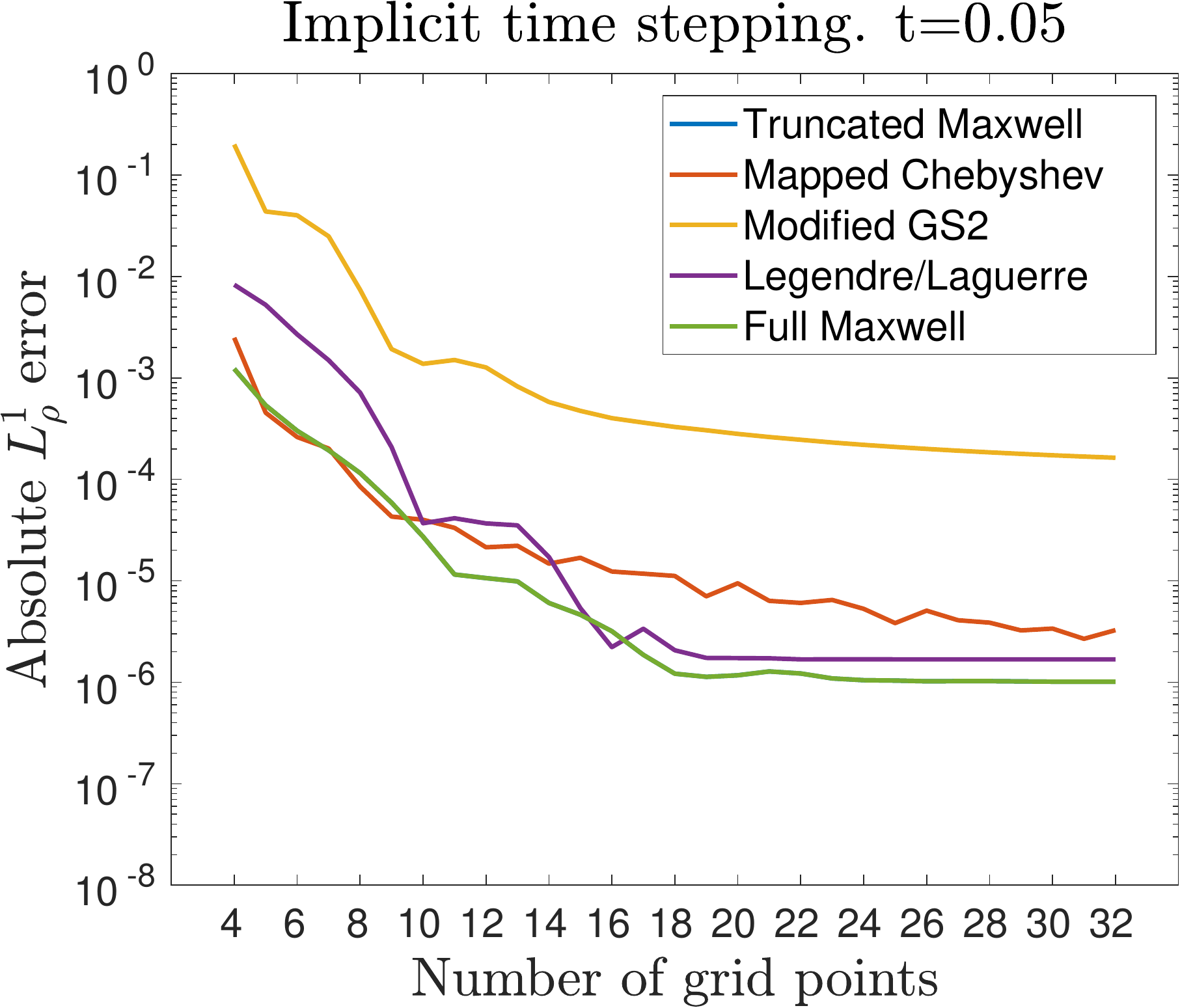} &
\includegraphics[width=.31\linewidth]{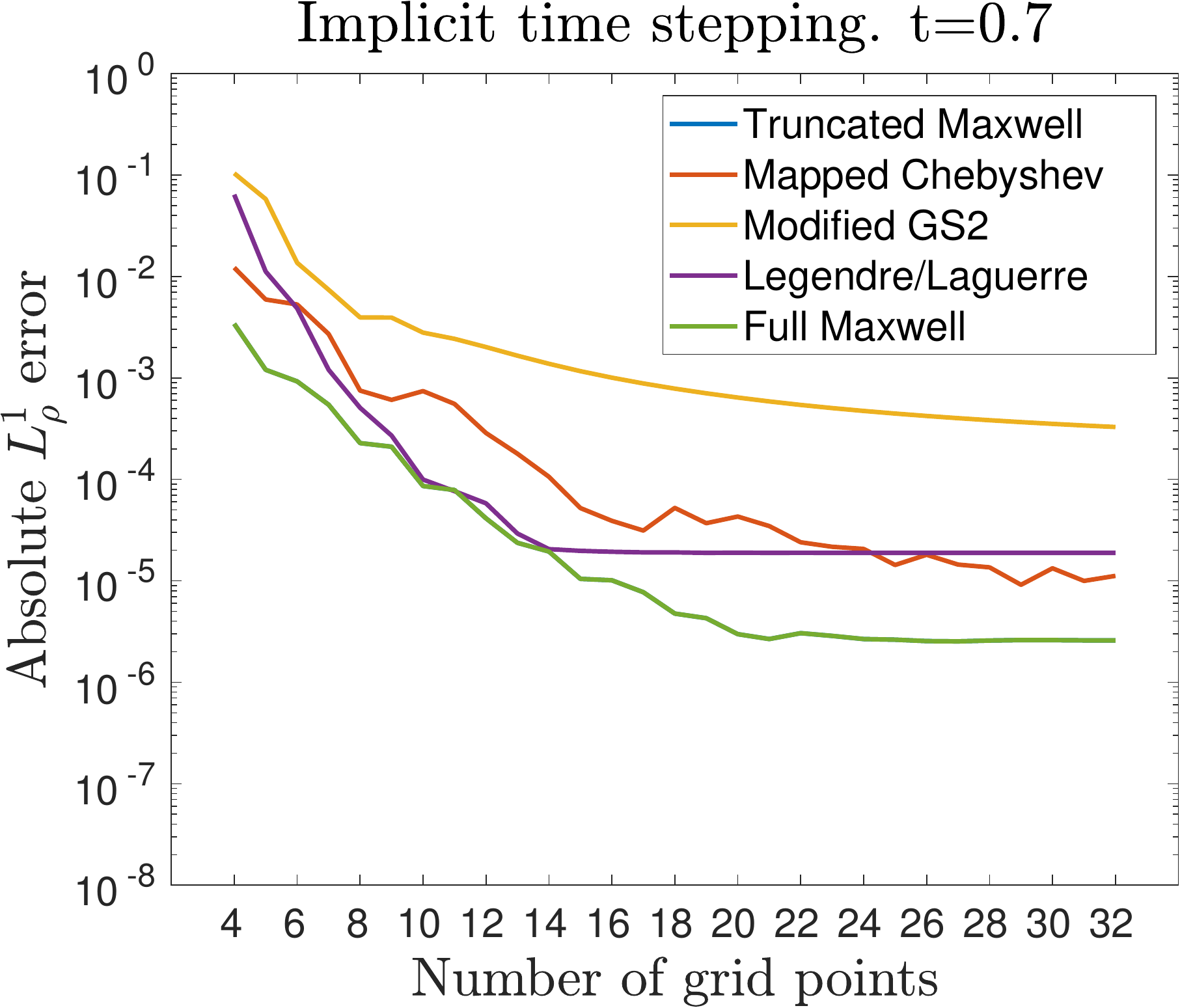} &
\includegraphics[width=.31\linewidth]{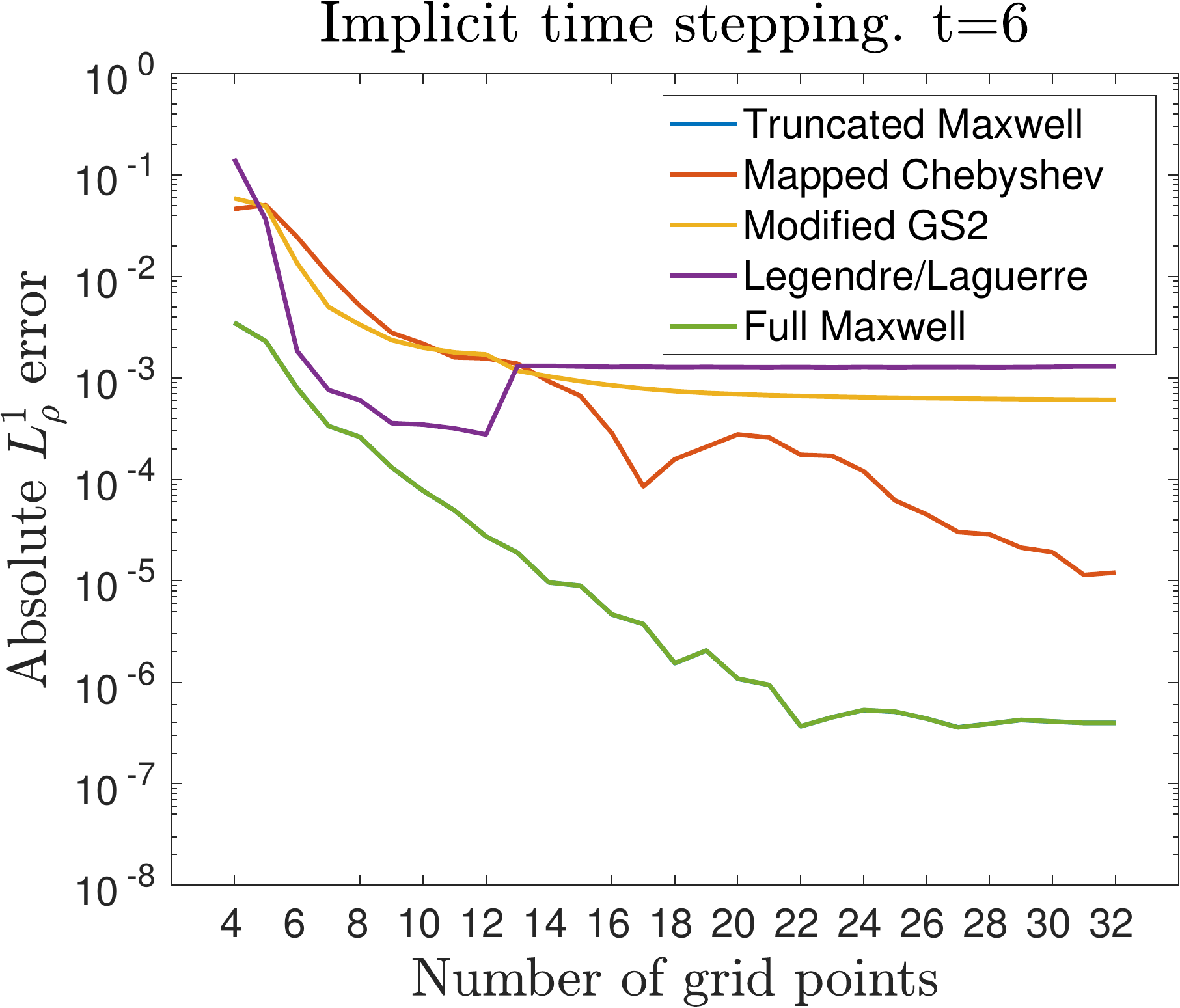} 
\end{tabular}
\caption{{\scriptsize Convergence plots in the $L^1_\rho$ norm for Example 1 with implicit time stepping. The behavior of the full and truncated versions of Maxwell nodes is virtually indistinguishable.}}\label{fig:E1ImplicitL1error}
\end{figure}

\begin{figure}[tb]
\begin{tabular}{ccc}
\includegraphics[width=.31\linewidth]{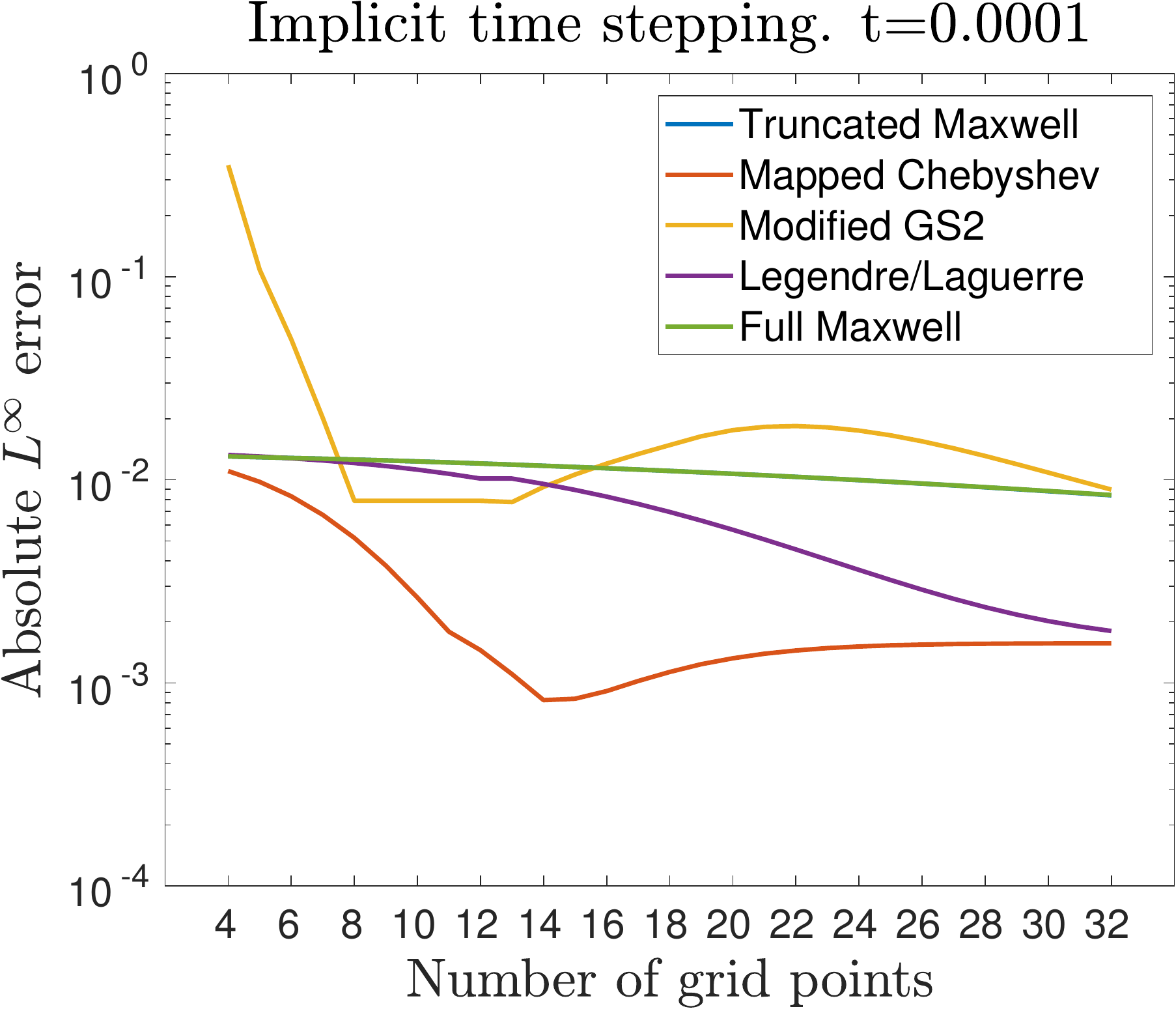} &
\includegraphics[width=.31\linewidth]{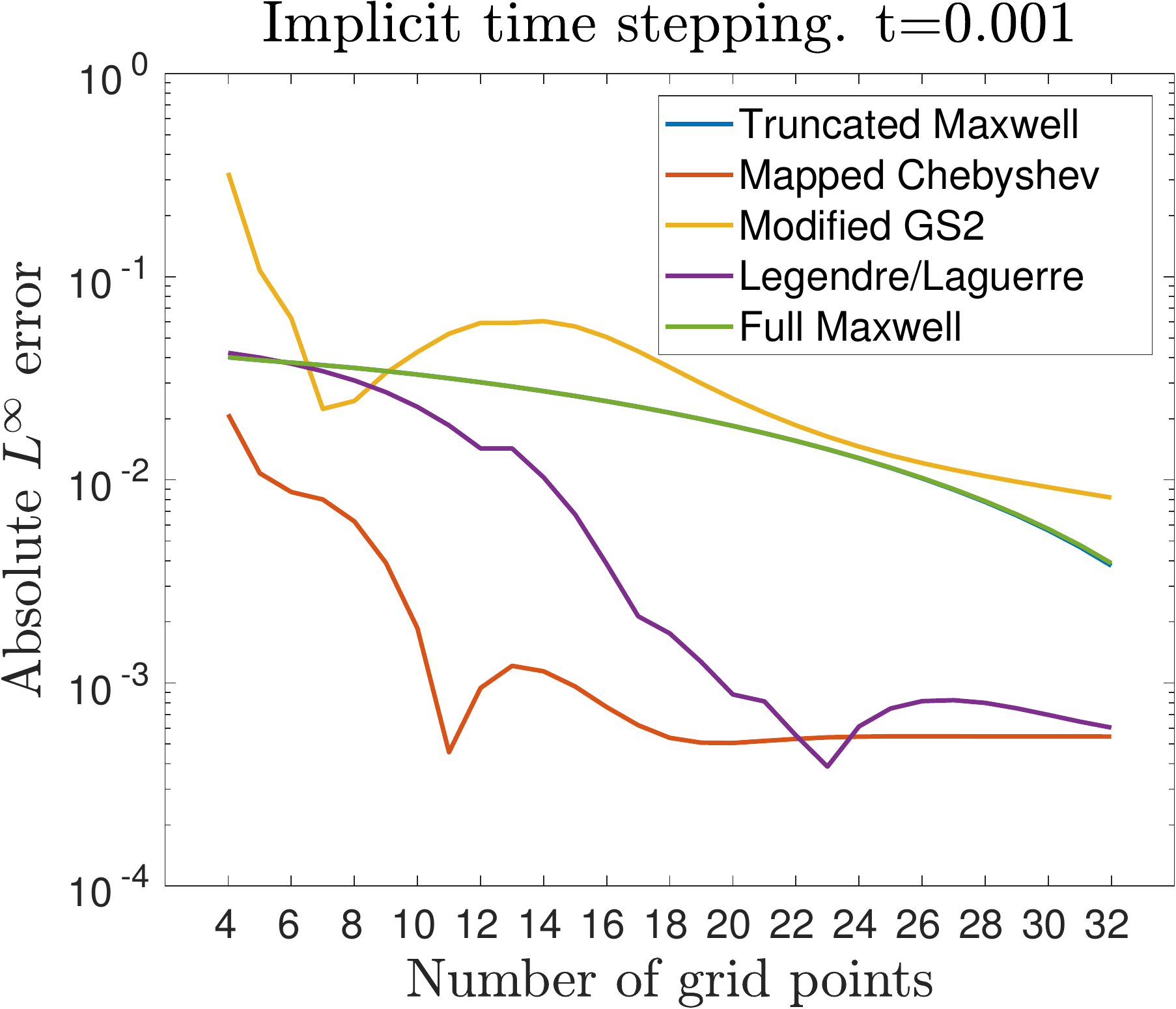} &
\includegraphics[width=.31\linewidth]{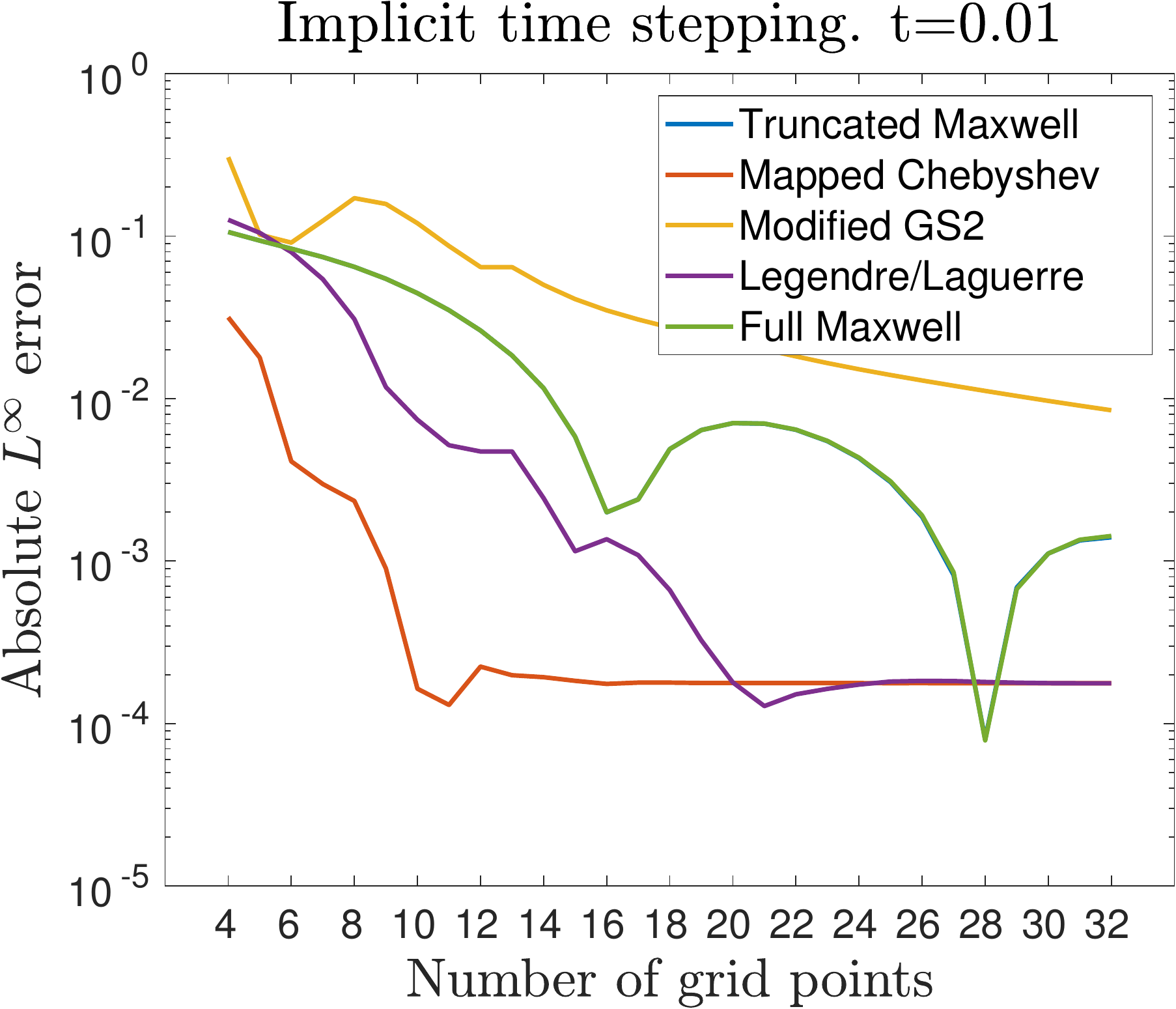} \\
\includegraphics[width=.31\linewidth]{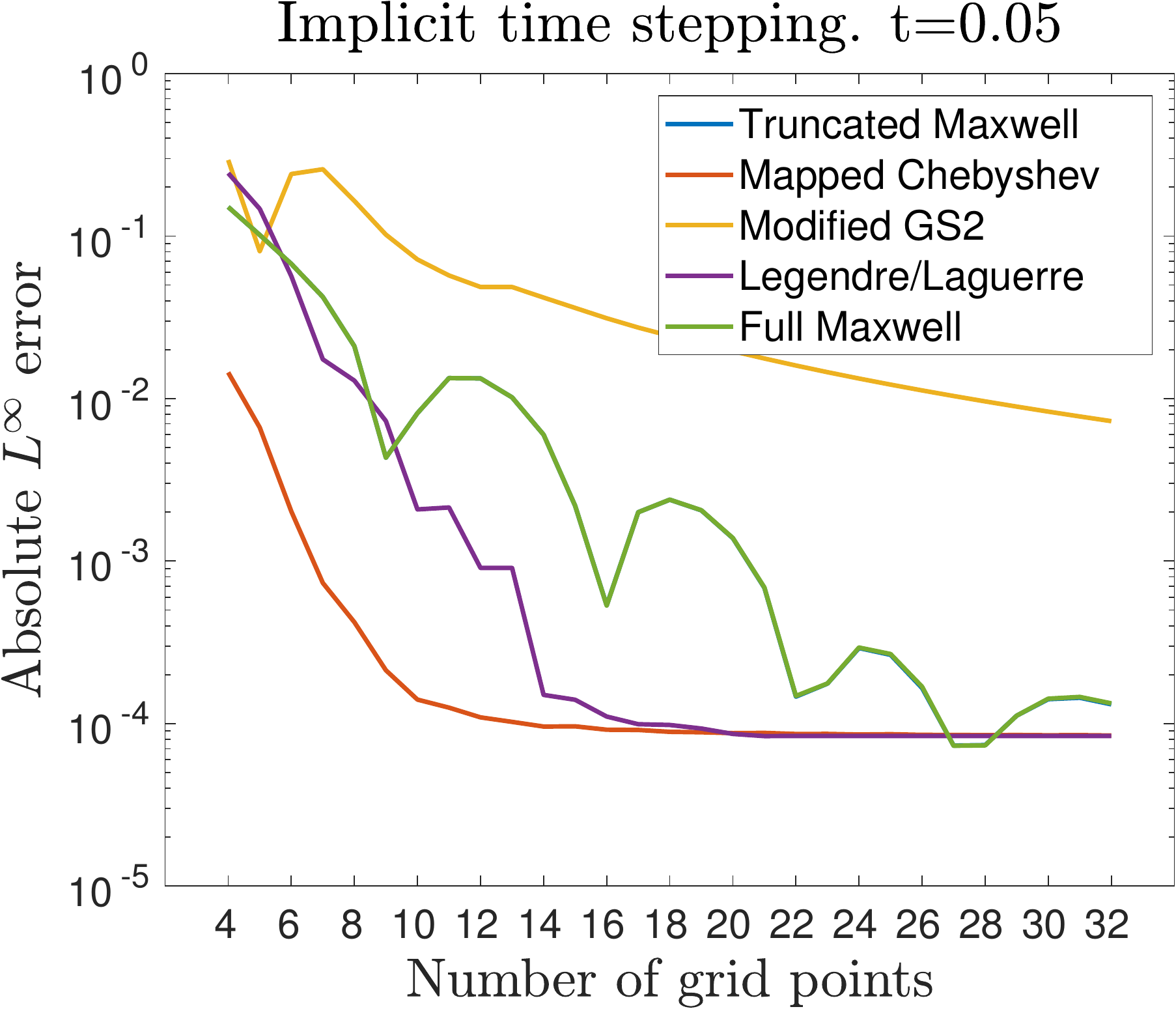} &
\includegraphics[width=.31\linewidth]{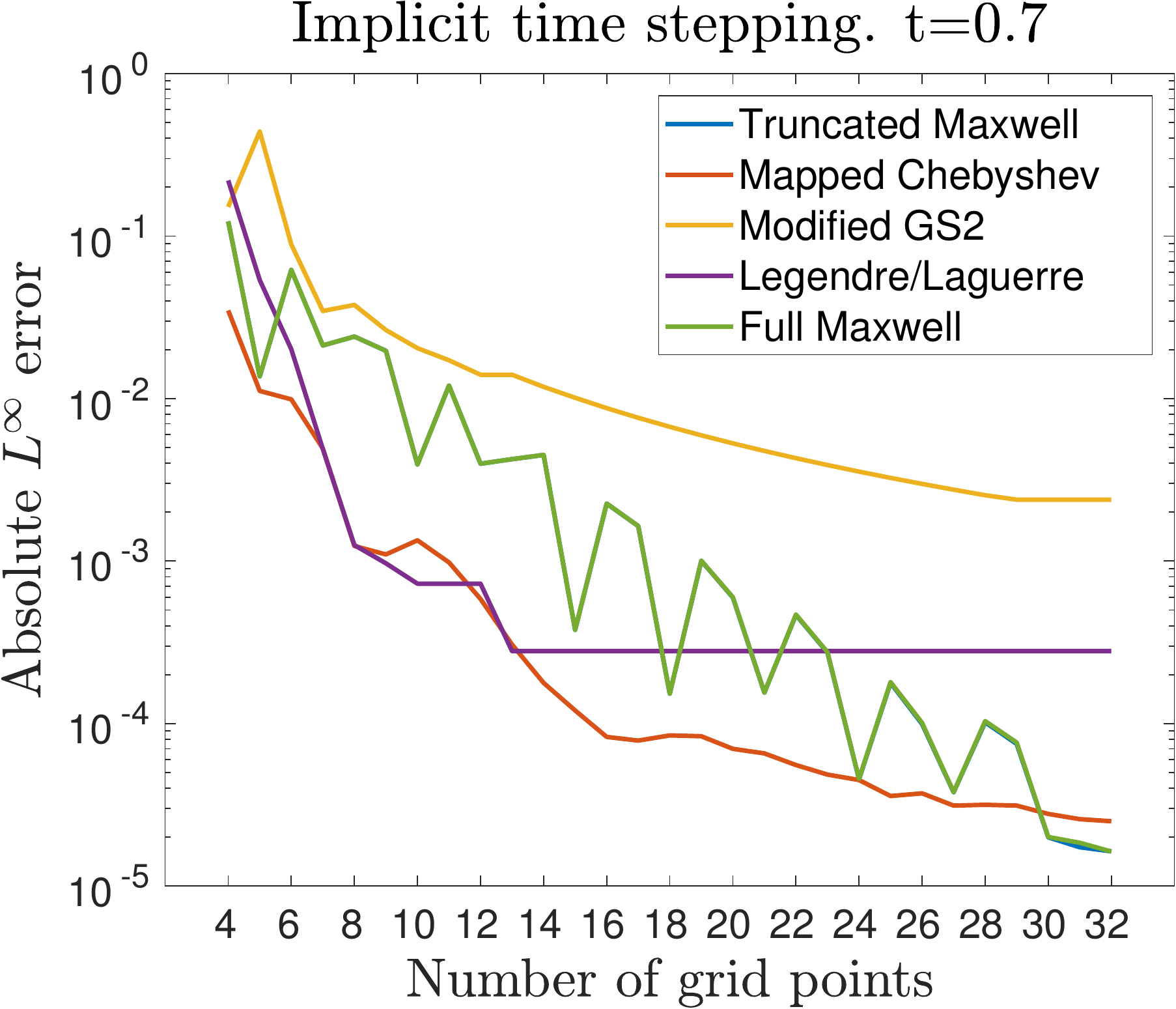} &
\includegraphics[width=.31\linewidth]{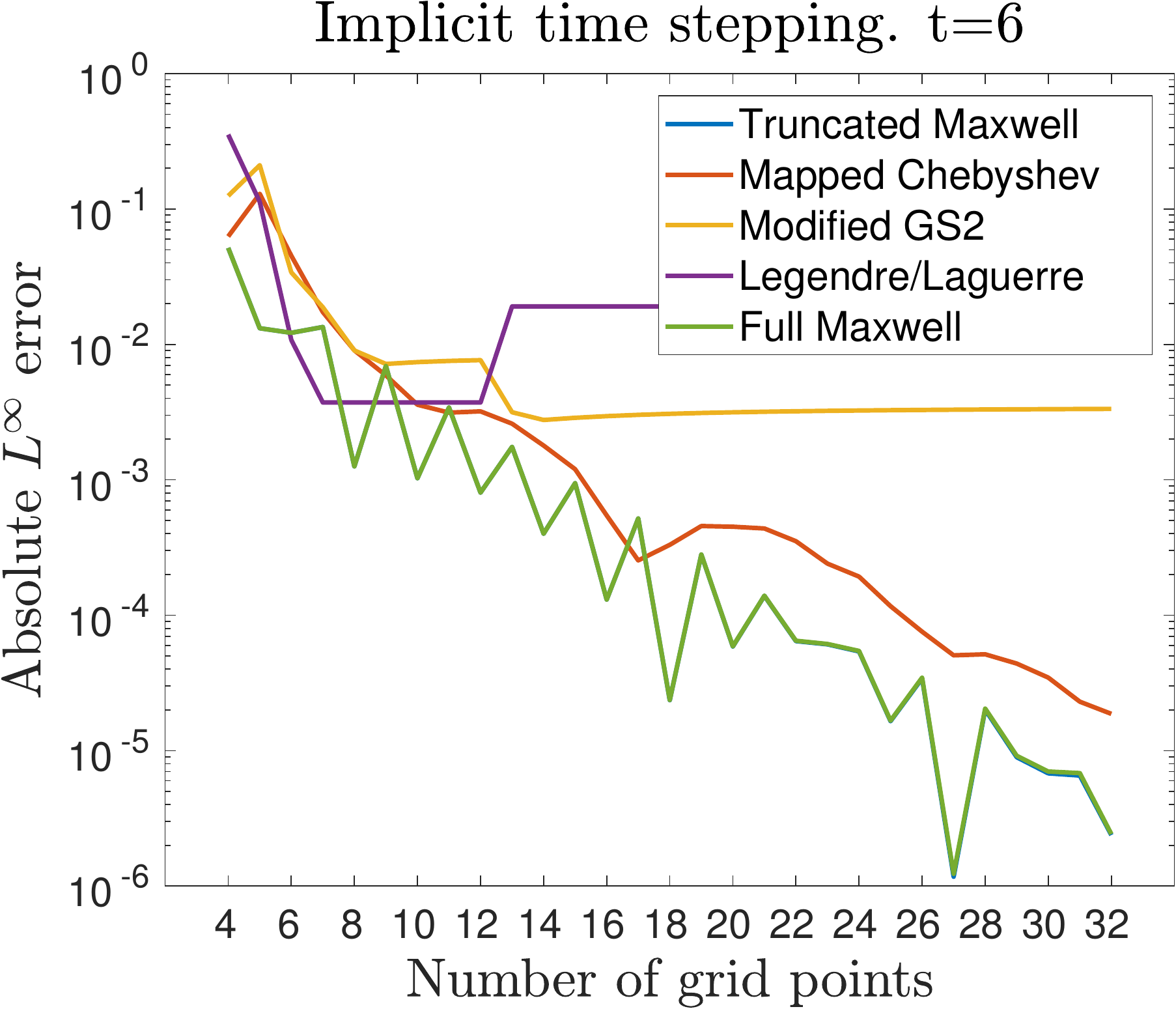} 
\end{tabular}
\caption{{\scriptsize Convergence plots in the $L^\infty$ norm for Example 1 with implicit time stepping. The behavior of the full and truncated versions of Maxwell nodes is virtually indistinguishable.}}\label{fig:E1ImplicitMAXerror}
\end{figure}

\begin{figure}[tb]
\begin{tabular}{ccc}
\includegraphics[width=.31\linewidth]{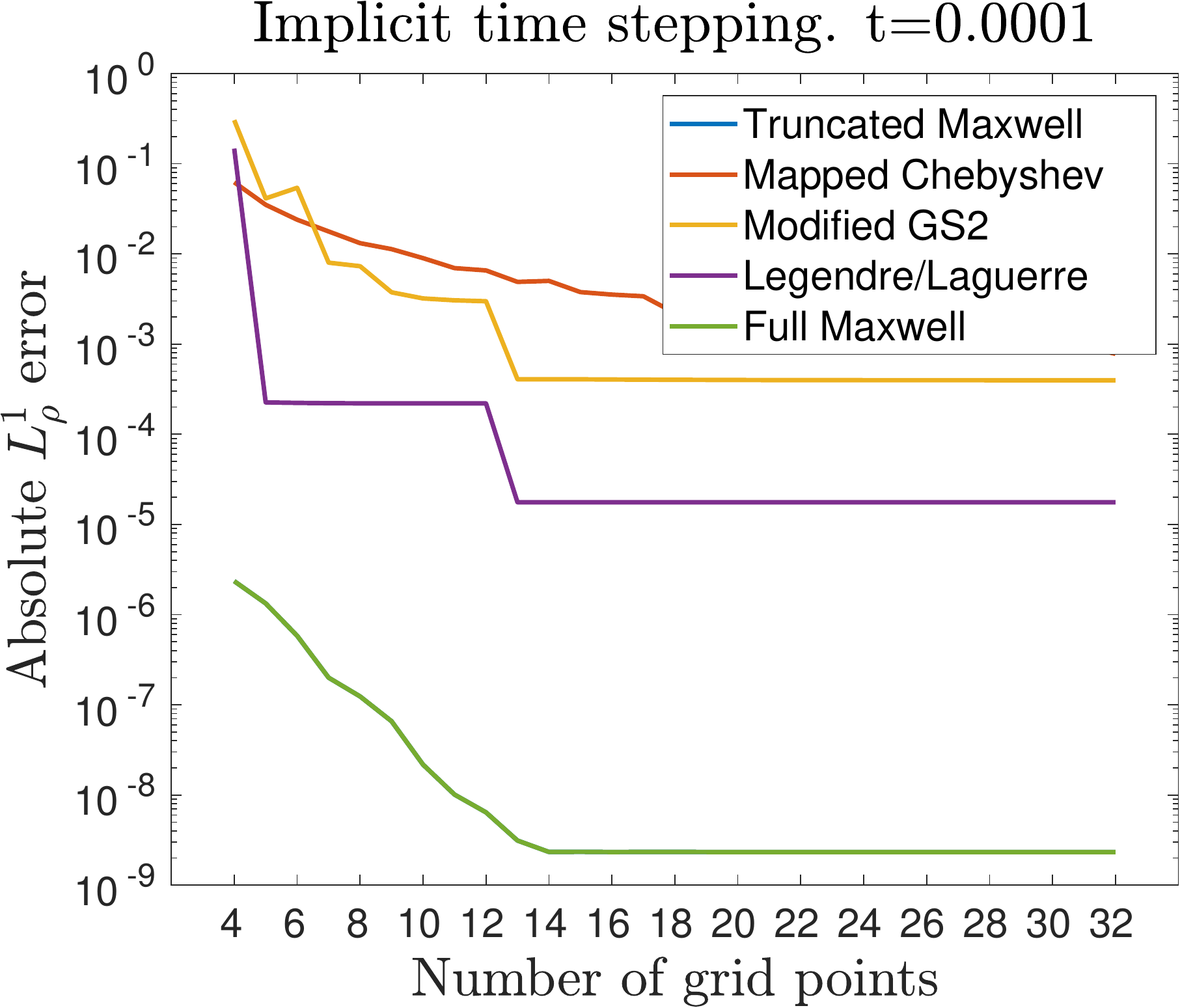} &
\includegraphics[width=.31\linewidth]{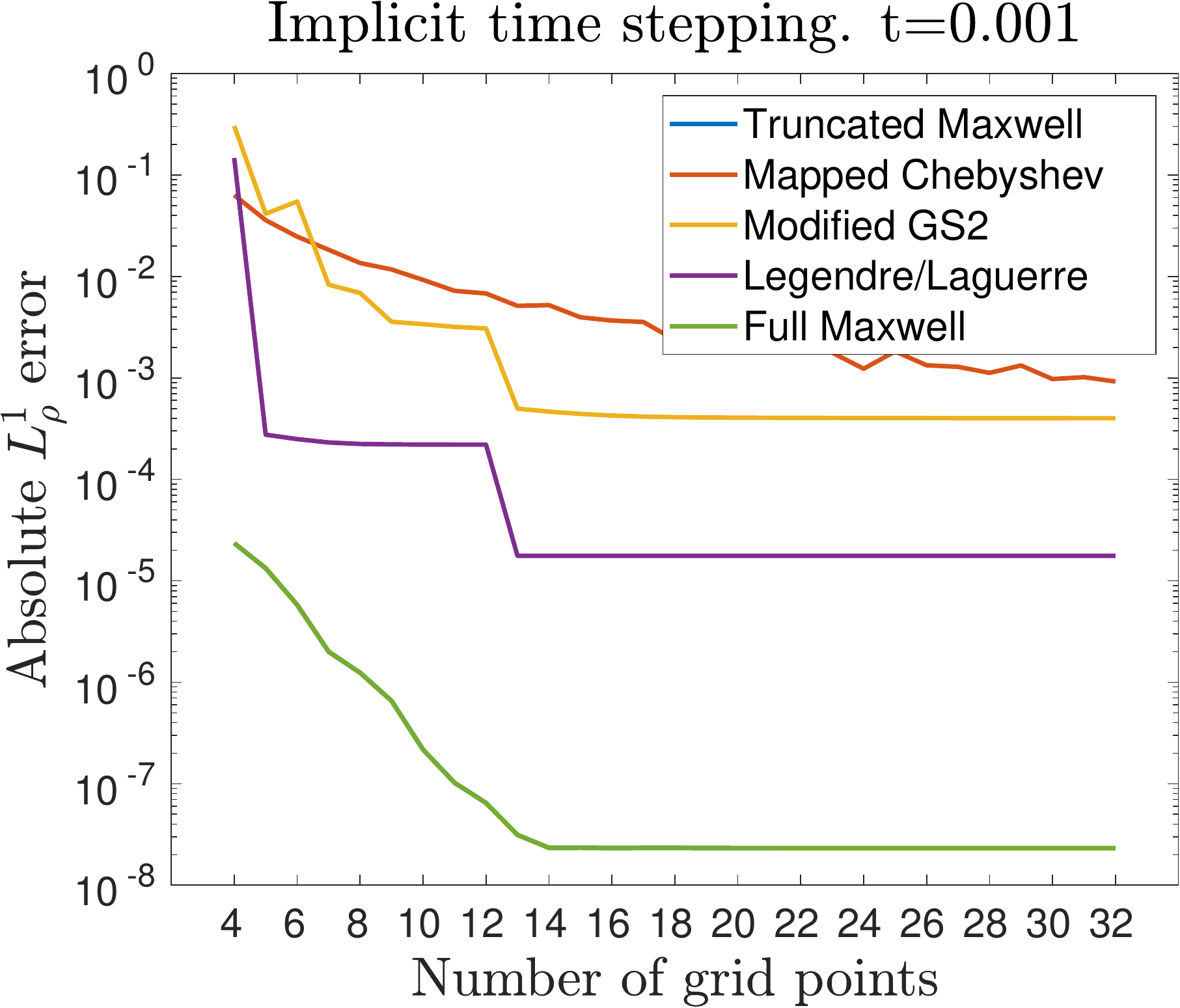} &
\includegraphics[width=.31\linewidth]{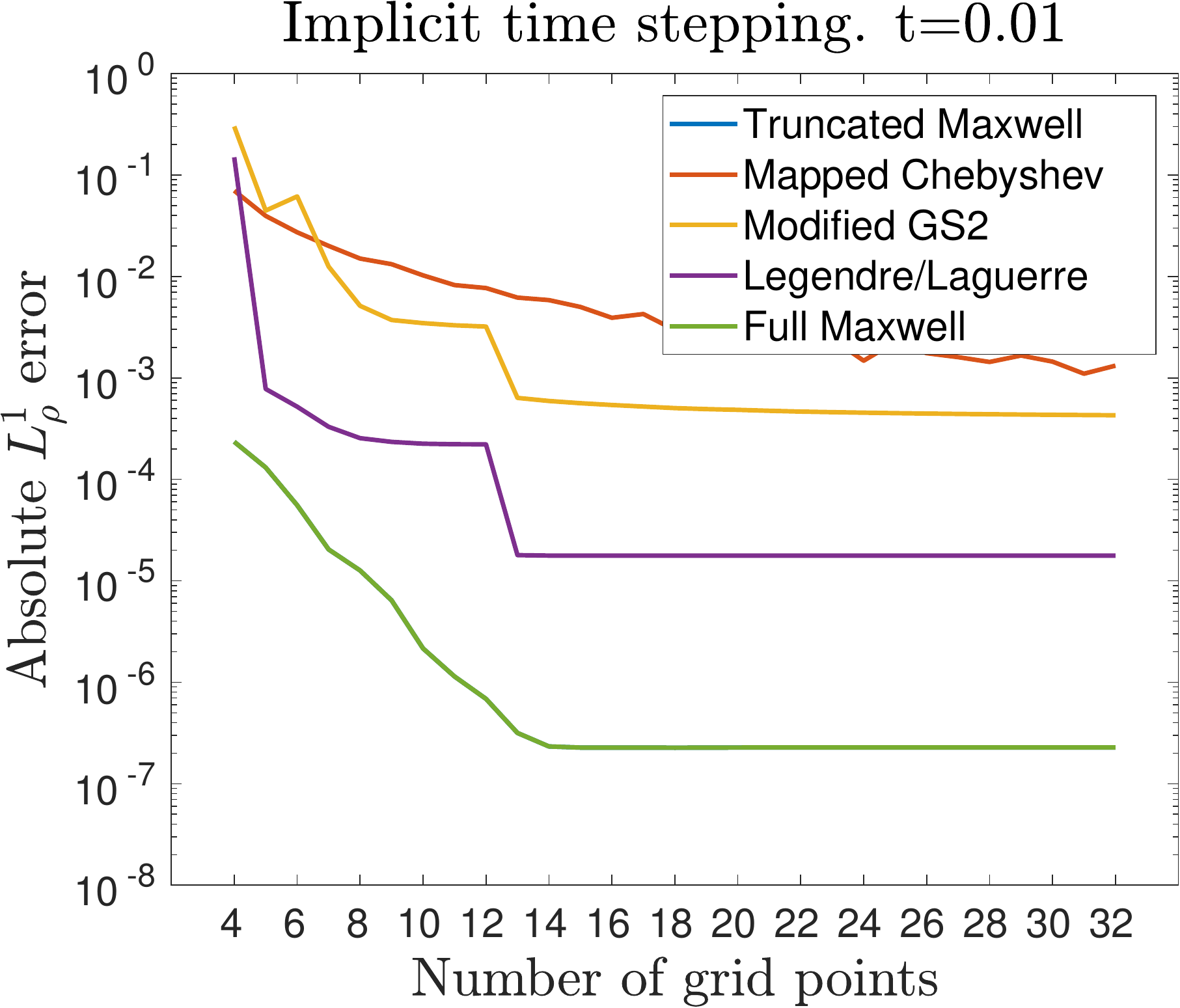} \\
\includegraphics[width=.31\linewidth]{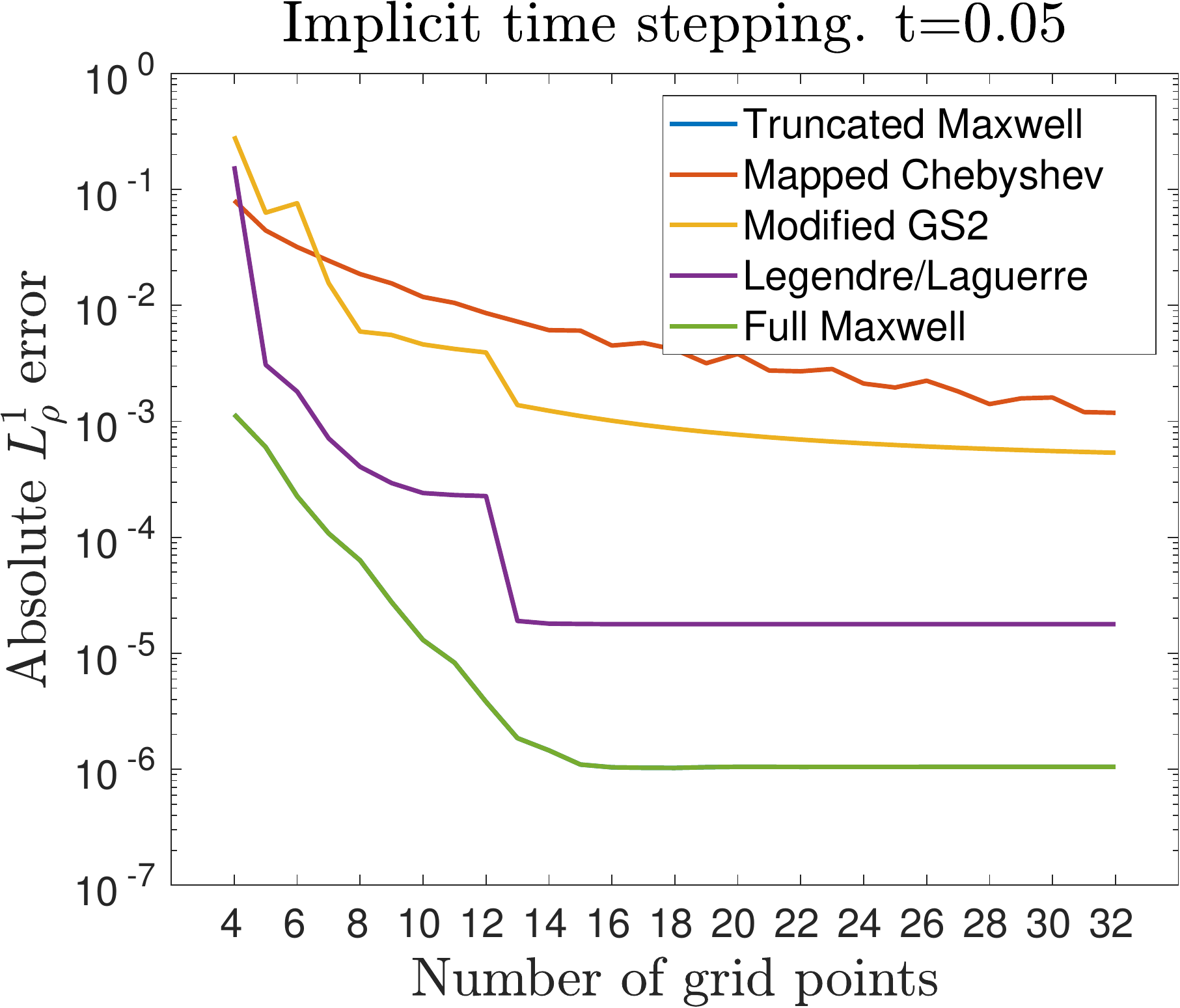} &
\includegraphics[width=.31\linewidth]{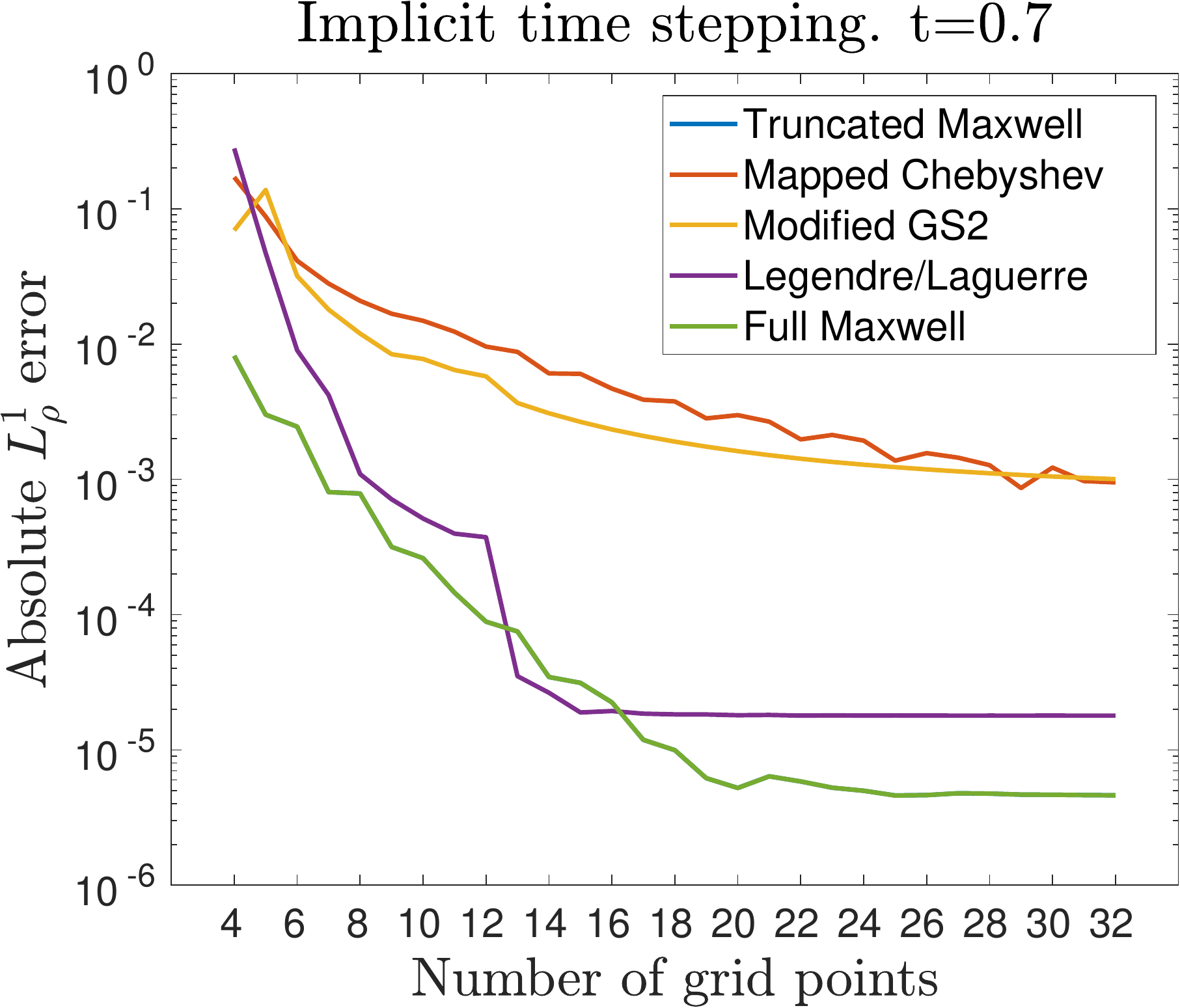} &
\includegraphics[width=.31\linewidth]{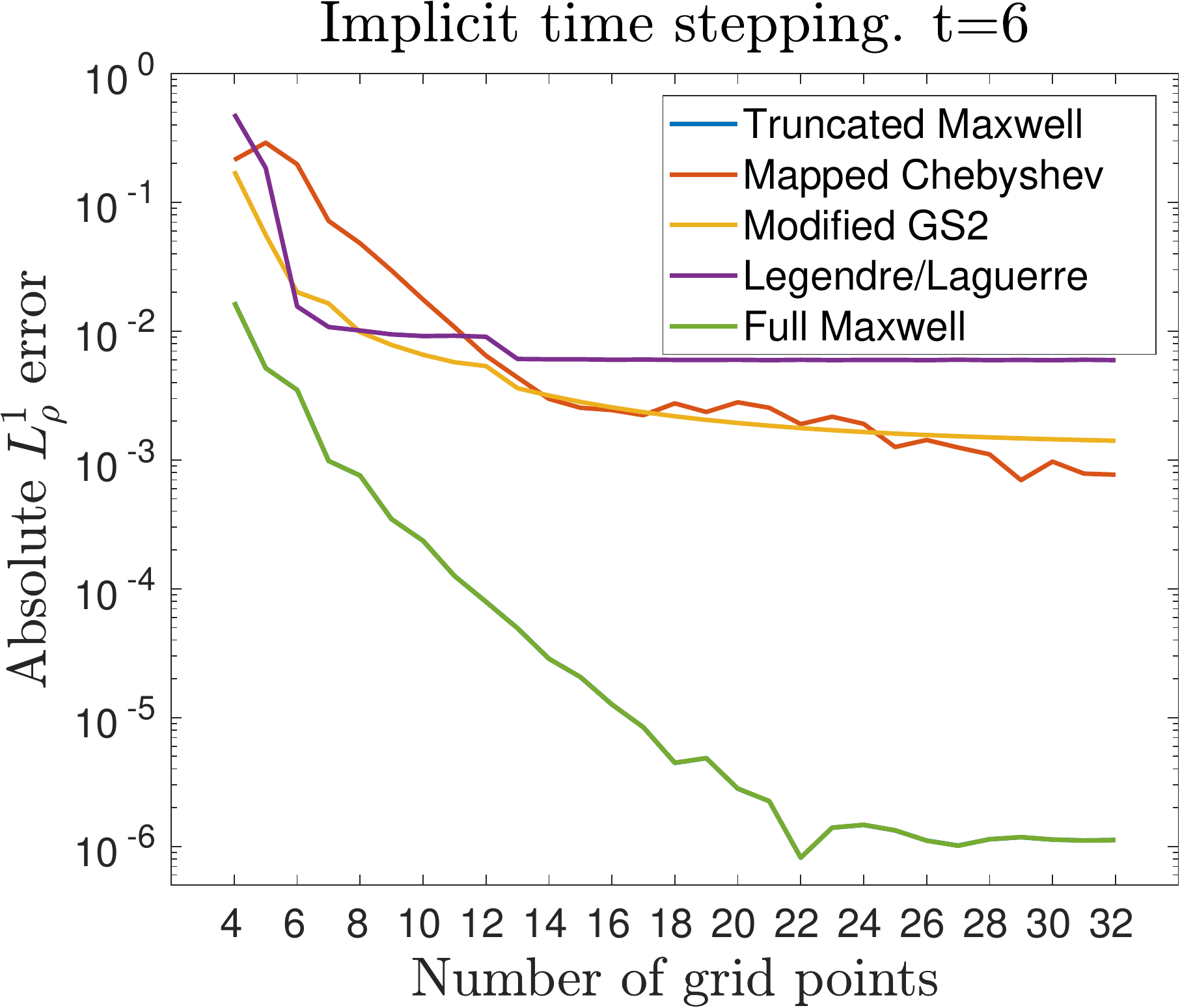} 
\end{tabular}
\caption{{\scriptsize Convergence plots in the $L^1_\rho$ norm for Example 2 with implicit time stepping. The behavior of the full and truncated versions of Maxwell nodes is virtually indistinguishable.}}\label{fig:E2ImplicitL1error}
\end{figure}

\begin{figure}[tb]
\begin{tabular}{ccc}
\includegraphics[width=.31\linewidth]{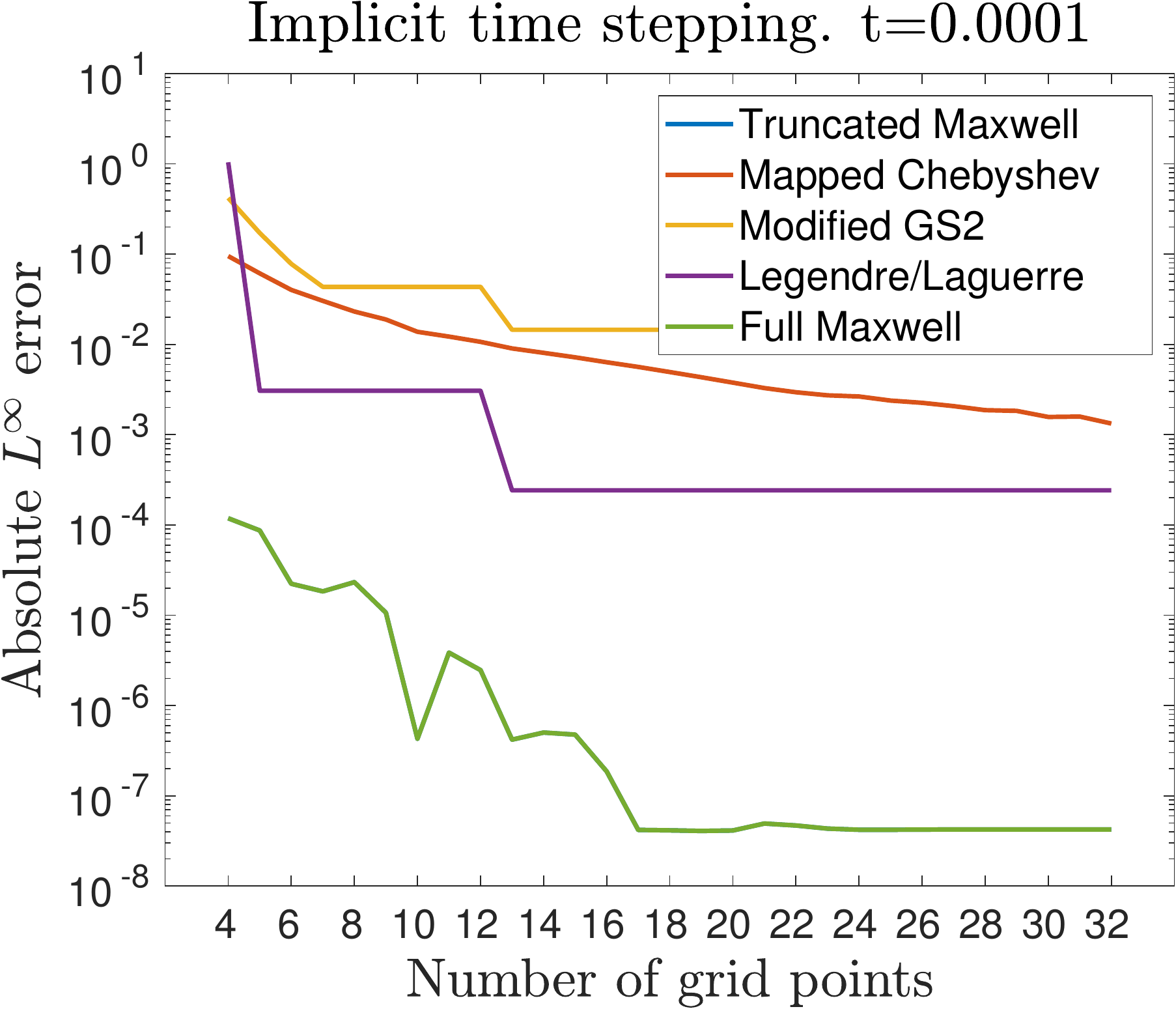} &
\includegraphics[width=.31\linewidth]{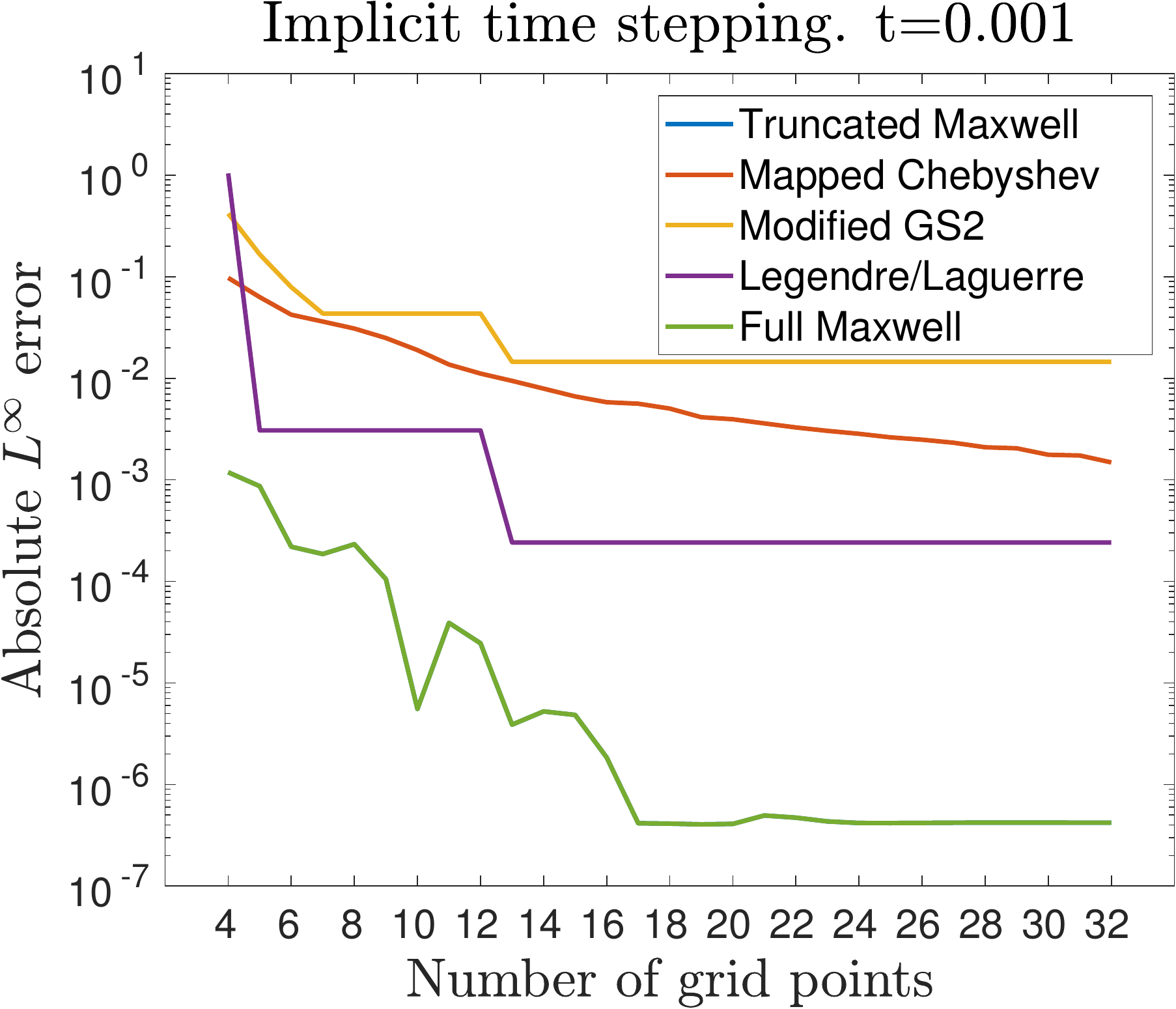} &
\includegraphics[width=.31\linewidth]{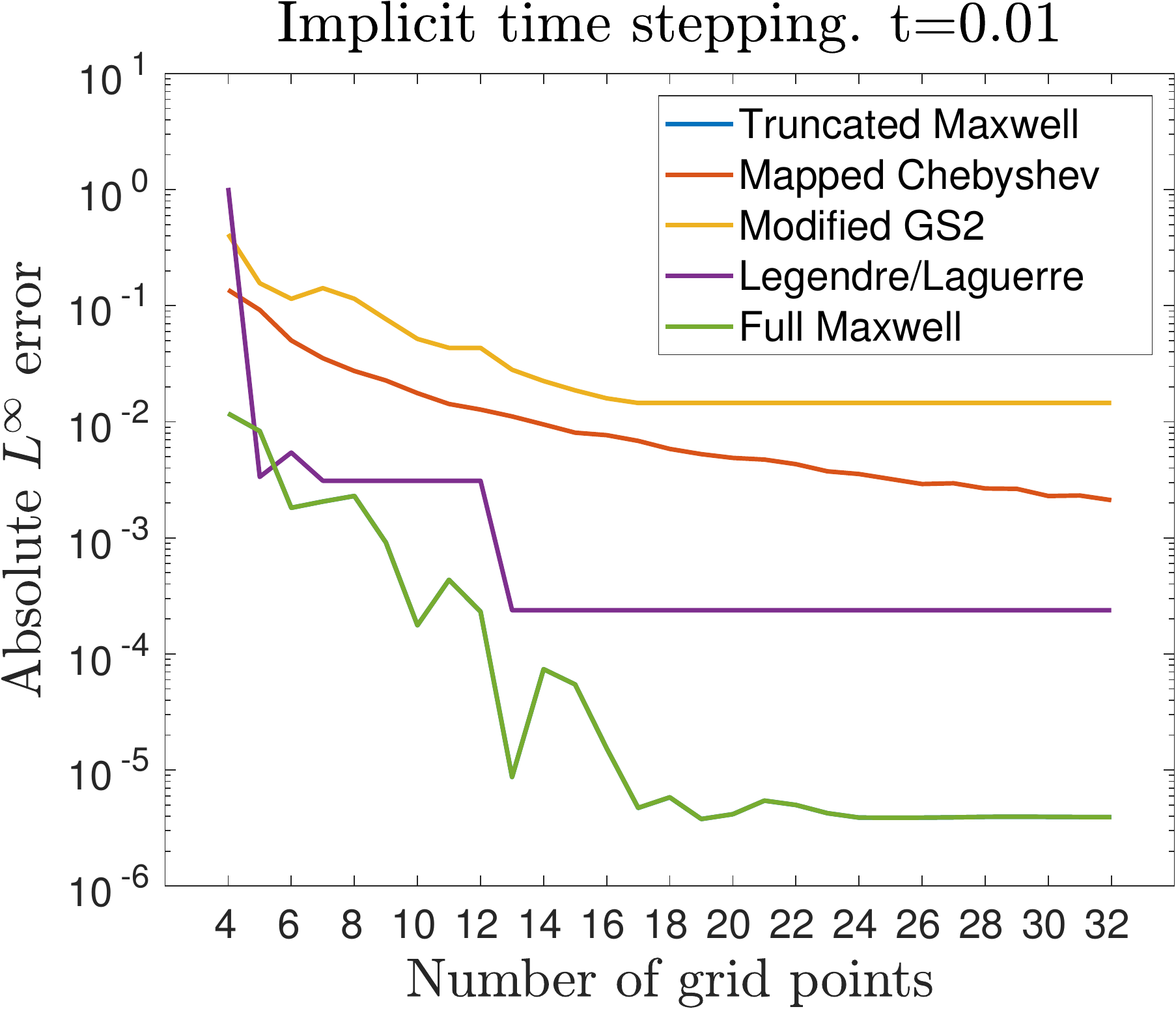} \\
\includegraphics[width=.31\linewidth]{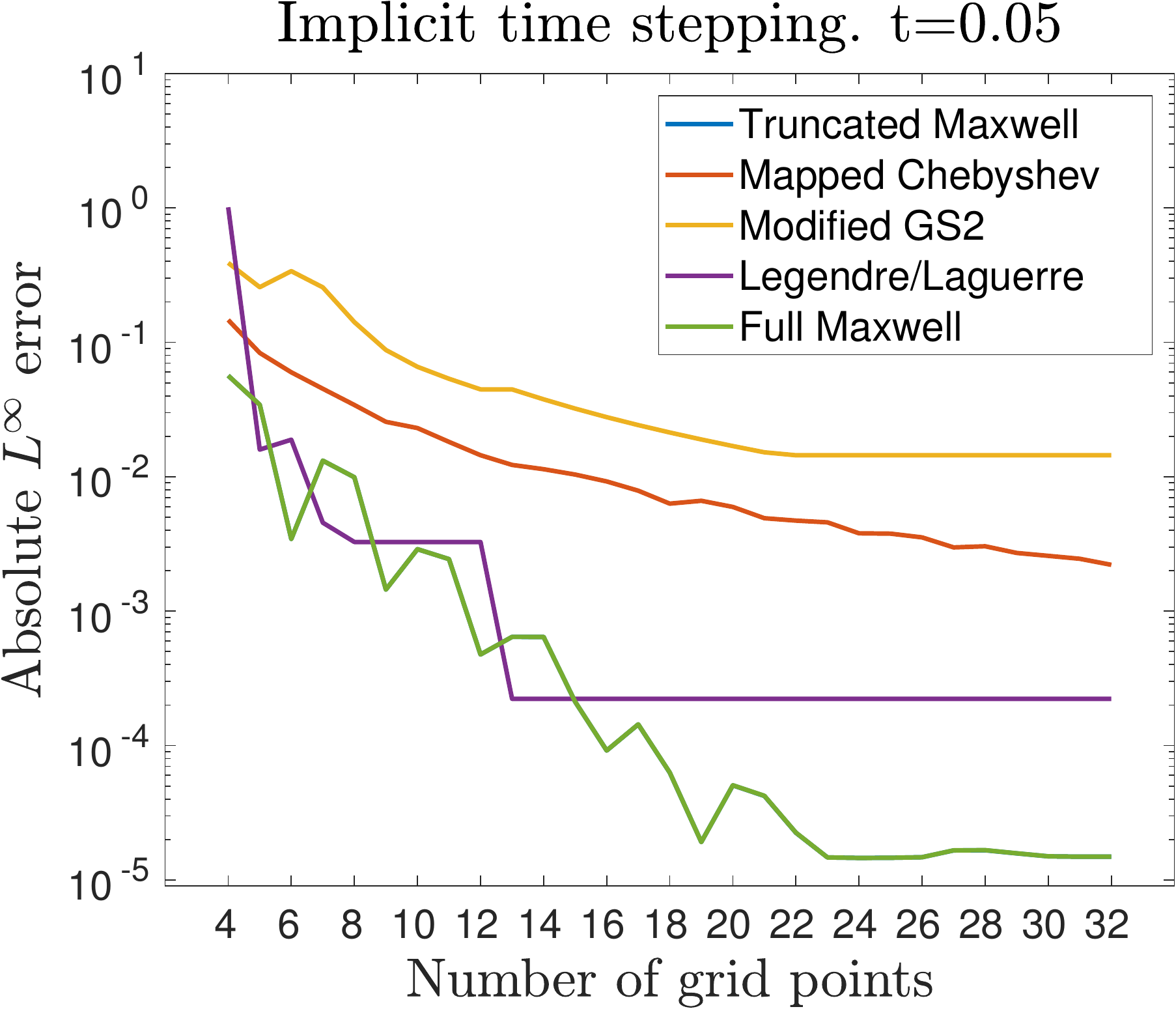} &
\includegraphics[width=.31\linewidth]{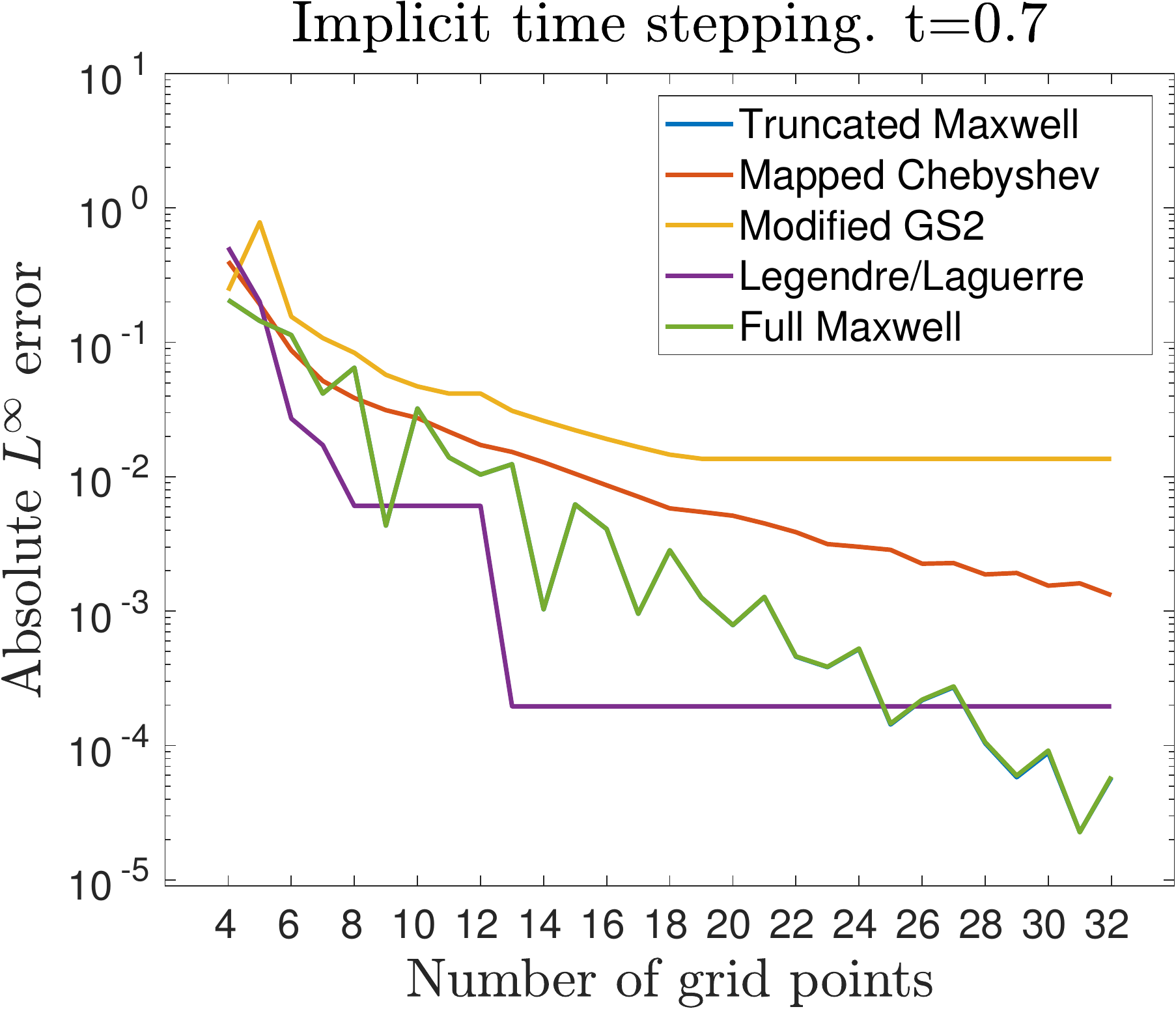} &
\includegraphics[width=.31\linewidth]{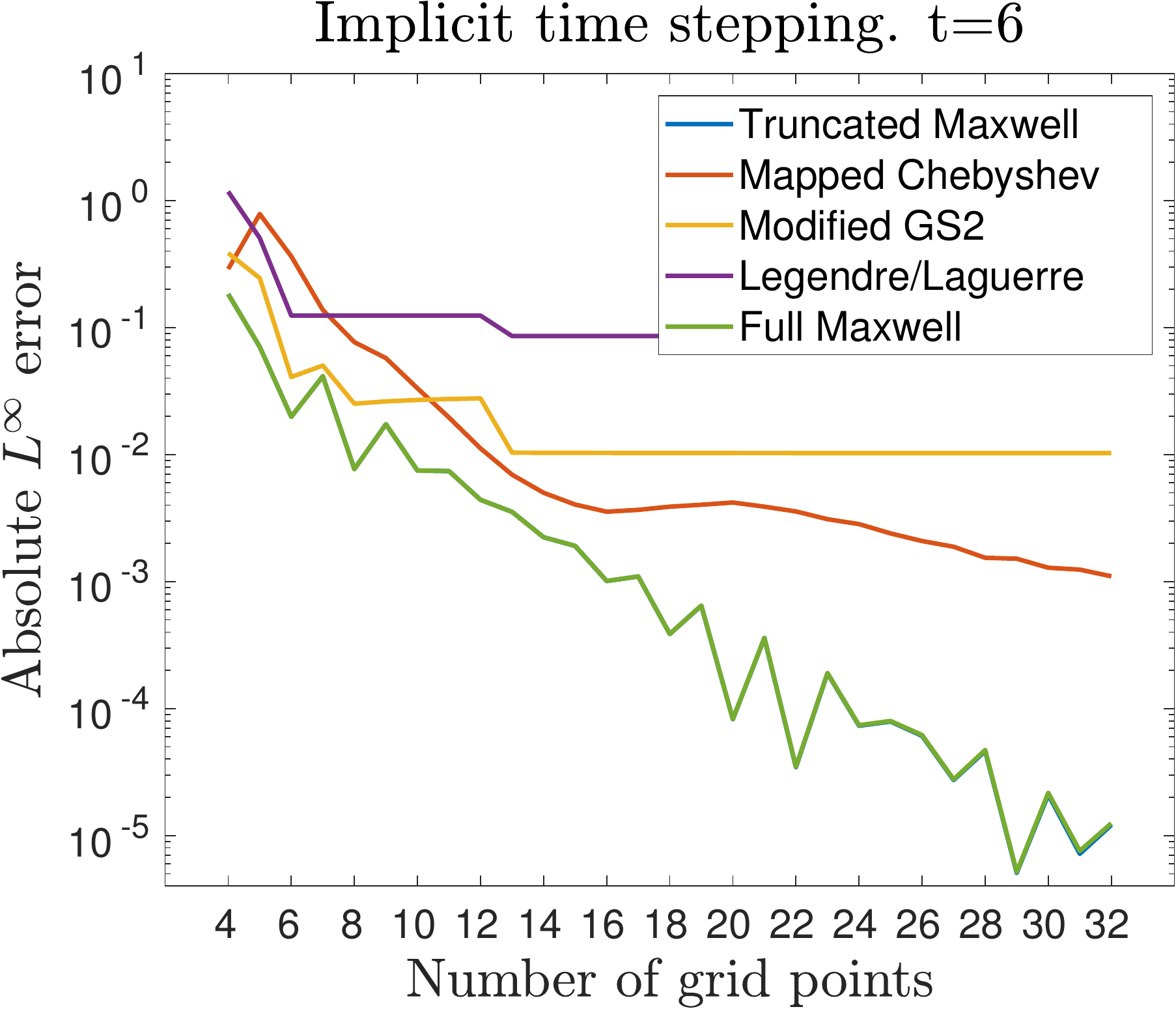} 
\end{tabular}
\caption{{\scriptsize Convergence plots in the $L^\infty$ norm for Example 2 with implicit time stepping. The behavior of the full and truncated versions of Maxwell nodes is virtually indistinguishable.}}\label{fig:E2ImplicitMAXerror}
\end{figure}

The results of our numerical experiments are shown in the convergence plots below. Figures \ref{fig:E1ExplicitL1error} and \ref{fig:E1ExplicitMAXerror} (resp. \ref{fig:E2ExplicitL1error} and \ref{fig:E2ExplicitMAXerror}) give the behavior of the approximation error for Example 1 (resp. Example 2) when second order predictor-corrector time stepping is used, while Figures \ref{fig:E1ImplicitL1error} and \ref{fig:E1ImplicitMAXerror} (resp. \ref{fig:E2ImplicitL1error} and \ref{fig:E2ImplicitMAXerror}) show the results for implicit second order backwards differentiation (BDF2). We chose to put all the figures at the end of the manuscript to facilitate comparisons between figures, and to avoid interrupting the flow of the article.

\begin{figure}[tb]
\begin{tabular}{ccc}
\includegraphics[width=.31\linewidth]{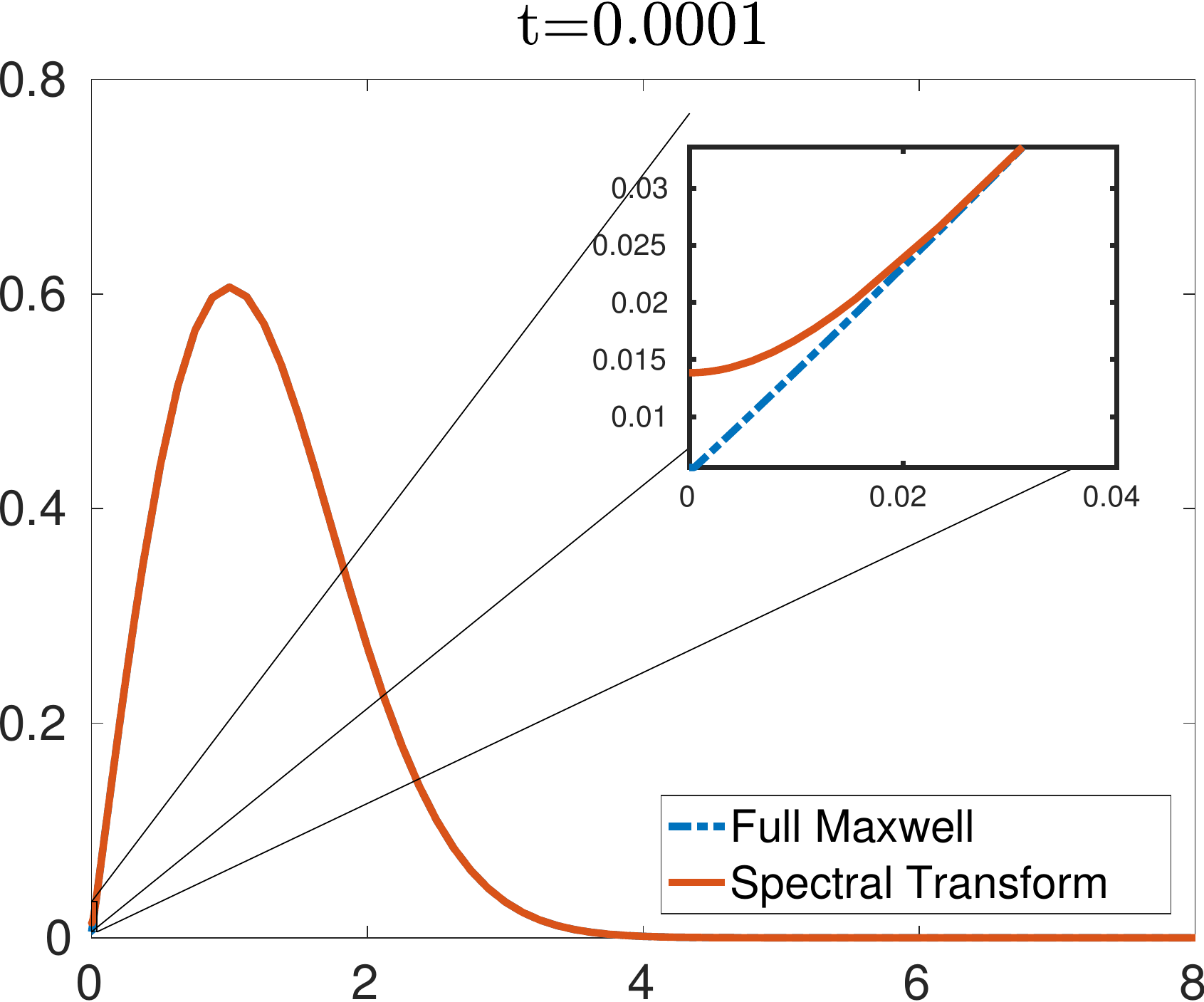} &
\includegraphics[width=.31\linewidth]{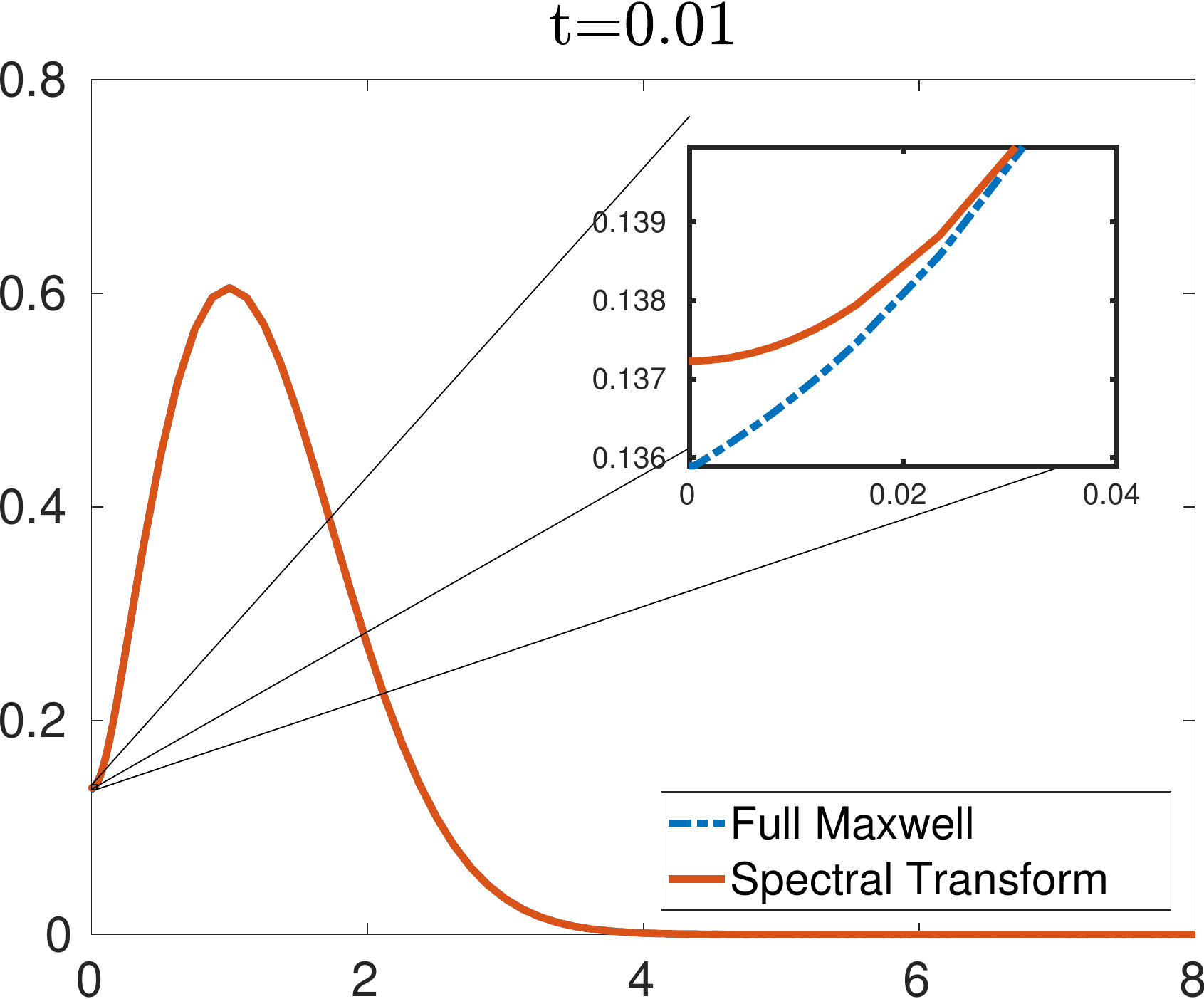} &
\includegraphics[width=.31\linewidth]{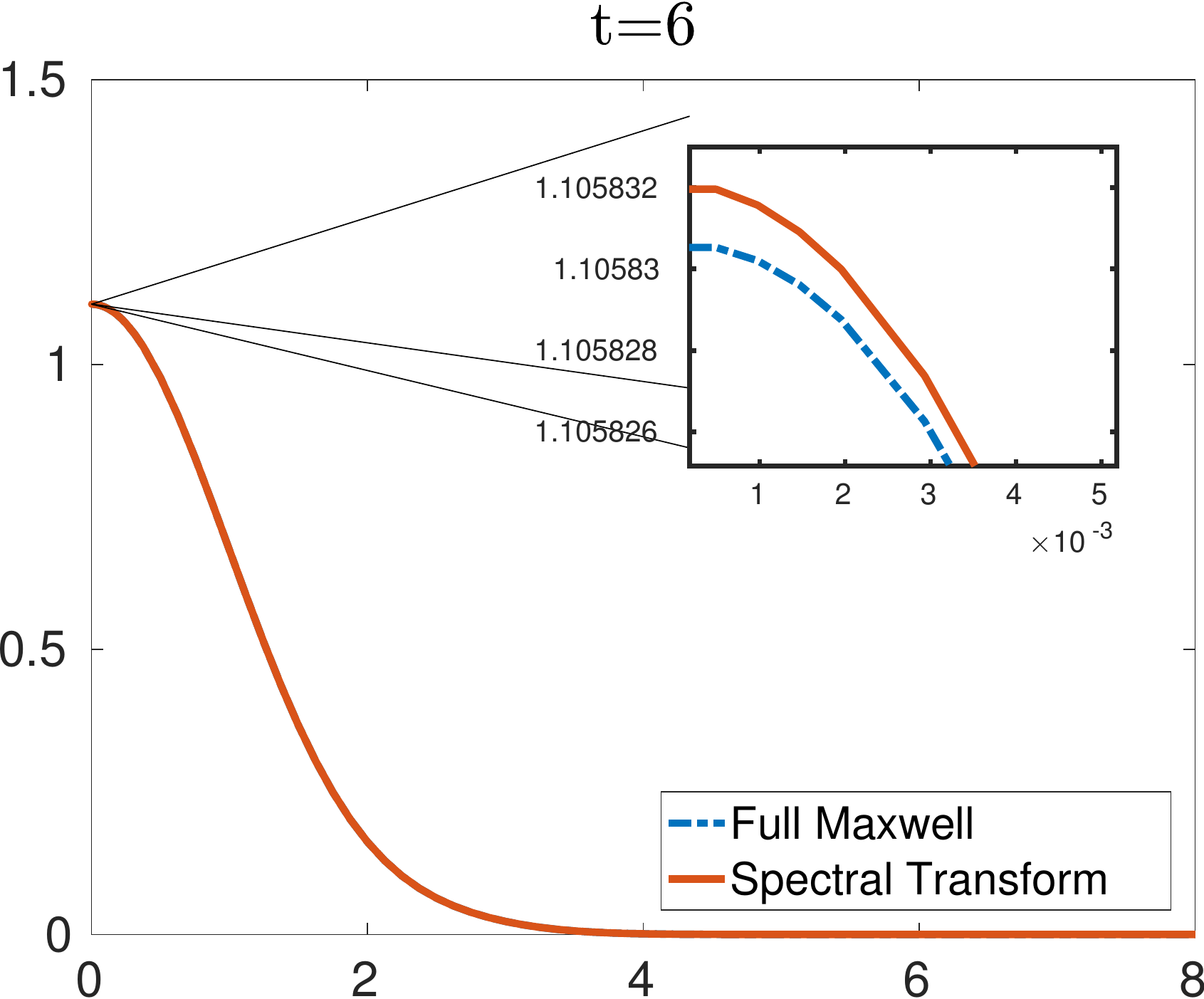}
\end{tabular}
\caption{{\scriptsize For Example 1, the approximation error concentrates at the origin, where the initial condition makes the right hand side of \eqref{eq:model} singular. The approximate solution is accurate in the rest of the domain and slowly converges at the origin as the system relaxes into a steady state Maxwellian distribution (note the scale on the zoomed-in axes). The figure depicts three time snapshots of the approximate solution obtained with 32 Maxwell nodes using explicit time stepping.}}\label{fig:zoom}
\end{figure}

We highlight a few points. First, we observe that Example 1, with its singularity at the origin, is the most challenging problem for all discretization schemes and for both implicit and explicit time stepping, as expected. As the initial singularity at the origin is smoothed out by the diffusive operator, the approximate solution trails behind the exact curve and converges towards the exact value as the distribution relaxes into a Maxwellian. The error, however, is highly localized and the solution is well approximated away from the origin, as can be seen in the snapshots presented in Figure \ref{fig:zoom}, obtained with Maxwell polynomials defined on the open interval $[0,\infty)$. This behavior explains the difference between the figures for the $L^\infty$  error and for the $L^1_\rho$ error. Most of the methods have poor accuracy and convergence properties in the $L^\infty$ norm for small $t$, and much better performance in the $L^1$ norm, which is much less sensitive to the behavior at the origin. In fact, as we will soon see in a slightly different context, for intermediate times the error in the $L^1$ norm for Example 1 can be smaller than the same error for Example 2.

The second point to note is that for smooth situations, such as Example 1 for large times and Example 2 for all times, and/or in the smoothing $L^{1}$ norm, both Maxwell approaches outperform the other methods, often by several orders of magnitude, and lead to high accuracy for a small number of grids points. While for conciseness we do not provide figures with these results in the article, we have verified that this is also true for smooth initial conditions which are more oscillatory, or which have a bump on the tail of the distribution, as long as these initial conditions have a Maxwellian envelope. These results confirm the desirability of discretization schemes based on Maxwell polynomials for kinetic calculations in plasma physics.  It is however important to stress that for the singular initial condition corresponding to example 1, the experiments indicate that the scheme based on a Chebyshev grid is more robust if one is interested in $L^\infty$ performance. This can be explained by the high density of Chebyshev nodes around the origin. 

The last point to highlight is the fact that the performances of the schemes based on Maxwell polynomials on the full interval and Maxwell polynomials on the truncated domain $[0,12]$ are comparable. Maxwell polynomials on the full interval sometimes outperform the polynomials on the truncated domain by a slight margin, which adds to the list of small advantages we gave previously in favor of using the full interval $[0,\infty)$.

\begin{figure}[tb]
\begin{tabular}{cc}
\includegraphics[width=.47\linewidth]{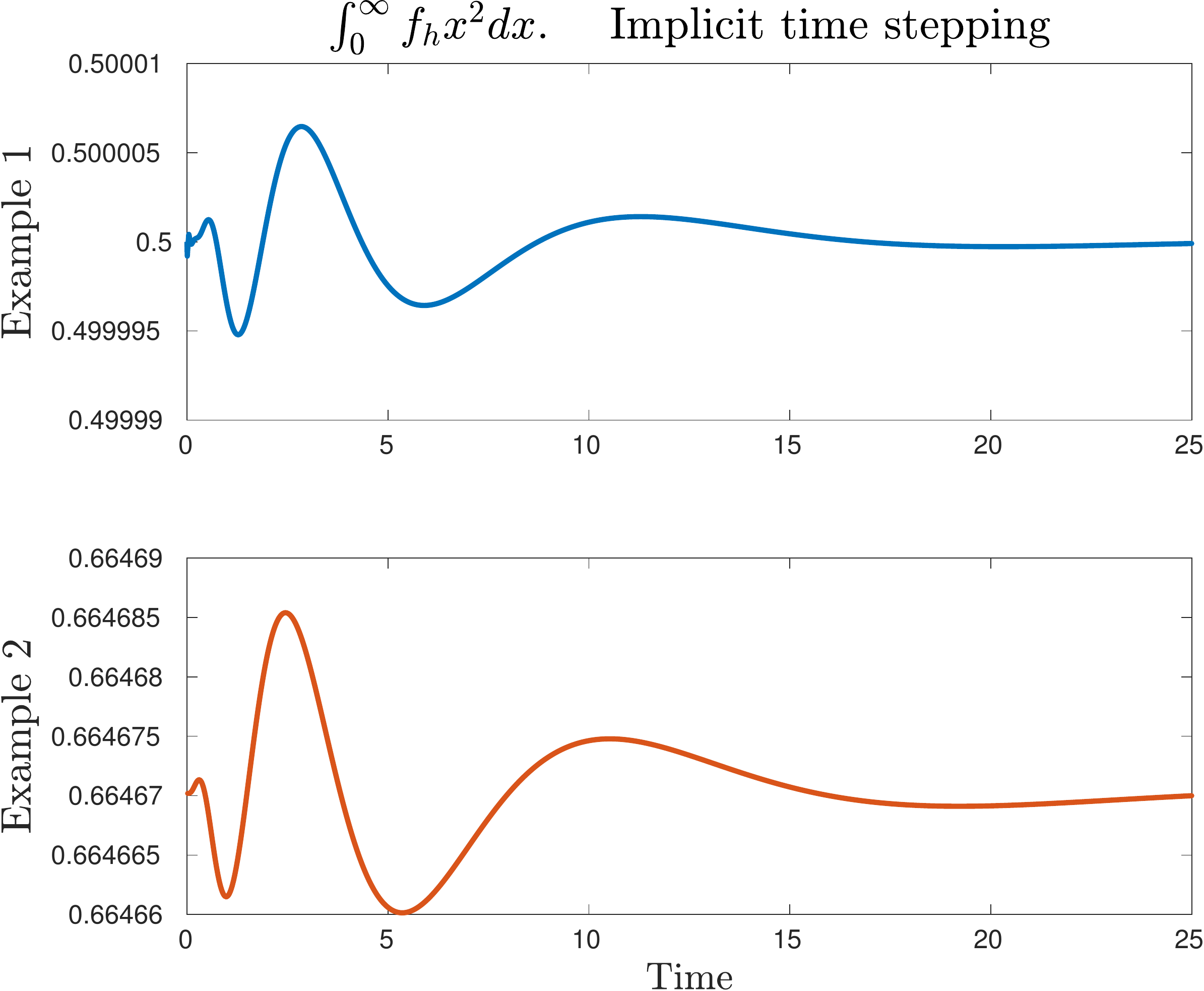} &
\includegraphics[width=.47\linewidth]{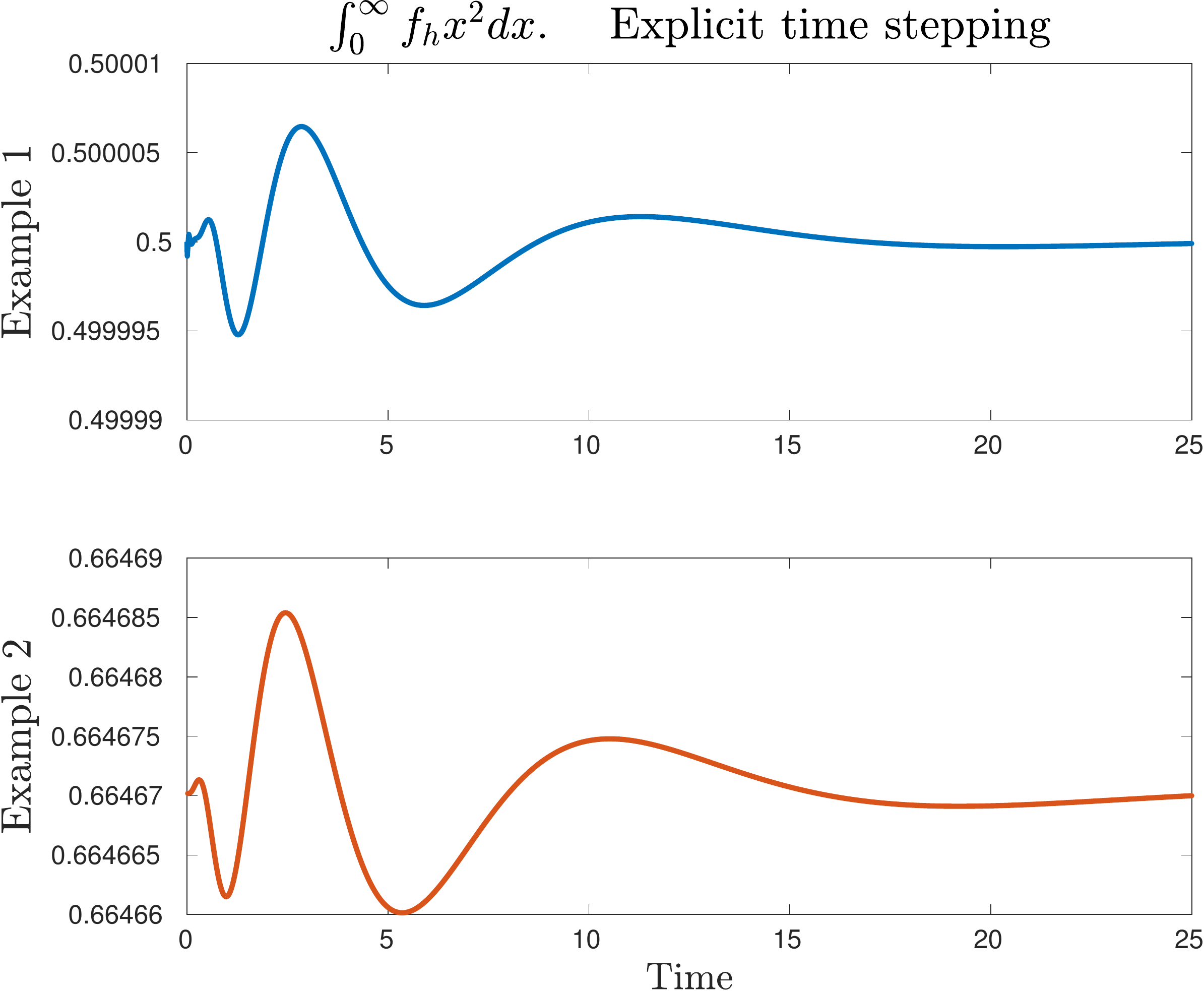}
\end{tabular}
\caption{{\scriptsize Variation in time of $\int_0^\infty f_hx^2\,dx$ for the numerical solution $f_{h}$ of Eq. \eqref{eq:model} using 16 Maxwell nodes. The integral is approximated by Gaussian quadrature at the grid points. Mass is conserved with high accuracy even with a relatively small grid.}}\label{fig:mass}
\end{figure}

As a physical measure of the good performance of the scheme based on Maxwell polynomials on the full interval, we demonstrate in Figure \ref{fig:mass} that with this scheme one conserves mass with high accuracy for a small number of grid points. Specifically, this figure shows the value of the integral $\int_0^\infty fx^2\,dx$ as a function of time for $t\in[0,25]$ for both implicit and explicit time stepping, and for both Example 1 and Example 2. As we already mentioned, one key reason why the pseudo-spectral approach with Maxwell polynomials works so well for mass conservation is that the location of the grid points is optimized for high order Gaussian quadrature of integrands with a Maxwellian behavior. One can observe that mass conservation is satisfied with a slightly lower accuracy for Example 2 than it is for Example 1. The reason for this is that for intermediate values of $t$, the departure of the solution from a Maxwellian behavior is more pronounced in Example 2 than in Example 1, making a representation with Maxwell polynomials less efficient. 

On the topic of mass conservation, we note that Wilkening \cite{Wilkening:2017} has recently proposed a symmetrized representation of the pseudo-spectral discrete operators based on Maxwell polynomials for Eq. \eqref{eq:operator} which leads to \textit{exact} mass conservation (to machine precision). Although promising, this representation was not considered in the present work because it remains to determine whether it is applicable in its present form to the full five-dimensional gyrokinetic codes, which are the motivation for our work. 

Finally, another relevant measure of the performance of our scheme is given by a comparison between the number of grid points needed to achieve a certain level of accuracy with our Maxwell based scheme and with the more standard GS2 scheme (in our improved version of it). The results are given in Table \ref{tab:performance}. The target accuracy we chose for all the tests is given by the minimum error our improved GS2 scheme can achieve in time with a number of grid points smaller than or equal to 32. For the size of time steps prescribed by \eqref{eq:time-step}, the smallest GS2 error was equal for both implicit and explicit time stepping. We can see that Maxwell polynomials consistently lead to smaller grid sizes for a given level of accuracy, and the number of grid points is smaller by a factor larger than 2 for the $L^{\infty}$ error and the smooth situation provided by Example 2.

\begin{table}[ht]\centering
\scalebox{0.75}{\!\centering
\begin{tabular}{c|c|c|c|c} \hline \hline
\multicolumn{1}{c|}{\textbf{Example 1}} & \multicolumn{1}{|c|}{Min. GS2 error} & \multicolumn{1}{|c|}{\# Modified GS2 pts.} & \multicolumn{1}{|c|}{\# Maxwell pts. (Implicit)} & \multicolumn{1}{|c}{\# Maxwell pts. (Explicit)} \\
\hline \hline
$L^\infty$ Error & 2.39 $\times 10^{-3}$ & 32 & 26 & 14 \\ \hline
$L^1_\rho$ Error 	 & 9.74 $\times 10^{-5}$ & 16 & 12 & 12 \\ \hline \hline
\multicolumn{1}{c|}{\textbf{Example 2}} & \multicolumn{1}{|c|}{Min. GS2 error} & \multicolumn{1}{|c|}{\# Modified GS2 pts.} & \multicolumn{1}{|c|}{\# Maxwell pts. (Implicit)} & \multicolumn{1}{|c}{\# Maxwell pts. (Explicit)} \\ \hline \hline
$L^\infty$ Error & 1.03 $\times 10^{-2}$ & 32 & 14 & 14 \\ \hline
$L^1_\rho$ Error 	 & 3.97 $\times 10^{-4}$ & 17 & 10 & 10 \\ \hline
\end{tabular}
}
\caption{{\scriptsize Minimum number of Maxwell grid points needed to reach errors with a magnitude below the smallest error obtained with our modified GS2 scheme uniformly in time. The number of modified GS2 points used is shown in the first column. For the time step used, given by \eqref{eq:time-step}, the reference minimum error attained by our modified GS2 scheme was the same for implicit and explicit time stepping. }}\label{tab:performance}
\end{table}
%
%
\section{Stability and discrete time evolution}\label{sec:stability}

After studying the performance of Maxwell polynomials with a focus on spatial discretization in Section \ref{sec:validation}, we now turn to the question of time stepping and stability. For this part of the study, we consider two pseudo-spectral discretizations of the operator $L$ based on Maxwell polynomials. The first one is a direct discretization of Eq. \eqref{eq:model}, which we call $\mathbf{L}_{a}$, and which we will show to be subject to numerical instabilities. The second discretization, constructed to address the limitations of $\mathbf{L}_{a}$, is obtained by rewriting Eq. \eqref{eq:model} as an equation for the quantity $x^2 f(x)$ instead of the distribution function $f(x)$ alone. We call the resulting discrete differential operator $\mathbf{L}_{b}$. We remind the reader that all the results discussed in Section \ref{sec:validation} were obtained using $\mathbf{L}_{b}$.

We present $\mathbf{L}_{a}$ and $\mathbf{L}_{b}$ in detail in Section \ref{sec:discrete}, and explain why a naive implementation of $\mathbf{L}_{a}$ is subject to an exponential instability which can be easily avoided with a better representation of the operator.  Even if so, we demonstrate in Section \ref{sec:spectra} that the improved discretization of $\mathbf{L}_{a}$ is subject to a non-modal instability, leading to non-physical transient behavior that produces significant error near the origin. The explanation for the unexpected behavior is found in the pseudo-spectrum of the discretization and its relation with the stability region of the time stepping procedure.

\subsection{Discrete representations for $L$}\label{sec:discrete}
To ensure the stability of the time discretization of a system of the form
\[
\partial_t \mathbf u = \mathbf M \mathbf u, \qquad \mathbf u\in \mathcal C^1([0,T],\mathbb R^N),
\]
where $\mathbf M$ is an $N\times N$ normal matrix, (i.e., $\mathbf M \mathbf M^* = \mathbf M^* \mathbf M$) it is enough to analyze the spectrum of $\mathbf M$ in relation to the stability region of the desired time-stepping scheme. If the spectrum of $\mathbf M$ is completely contained in the stability region of the method, the full space-time discretization will be stable. This requirement  imposes restrictions on the size of the time step $\Delta t$ for both implicit and explicit time-stepping. It is common knowledge that implicit methods have wider stability regions which allow larger values of $\Delta t$, but come at the expense of solving a linear system per time step. Conversely, explicit methods often require much smaller time steps but provide a simple updating strategy which is often more accurate and requires less computations per time step.

For our present purposes, there are two elements which complicate this simple approach to stability analysis. First, the spectrum of $\mathbf{L}$ depends on the specific form of the pseudo-spectral discrete representation. Second, with the most straightforward pseudo-spectral collocation implementations, $\mathbf{L}$ is not a normal matrix. This means that the criterion on the spectrum given above for normal matrices is a necessary condition for stability, but no longer a sufficient condition \cite[Sec. 4.6--4.8]{MoRi:1967}. This fact is often overlooked in practical implementations, perhaps due to the prevalence of symmetric matrices  stemming from finite difference and finite element discretizations in combination with the practical simplicity of the eigenvalue test for stability. These subtleties however lead us to consider the following two natural yet different pseudo-spectral discretizations of $L$ in this article, which will be used to discuss the aforementioned complications in detail in section \ref{sec:spectra}.

The first discrete representation follows directly from Eq.\eqref{eq:model}, repeated below for convenience:
\begin{equation}\label{eq:f6}
\partial_t f = \frac{1}{x^2}\,\partial_x\left[w(x)\psi(x)\partial_x(e^{x^2}f)\right].
\end{equation}
The second representation follows from the idea that one can expand the innermost derivative in the right-hand side of this expression to get rid of the exponential factors, and multiply through by $x^2$ to try to avoid -- at least formally -- the $1/x^2$ singularity, solving for $x^2 f$ instead. Writing $f=(x^2 f)/x^2$, one obtains
\begin{equation}\label{eq:f4}
\partial_t(x^2 f) = \partial_x\left\{\psi(x)\left[2(x-1/x)x^2f+\partial_x(x^2f)\right]\right\}.
\end{equation}
The right-hand sides of the above expressions lead to different discrete operators. We make use of the following notation to denote the associated matrices. Given $\{x_i\}_{i=1}^{N}$ collocation nodes we let $\mathbf{diag}(g)$ be a diagonal matrix with entries 
\[
\mathbf{diag}(g)_{ii}= g(x_i)\,,\quad  i=1,\ldots N.
\] 
As in Section \ref{sec:Collocation}, $\mathbf D^{(\ell)}$ is the $N \times N$ spectral approximation to the differential operator $\frac{d^{\ell}}{dx^{\ell}}$.  The discrete counterparts of the operators in the right-hand side of \eqref{eq:f6} and \eqref{eq:f4} thus yield the following matrices:
\begin{alignat}{6}
\label{eq:La}
\mathbf L_{a} := \, & -\diag{x^{-2}}*\mathbf D^{(1)}*\diag{w(x)\psi(x)}*\widehat{\mathbf D}^{(1)}*\diag{e^{x^2}} \quad&& (\hbox{unknown } f ),\\
\label{eq:Lb}
\mathbf L_{b} := \, & -\mathbf D^{(1)}*\diag{\psi(x)}*\left(\diag{2(x-1/x)}+\mathbf D^{(1)}\right) \quad&& (\hbox{unknown } x^2f ).
\end{alignat}

The reader may have noticed that in equation \eqref{eq:La} we have distinguished between the two differentiation matrices involved by the introduction of the notation $\widehat{\mathbf D}^{(1)}$ at the right end of the expression. We explain the reason for this distinction in what follows. As was mentioned in Section \ref{sec:Collocation}, in order to exploit the additional knowledge that the solution of \eqref{eq:model} has a Maxwellian envelope, the envelope of the interpolant $\pi(x)$ in equation \eqref{eq:interpolant} has been chosen to be precisely $\alpha(x)=e^{-x^2}$. This choice for the envelope carries over to the definition of the entries of the differentiation matrices $\mathbf D^{(\ell)}$ through the four equations in \eqref{eq:welfert}. This means that the differentiation matrix  $\mathbf D^{(1)}$ is expecting to act on vectors of the form $(e^{-x_1^2}h(x_1),\ldots,e^{-x_N^2}h(x_N))^\top$, and will not produce a correct approximation to the desired derivative otherwise. However, since the unknown $f$ in the right-hand side of the continuous formulation \eqref{eq:f6} giving rise to the operator $\mathbf L_a$ is multiplied by $e^{x^2}$, the Maxwellian envelope assumed to be included in $f$ is canceled out by the exponential factor, and the matrix $\mathbf D^{(1)}$ built with the envelope $\alpha(x)=e^{-x^2}$ would yield an erroneous approximation. We must therefore replace the rightmost differentiation matrix of \eqref{eq:La} by the one obtained by using the envelope $\widehat\alpha(x)=1$ in \eqref{eq:welfert}, which is denoted $\widehat{\mathbf D}^{(1)}$. The last row in Figure \ref{fig:spectraAndpseudospectra} provides a graphical explanation for the need to introduce $\widehat{\mathbf D}^{(1)}$ in place of $\mathbf D^{(1)}$ in \eqref{eq:La}: if $\mathbf D^{(1)}$ is used twice, as we did to create the figures for that last row, the eigenvalues of the discrete operator lie on both sides of the imaginary axis and the discretization is unconditionally unstable. In contrast, due to the presence of the intermediate factor $w(x)=x^2e^{-x^2}$, the second instance of the differentiation matrix on the left of \eqref{eq:La} will indeed be acting upon a function with a Maxwellian envelope and thus $\alpha(x)=e^{-x^2}$ must be used in the definition of the second differentiation matrix. 

Now, as we stated above, for the non-normal matrices $\mathbf{L}_{a}$ and $\mathbf{L}_{b}$, exponential stability is not sufficient to guarantee stability. To investigate the question of the stability of the discrete representations with $\mathbf{L}_{a}$ and $\mathbf{L}_{b}$, we turn to the question of the pseudo-spectra of these operators (and thus the remaining plots in Figure \ref{fig:spectraAndpseudospectra}) in the next section.

\subsection{Pseudo-spectra and non-modal instability}\label{sec:spectra}

The farther from normality a matrix is -- i.e. the farther its eigenvectors depart from orthogonality  -- the less likely it is that the information provided by a simple eigenvalue analysis is sufficient, and additional information is needed to verify the stability of the method \cite{Trefethen:1997}. This is in particular the case for our matrices $\mathbf{L}_{a}$ and $\mathbf{L}_{b}$. An improved, yet still not entirely robust, stability criterion \cite{ReTr:1992} for a non-normal matrix $\mathbf{M}$ extends the analysis to the $\epsilon$-pseudo spectrum of the matrix, $\Lambda_\epsilon(\mathbf M)$, defined as
\[
\Lambda_\epsilon(\mathbf M) = \left\{ z\in \mathbb C : \|(\mathbf M - z\mathbf I)^{-1}\|_2 \geq 1/\epsilon\right\}.
\]
The distribution of $\Lambda_\epsilon(\mathbf M)$ with respect to the stability regions of the chosen time-discretization provides a better estimate of the dynamics of the discrete system. It is well-known that $\Lambda_\epsilon(\mathbf M)$ is highly sensitive to the particular form of the continuous equation that is discretized and the way in which the discrete differential operator is built. This is precisely the case for our operators $\mathbf{L}_{a}$ and $\mathbf{L}_{b}$.

\begin{figure}[htb]
\begin{tabular}{cccc}
\includegraphics[width=.31\linewidth]{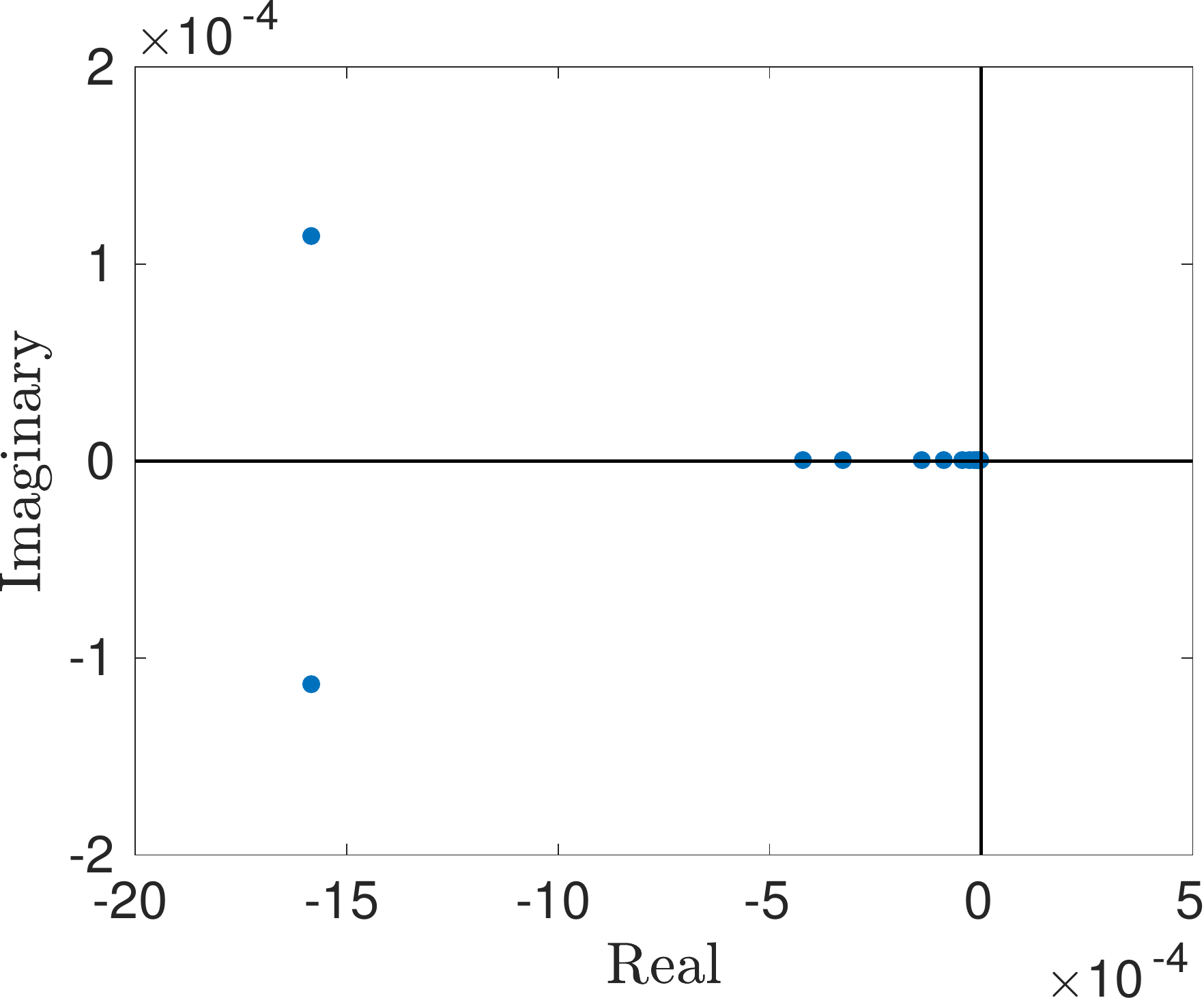} &
\includegraphics[width=.31\linewidth]{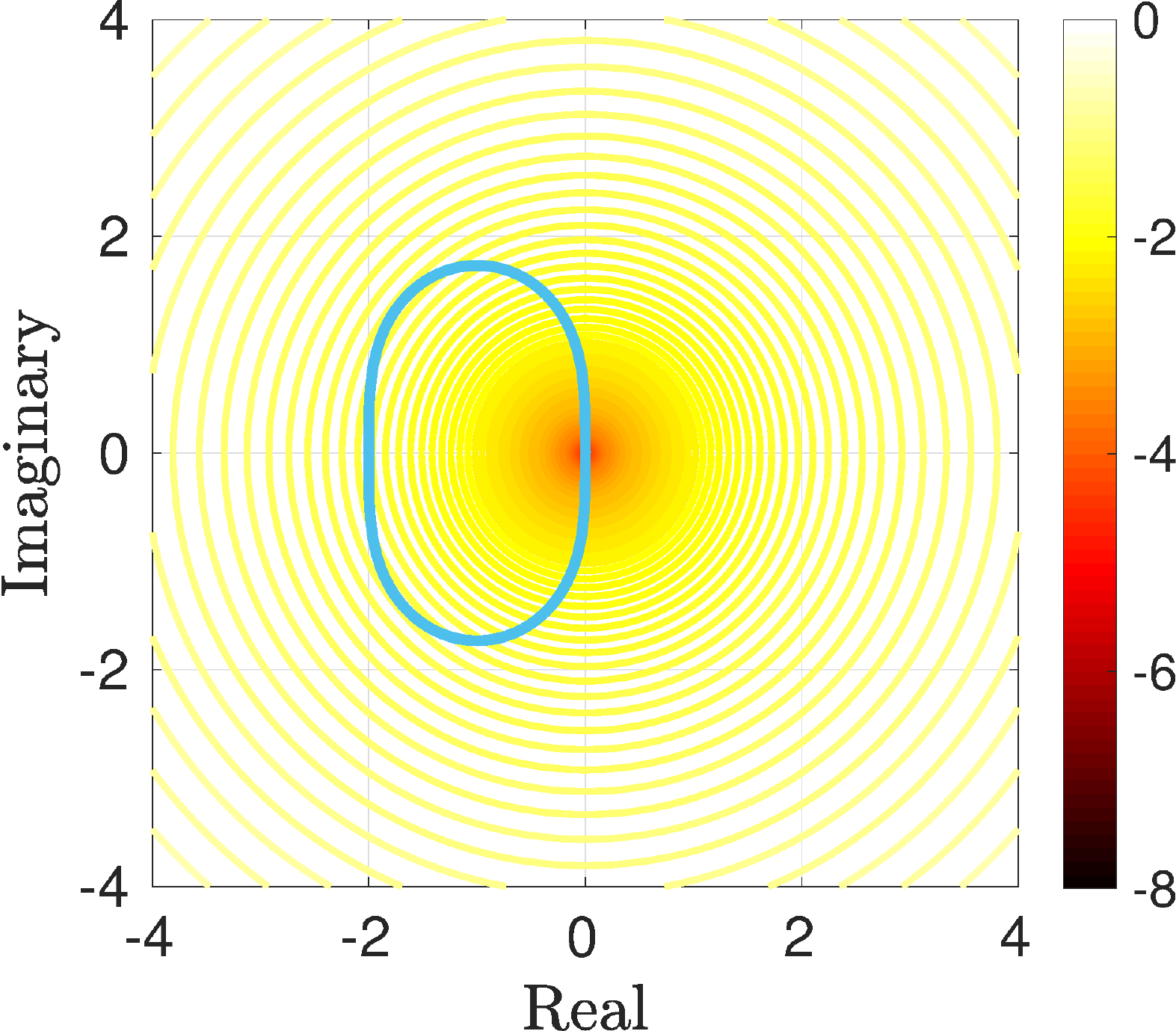} &
\includegraphics[width=.31\linewidth]{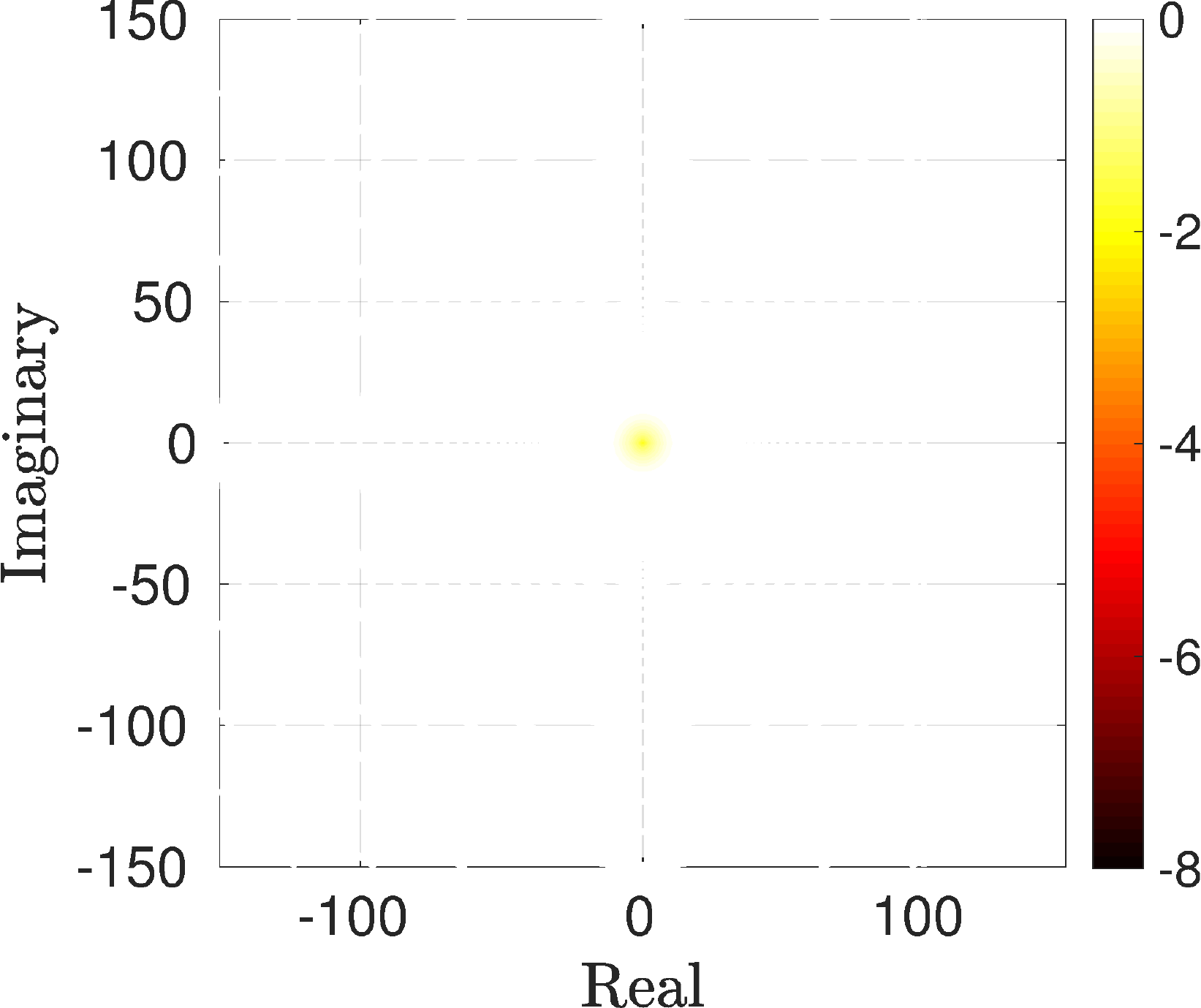} \\
\includegraphics[width=.31\linewidth]{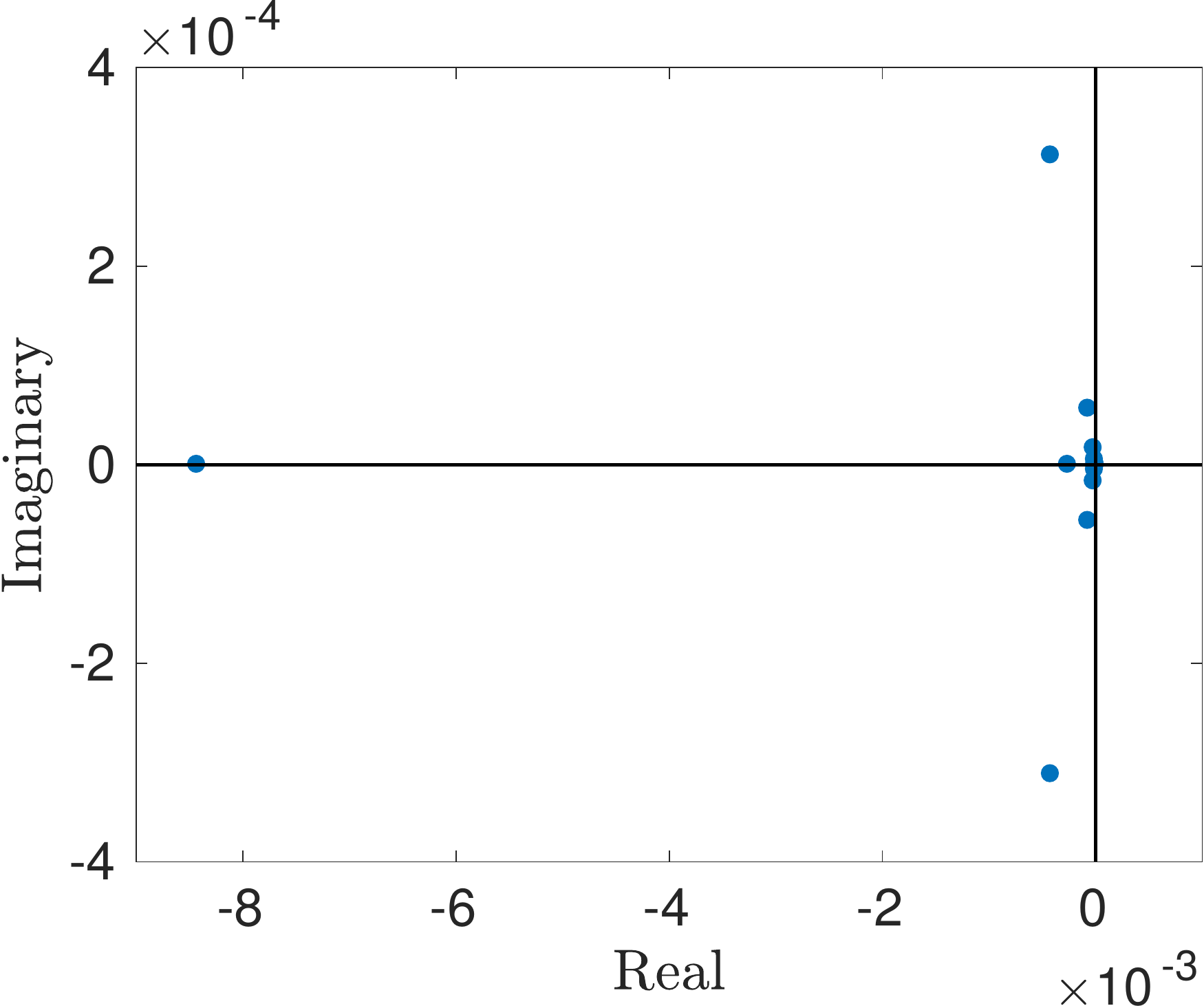} &
\includegraphics[width=.31\linewidth]{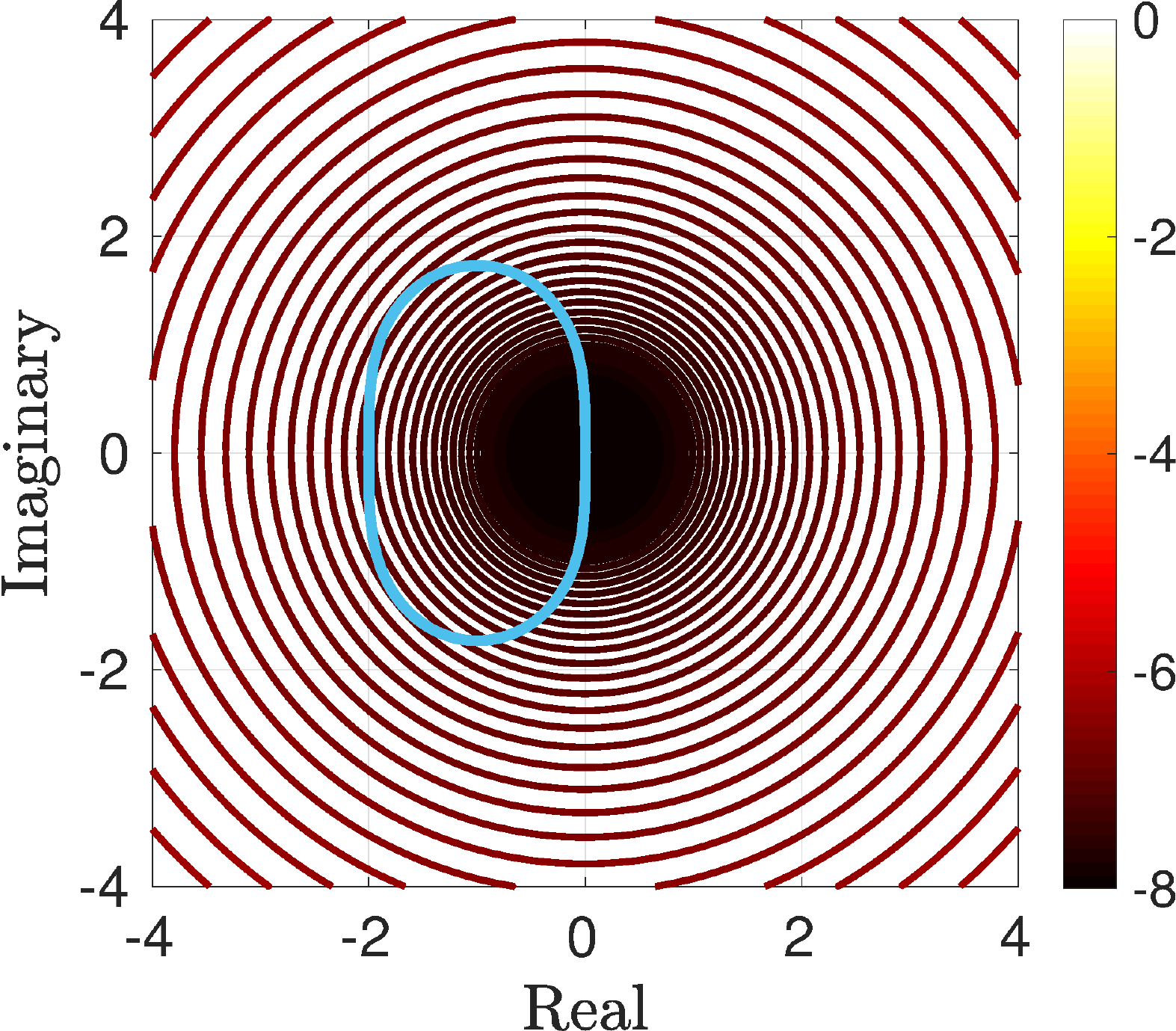} &
\includegraphics[width=.31\linewidth]{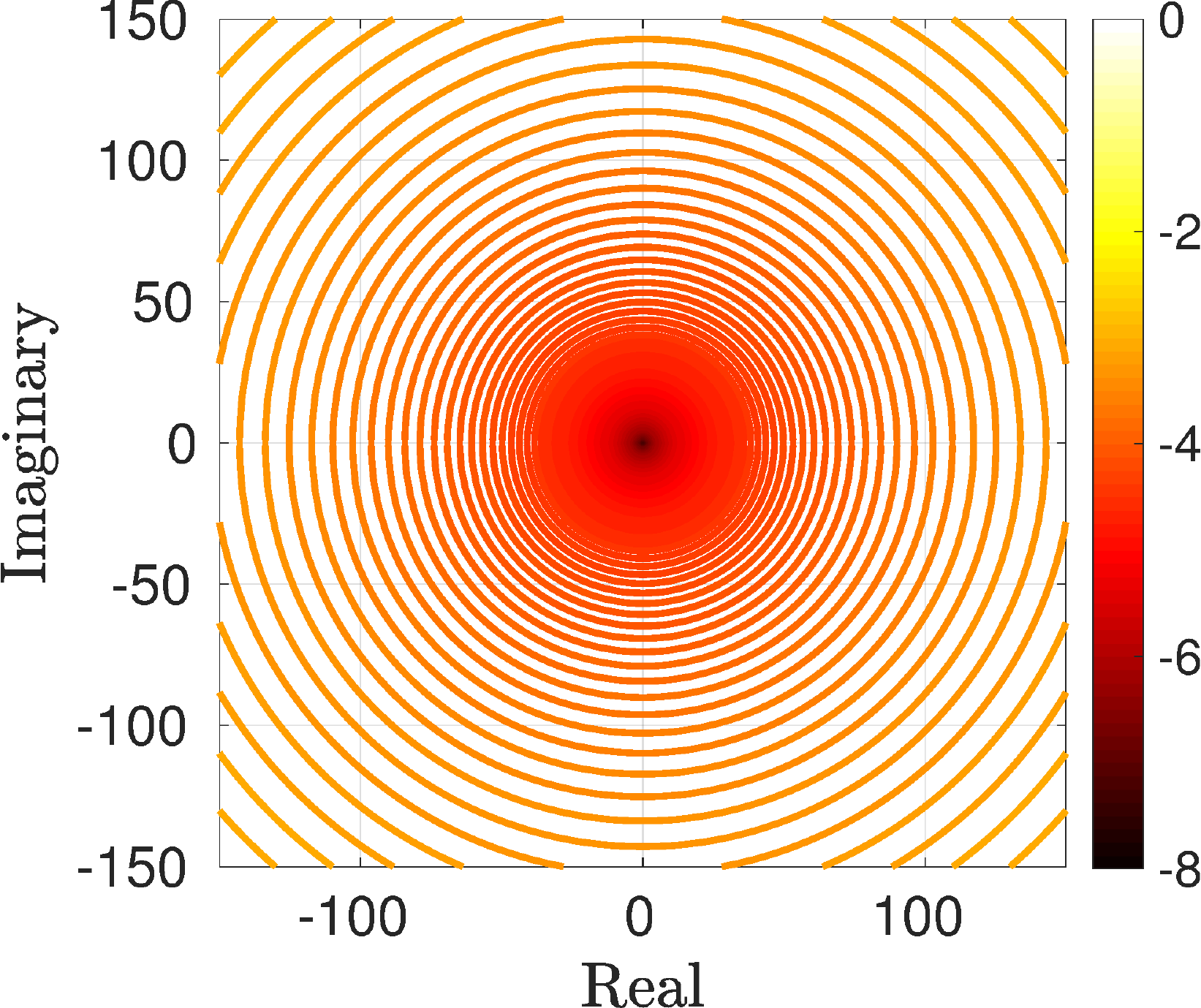} \\
\includegraphics[width=.31\linewidth]{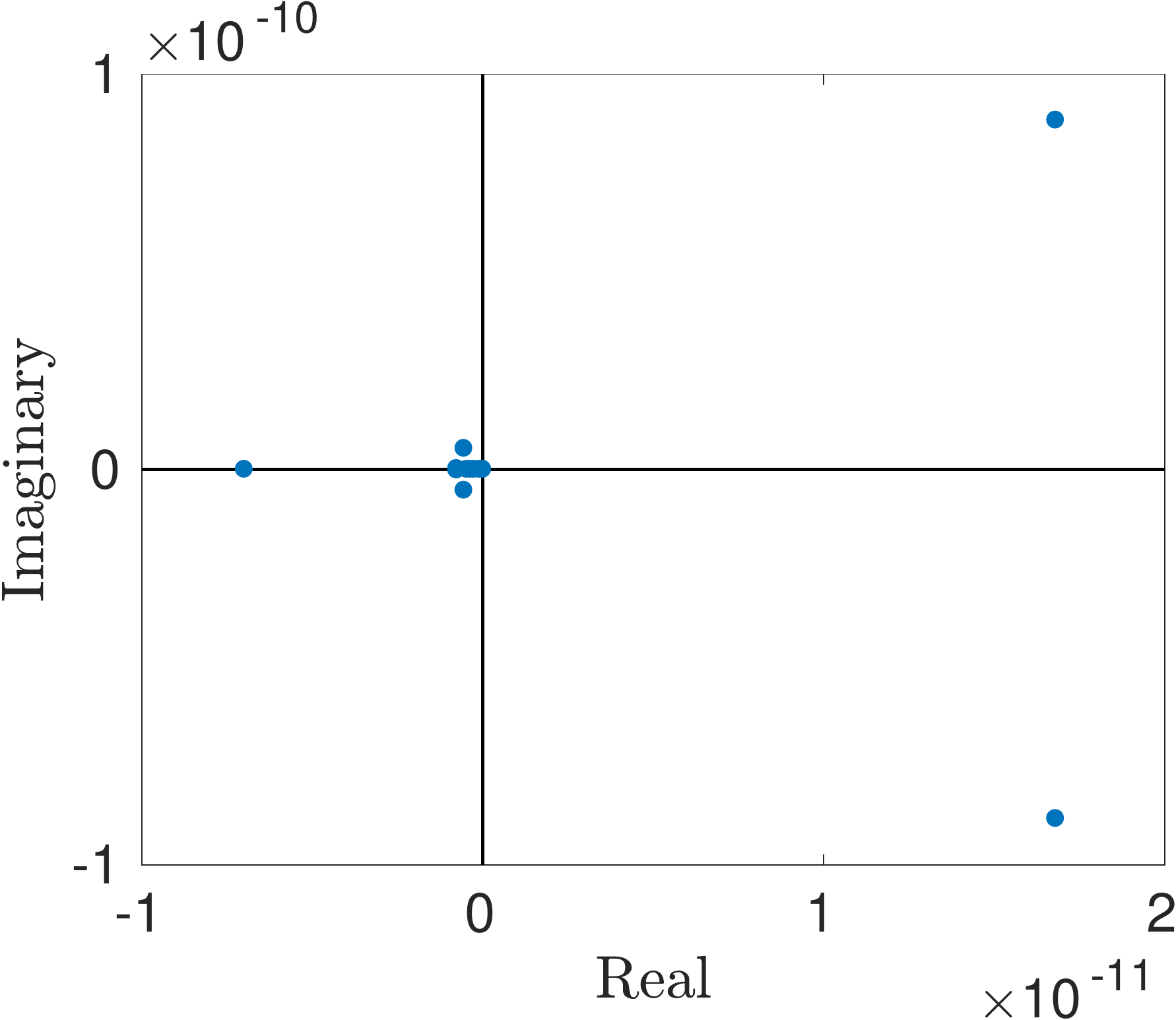} &
\includegraphics[width=.31\linewidth]{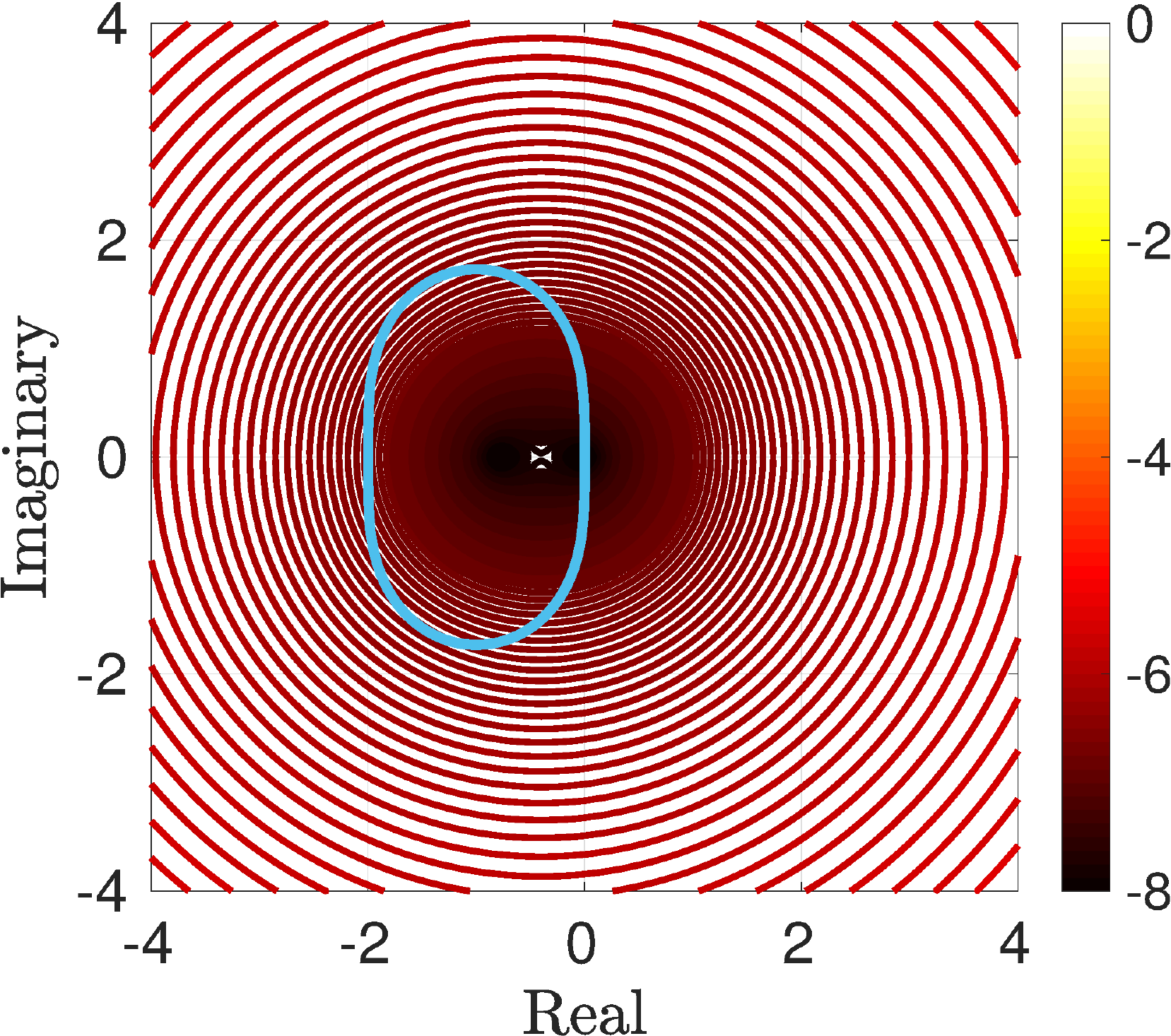} &
\includegraphics[width=.31\linewidth]{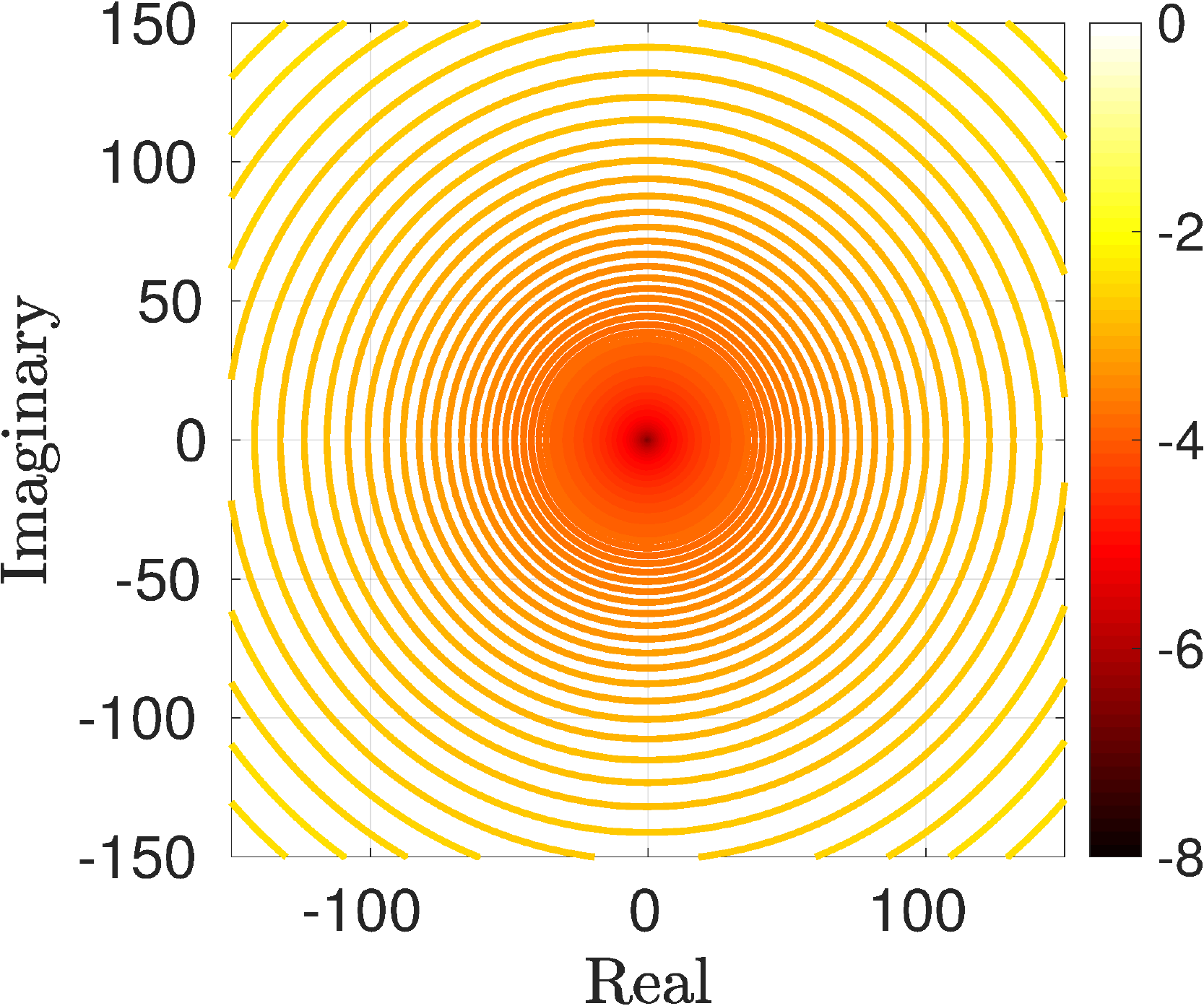}  
\end{tabular}
\caption{{\scriptsize Spectra and pseudo spectra of three 16 point discretizations of \eqref{eq:model} and Maxwell polynomials defined on the full interval. In the pseudo-spectral plots (center and right columns), the contour lines are the complex numbers $z$ such that $\|(\Delta t\,\mathbf L+z\mathbf I)^{-1}\|_2 \geq 1/10^{M}$ with M ranging from $0$  to $-8$, and the stability region of the predictor/corrector method is enclosed by the curve in light blue. Top row:  The operator $-\Delta t\,\mathbf L_b$ in \eqref{eq:Lb}, whose eigenvalues are safely within the stability region (left), also has localized pseudo spectrum (center and right). Center row: The eigenvalues of the operator $-\Delta t\,\mathbf L_{a}$ in \eqref{eq:La} lie safely within the stability region of the method (left). The pseudospectrum, on the other hand, extends far beyond the stability region. Bottom row: Spectra (left) and pseudo spectra (center and right) of the 16 point discretization of $-\Delta t\, \mathbf L_a$ obtained using $\mathbf D^{(1)}$ in both instances of \eqref{eq:La}. The eigenvalues of the matrix lie on both positive and negative complex half planes.}}\label{fig:spectraAndpseudospectra}
\end{figure}

\begin{figure}[htb]
\begin{tabular}{ccc}
\includegraphics[width=.31\linewidth]{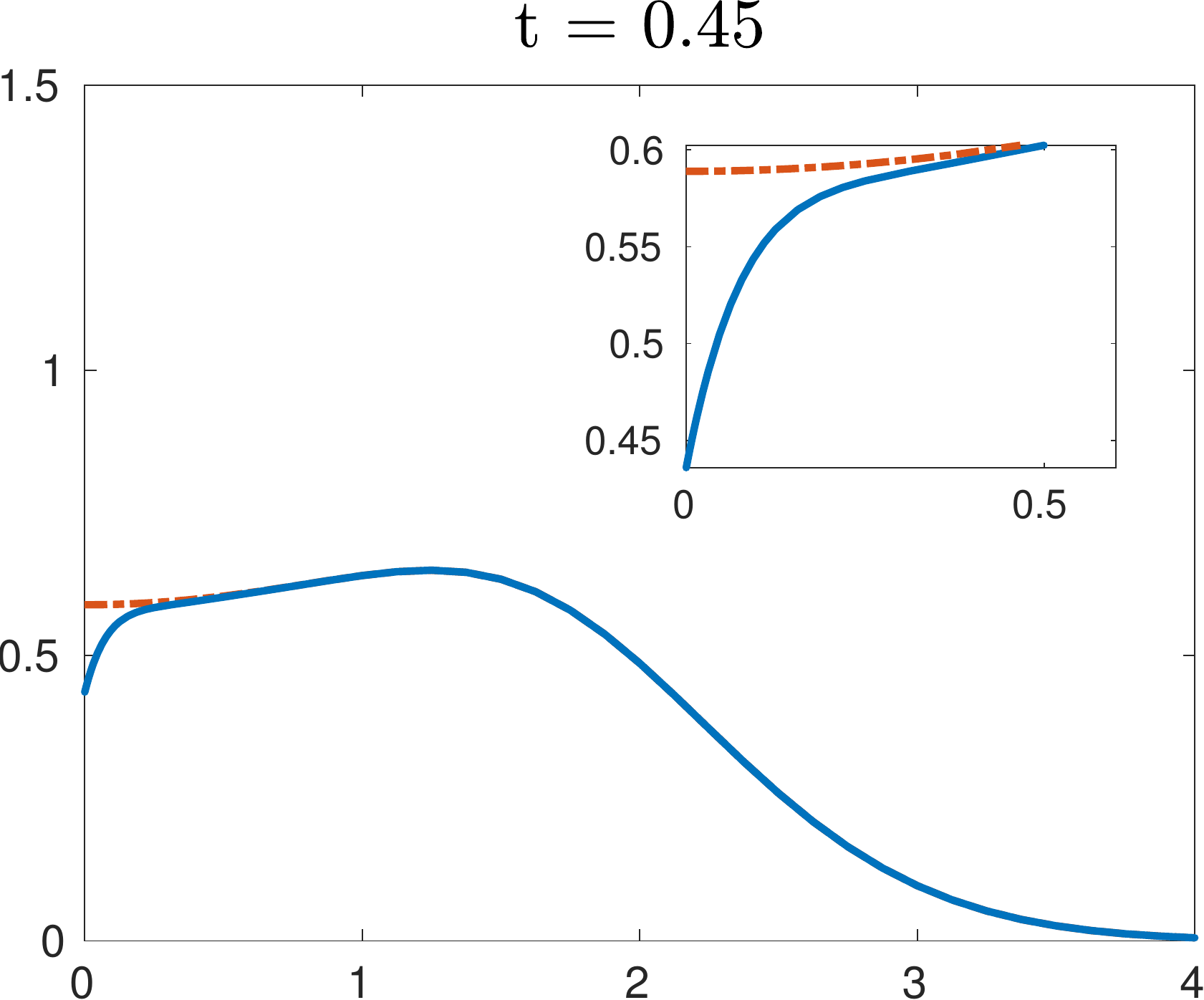} &
\includegraphics[width=.31\linewidth]{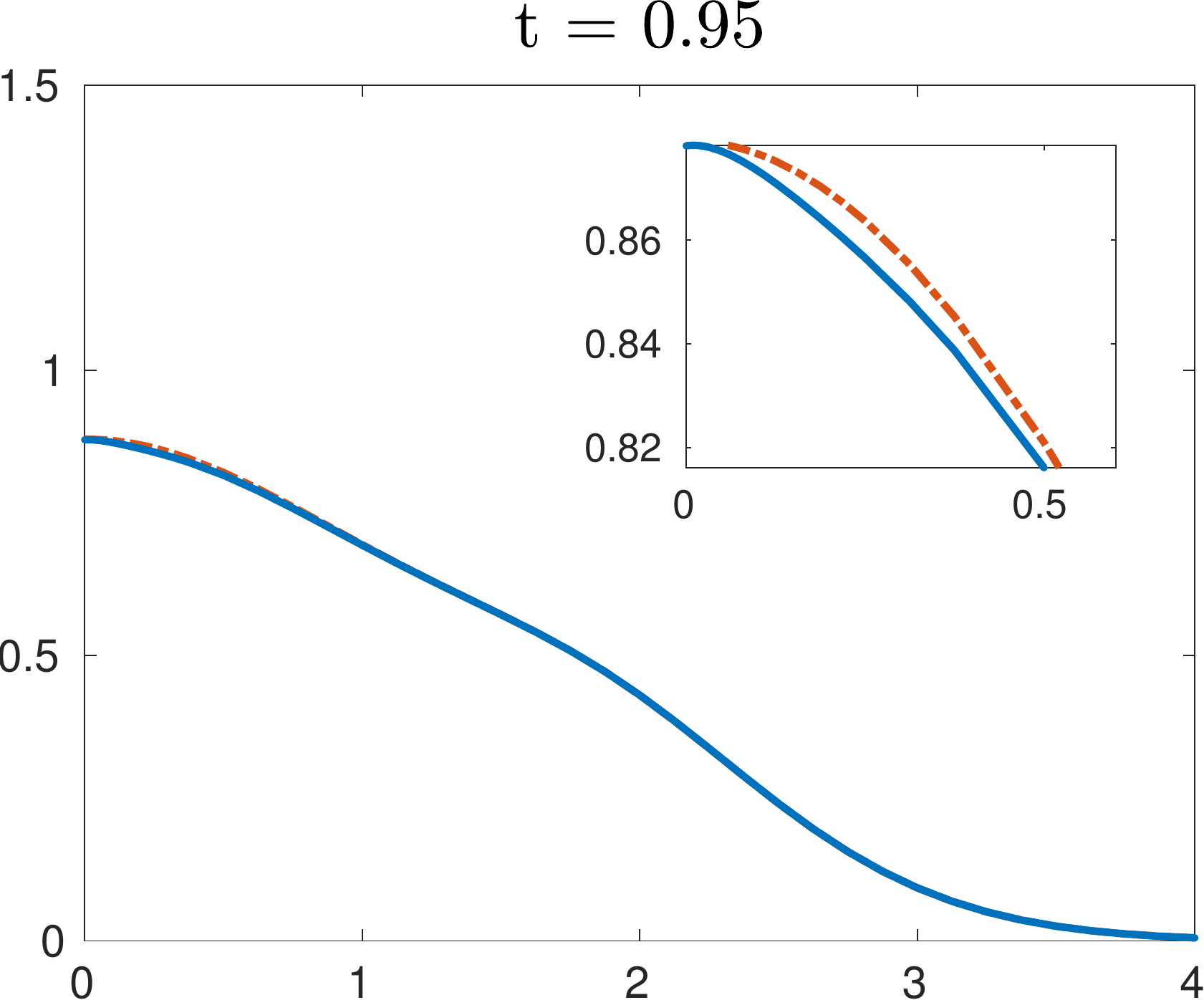} &
\includegraphics[width=.31\linewidth]{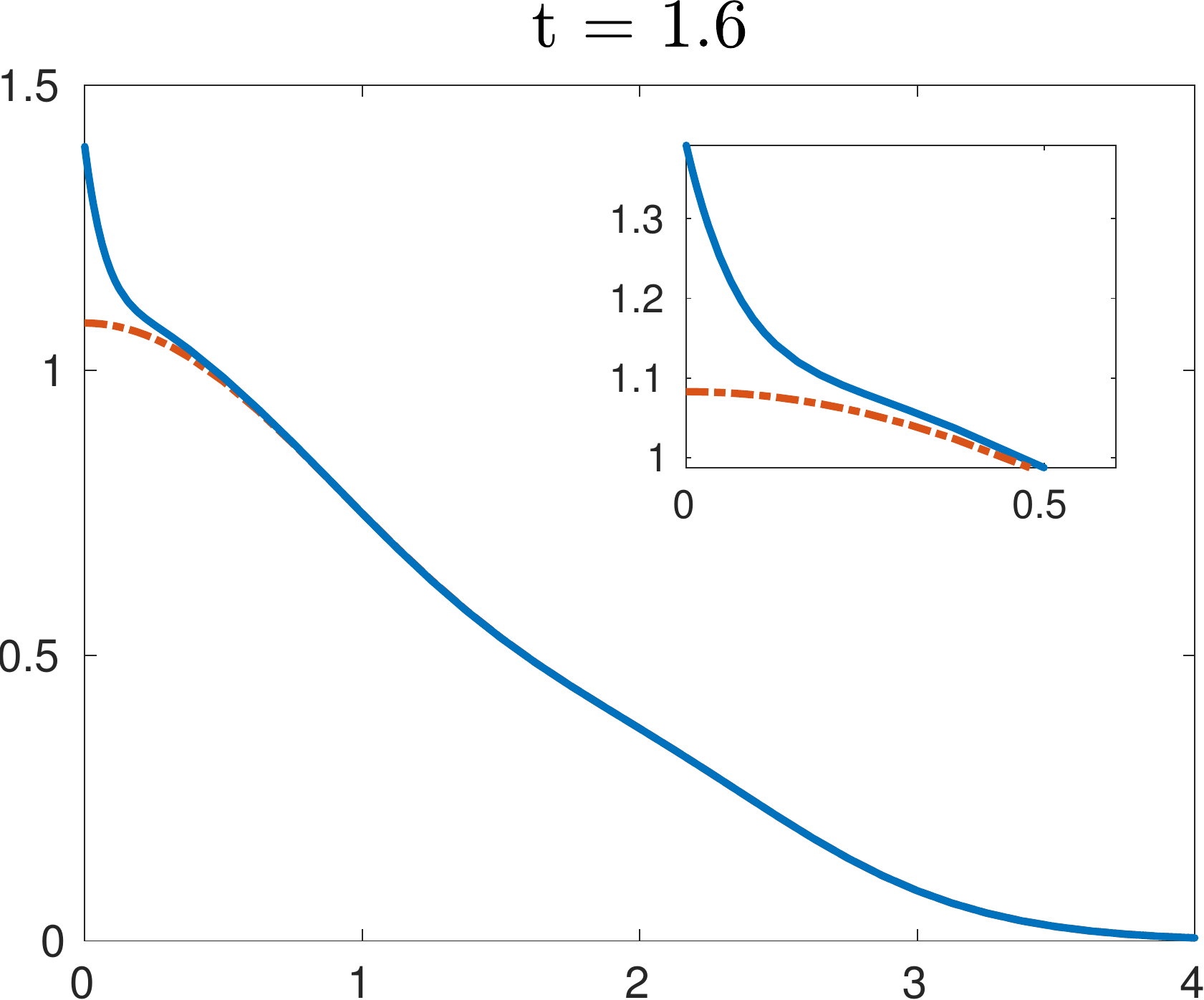} \\
\includegraphics[width=.31\linewidth]{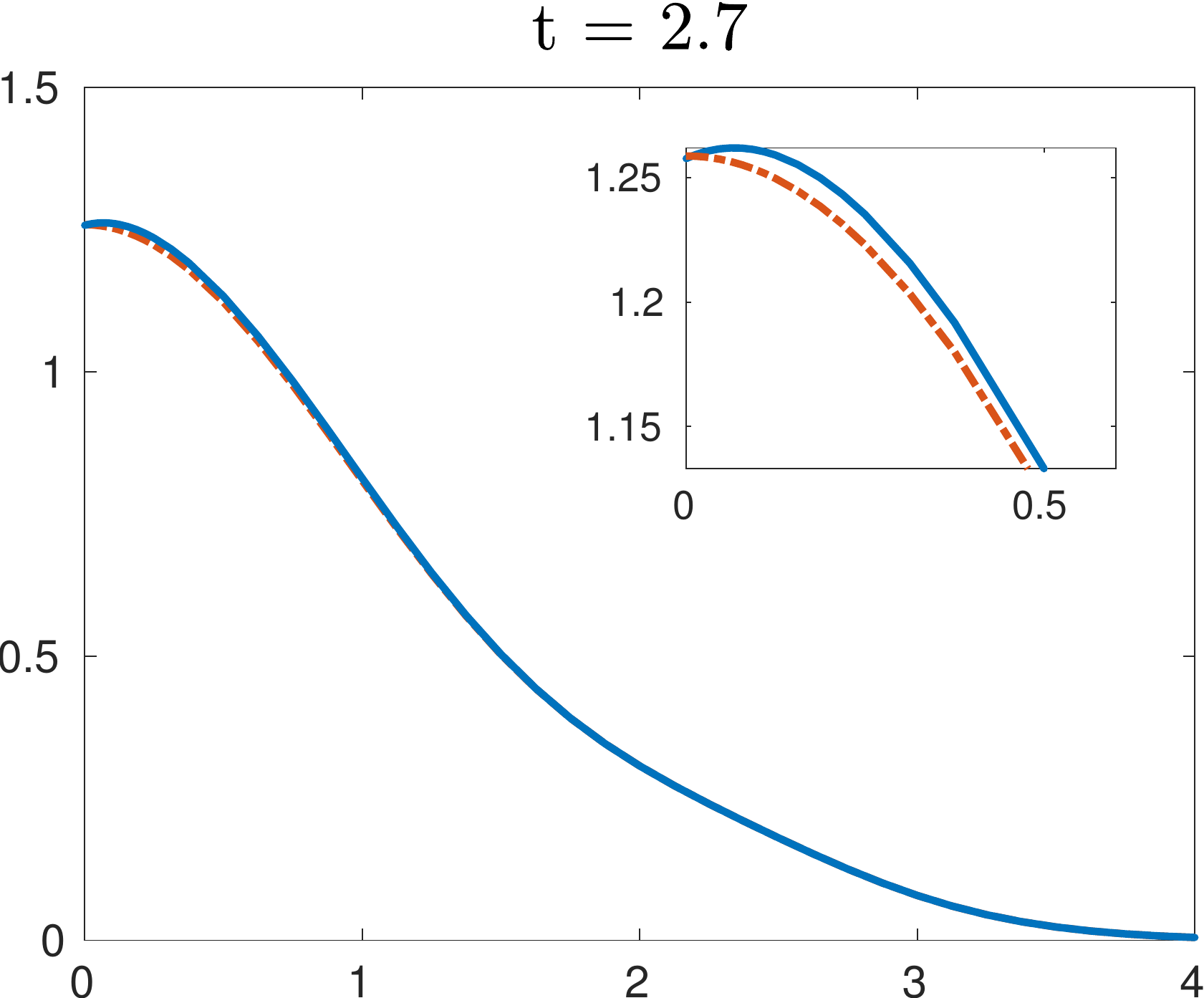} &
\includegraphics[width=.31\linewidth]{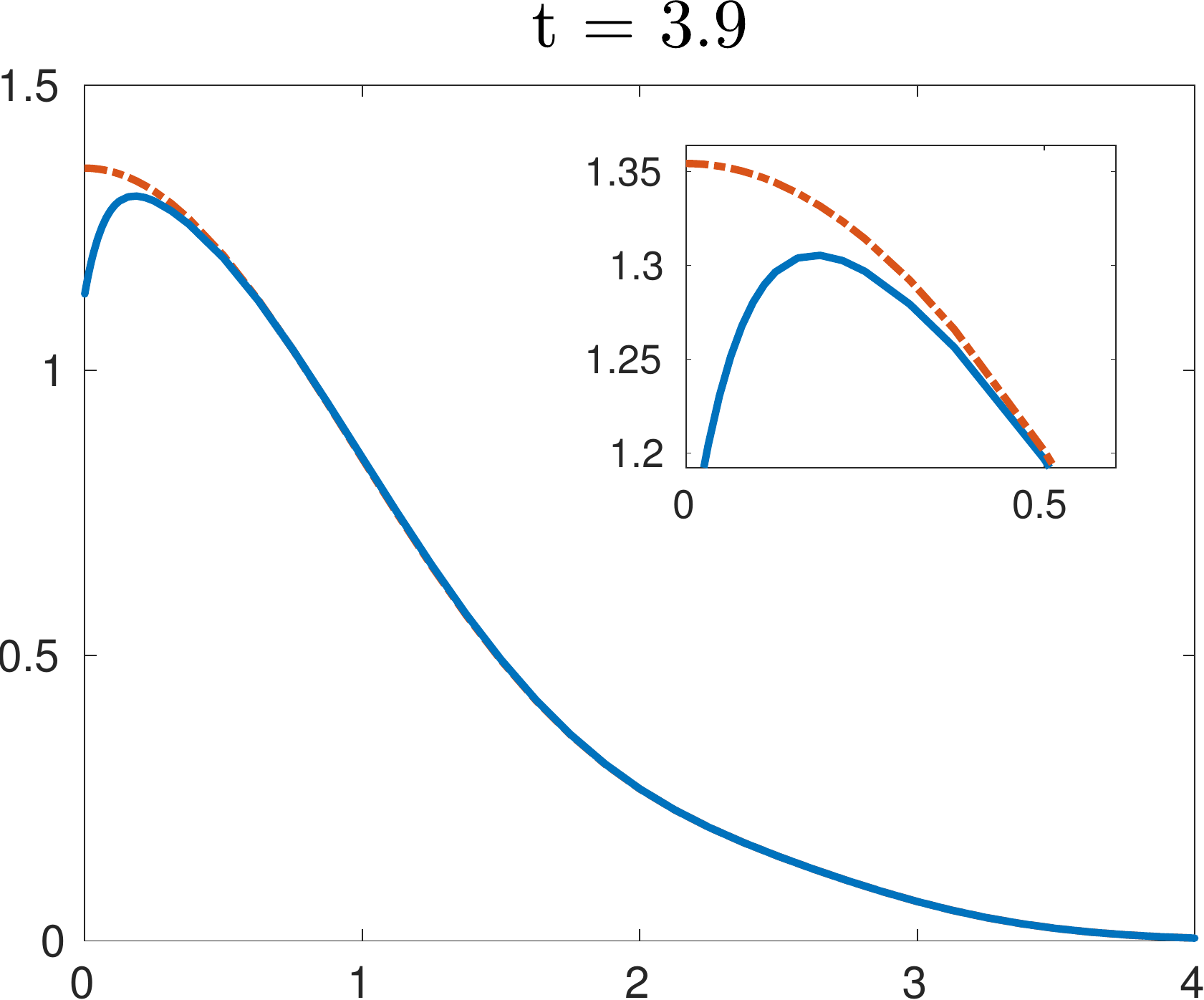} &
\includegraphics[width=.31\linewidth]{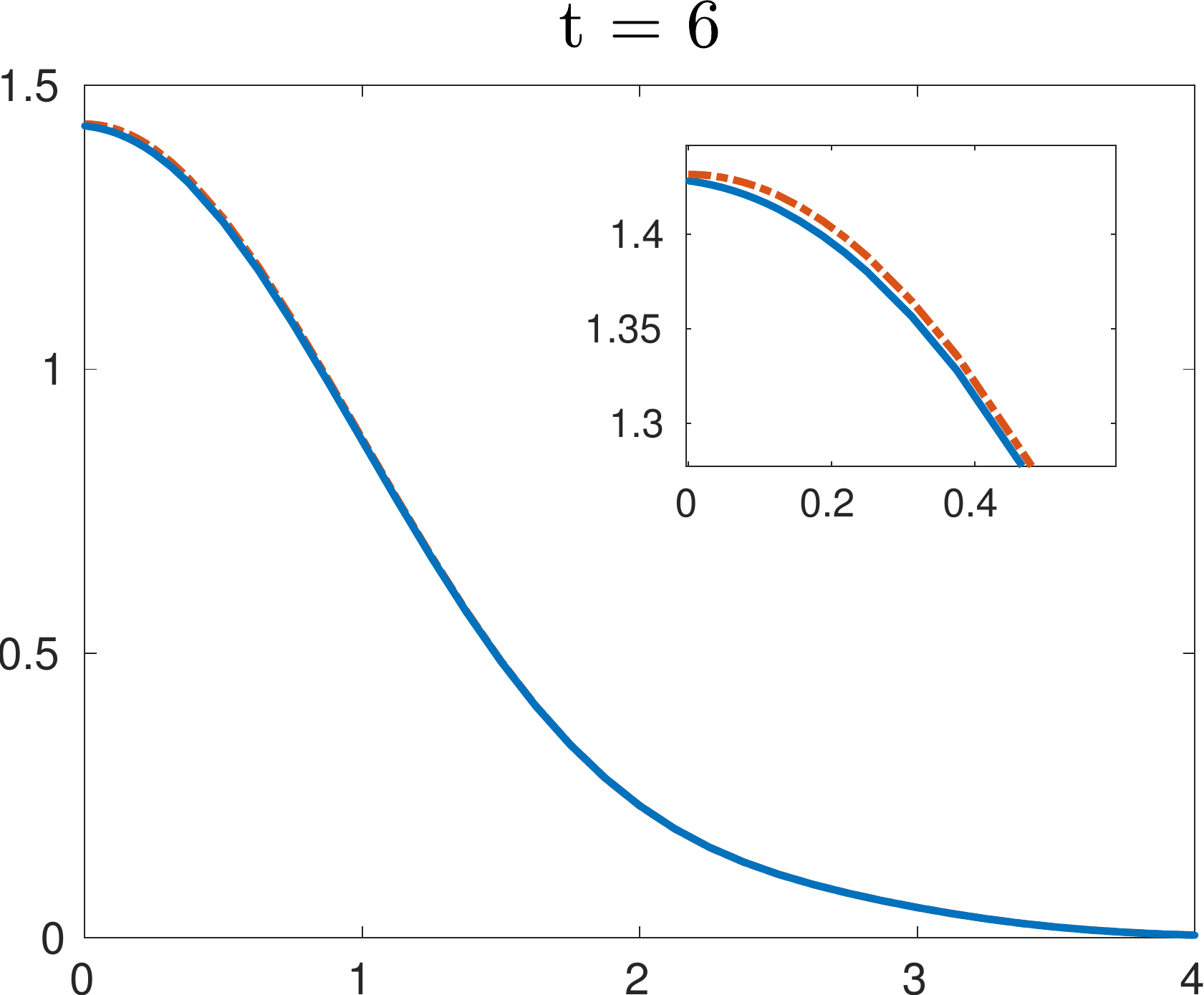}
\end{tabular}
\caption{{\scriptsize Time evolution of the solution with initial condition corresponding to Example 2 obtained with the discrete operator $-\mathbf L_a$  -- given in \eqref{eq:La} -- is shown in solid blue, while the one from the operator $\mathbf L_b$ -- as defined by \eqref{eq:Lb} -- is shown in dashed red. The operators were discretized with 16 grid points corresponding to the nodes of Maxwell polynomials defined on the half line. Second order predictor corrector time stepping with a time step $\Delta t= 5\times 10^{-5}$ was used for the time evolution. The solution obtained with the operator $\mathbf L_a$ presents unphysical oscillations at the origin that remain bounded over time. This non-modal instability is not predicted by pure eigenvalue stability analysis. A similar non-modal instability at the origin is observed for the initial condition corresponding to Example 1 when the discretization $\mathbf L_a$ is used.}}\label{fig:unstabledynamics}
\end{figure}

The last two columns in Figure \ref{fig:spectraAndpseudospectra} show the pseudo-spectra of $-\Delta t\mathbf{L}_{a}$, in the middle row, and $-\Delta t\mathbf{L}_{b}$, in the top row, for a series of values of $\epsilon$. Note that the middle column corresponds to much finer scales than the right column. We observe that the pseudo spectrum of $-\Delta t\mathbf{L}_{b}$ is well localized, and overlaps fairly well with the stability region of the predictor-corrector scheme. In contrast, the pseudo spectra of $-\Delta t\mathbf{L}_{a}$ extends considerably beyond the stability region. As a result, the dynamics of the discrete system obtained with $\mathbf{L}_{a}$ presents a non-modal instability at the origin, as shown in Figure \ref{fig:unstabledynamics}, where the time evolution of the solutions of the discretizations associated with \eqref{eq:La} and \eqref{eq:Lb} are plotted for the initial conditions corresponding to Example 2. The same behavior is observed for Example 1. We carried out the same analysis for Maxwell polynomials on a truncated interval, and obtained figures which were nearly identical to Figure \ref{fig:spectraAndpseudospectra} and Figure \ref{fig:unstabledynamics} 

These results demonstrate the superiority of using $\mathbf L_b$ for the solution of \eqref{eq:model}, and all the convergence studies presented in Section \ref{sec:validation} have consequenly been carried out with $\mathbf L_b$ as defined in \eqref{eq:Lb}. The key role played by the pseudo-spectra of the operators is yet another argument -- aside from exact mass conservation -- for the development of symmetrized Maxwell based pseudo-spectral operators \cite{Wilkening:2017,Shizgal:1992}, whose stability properties would be more predictable and easier to understand.

%

\section{Conclusions}\label{sec:conclusions}
We used the semi-analytic solution to the energy diffusion equation \eqref{eq:model} constructed by Wilkening and Cerfon to test the performance of pseudo-spectral schemes based on Maxwell polynomials for the discretization of the speed variable in time-dependent problems with Fokker-Planck collisions. Our numerical experiments confirm that these schemes provide a very competitive alternative for problems with non-singular initial conditions, often being orders of magnitude superior to other methods. Chebyshev nodes seem to be more robust for situations with singular initial conditions, but have moderate performance otherwise. Full and truncated Maxwell polynomials are equally satisfactory in terms of accuracy, but the polynomials defined on the full interval may be preferred for their flexibility, in the sense that no prior knowledge of an upper bound on the speed is required.

We discovered that the most straightforward pseudospectral discretization of the diffusion operator is subject to a non-modal instability not predicted by eigenvalue analysis, leading to significant numerical error in the vicinity of the origin. We proposed a simple discrete operator which is not subject to this instability, and which is the one we have relied upon for all the convergence studies in this article. 

With the goal of running kinetic simulations with large grid sizes on future computer architectures in mind, we see two key directions for improvement. First, symmetrization of the pseudo-spectral discrete operators \cite{Shizgal:1992} is likely to be a more robust answer to the non-modal instability we observed, to improve the overall stability properties of the operators, and to guarantee exact mass conservation. This is the subject of ongoing work by Wilkening \cite{Wilkening:2017}. Second, the development of stable and fast algorithms for the generation of the Maxwellian nodes as well as the associated Gaussian quadrature rules and barycentric weights, similar to those available for classical orthogonal polynomials, would be highly desirable for high performance solvers, but remains an open challenge.

\section*{Acknowledgments}
The authors would like to thank M. Landreman (U. Maryland) for providing the motivation for this work and for stimulating discussions, as well as Jon Wilkening (U.C. Berkeley) for his valuable comments and insights, and for providing the spectral transform data that made the convergence studies in this article possible. T.S-V and A.J.C. were supported by the U.S. Department of Energy, Office of Science, Fusion Energy
Sciences under Award Numbers DE-FG02-86ER53223 and DE-SC0012398.
\clearpage
\bibliographystyle{iopart-num}
\bibliography{ReferencesMaxwell}

\end{document}